\documentclass{ptephy_v1}
\preprintnumber{XXXX-XXXX} 

\makeatletter
\def\lastpage@putlabel{}
\makeatother
\usepackage{lastpage}
\usepackage[backref=page]{hyperref}

\hypersetup{
    colorlinks=true,
    linkcolor=blue,
    filecolor=magenta,      
    urlcolor=cyan,
    pdftitle={Cosmological gravity probes},
    pdfpagemode=FullScreen,
    }

\newcommand{\hiMpc}{h^{-1}{\rm Mpc}}
\newcommand{\simgt}{\lower.5ex\hbox{$\; \buildrel > \over \sim \;$}}
\newcommand{\simlt}{\lower.5ex\hbox{$\; \buildrel < \over \sim \;$}}

\begin{document}

\title{Cosmological gravity probes: connecting recent theoretical developments to forthcoming observations}


%
%
\author[1]{Shun Arai }
\author[2]{Katsuki Aoki}
\author[3,4]{Yuji Chinone}
\author[5]{Rampei Kimura}
\author[6]{Tsutomu Kobayashi}
\author[1,4,7]{Hironao Miyatake}
\author[8]{Daisuke Yamauchi}
\author[1,4]{Shuichiro Yokoyama}
\author[9]{Kazuyuki Akitsu}
\author[6]{Takashi Hiramatsu}
\author[10]{Shin'ichi Hirano}
\author[11]{Ryotaro Kase}
\author[12]{Taishi Katsuragawa}
\author[13]{Yosuke Kobayashi}
\author[4]{Toshiya Namikawa}
\author[2,4]{Takahiro Nishimichi}
\author[4,14]{Teppei Okumura}
\author[15]{Maresuke Shiraishi}
\author[16,17]{Masato Shirasaki}
\author[13,18]{Tomomi Sunayama}
\author[2]{Kazufumi Takahashi}
\author[2,4]{Atsushi Taruya}
\author[19,20]{Junsei Tokuda}
%
%
\affil[1]{Kobayashi-Maskawa Institute, Nagoya University, Nagoya 464-8602, Japan \email{shunarai@kmi.nagoya-u.ac.jp}}
\affil[2]{Center for Gravitational Physics and Quantum Information, Yukawa Institute for Theoretical Physics, Kyoto University, Kyoto 606-8502, Japan}
\affil[3]{QUP (WPI), KEK, Oho 1-1, Tsukuba, Ibaraki 305-0801, Japan}
\affil[4]{Kavli Institute for the Physics and Mathematics of the Universe (WPI), UTIAS, The University of Tokyo, Kashiwa, Chiba, 277-8583, Japan}
\affil[5]{Waseda Institute for Advanced Study, Waseda University,
1-6-1 Nishi-Waseda, Shinjuku, Tokyo 169-8050, Japan}
\affil[6]{Department of Physics, Rikkyo University, Toshima, Tokyo 171-8501, Japan}
\affil[7]{Institute for Advanced Research, Nagoya University, Nagoya 464-8601, Japan}
\affil[8]{Faculty of Engineering, Kanagawa University, Kanagawa, 221-8686, Japan}
\affil[9]{School of Natural Sciences, Institute for Advanced Study, 1 Einstein Drive, Princeton, NJ 08540, USA}
\affil[10]{Department of Physics, Tokyo Institute of Technology, Tokyo 152-8551, Japan}
\affil[11]{Department of Physics, Faculty of Science, Tokyo University of Science,
1-3, Kagurazaka, Shinjuku-ku, Tokyo 162-8601, Japan}
\affil[12]{Institute of Astrophysics, Central China Normal University, Wuhan 430079, China}
\affil[13]{Department of Astronomy/Steward Observatory, University of Arizona, 933 N Cherry Ave, Tucson, AZ 85721, USA}
\affil[14]{Academia Sinica Institute of Astronomy and Astrophysics (ASIAA), No. 1, Section 4, Roosevelt Road, Taipei 10617, Taiwan}
\affil[15]{School of General and Management Studies, Suwa University of Science, Chino, Nagano 391-0292, Japan}
\affil[16]{Research Enhancement Strategy Office, National Astronomical Observatory of Japan, Mitaka, Tokyo, 181-8588, Japan}
\affil[17]{School of Statistical Thinking, The Institute of Statistical Mathematics, Tachikawa, Tokyo, 190-8562, Japan}
\affil[18]{Division of Particle and Astrophysical Science, Graduate School of Science,
Nagoya University, Chikusa, Nagoya 464-8602, Japan}
\affil[19]{Center for Theoretical Physics of the Universe, Institute for Basic Science (IBS), Daejeon, 34126, Korea}
\affil[20]{Department of Physics, Kobe University, Kobe 657-8501, Japan}

\begin{abstract}%
Since the discovery of the accelerated expansion of the present Universe, significant theoretical developments have been made in the area of modified gravity. In the meantime, cosmological observations have been providing more high-quality data, allowing us to explore gravity on cosmological scales. To bridge the recent theoretical developments and observations, we present an overview of a variety of modified theories of gravity and the cosmological observables in the cosmic microwave background and large-scale structure, supplemented with a summary of predictions for cosmological observables derived from cosmological perturbations and sophisticated numerical studies. We specifically consider scalar-tensor theories in the Horndeski and DHOST family, massive gravity/bigravity, vector-tensor theories, metric-affine gravity, and cuscuton/minimally-modified gravity, and discuss the current status of those theories with emphasis on their physical motivations, validity, appealing features, the level of maturity, and calculability. We conclude that the Horndeski theory is one of the most well-developed theories of modified gravity, although several remaining issues are left for future observations. The paper aims to help to develop strategies for testing gravity with ongoing and forthcoming
cosmological observations.
\end{abstract}

\subjectindex{xxxx, xxx}

\maketitle

\tableofcontents

\section{Introduction}

In the past two decades,
the ``standard model'' of cosmology has been established
by precise measurements of the statistical properties of
the cosmic microwave background (CMB) anisotropies
and large-scale structure (LSS) of the Universe, as well as the cosmic expansion rate and the geometry with Type-Ia supernovae and baryon acoustic oscillations (BAO). 
The spatially flat $\Lambda$-cold-dark-matter ($\Lambda$CDM) model has been the concordance model of the late-time Universe, i.e., the epoch after the CMB last scattering, well explaining the interplay of cosmic expansion and structure formation.
The $\Lambda$CDM model particularly features the present accelerated expansion of the Universe
inferred from the observations of Type-Ia supernovae~\cite{SupernovaSearchTeam:1998fmf,SupernovaCosmologyProject:1998vns}, provided that it is driven by the cosmological constant.
The remarkable success of the measurements of the CMB anisotropies 
by the WMAP~\cite{WMAP:2012fli,WMAP:2012nax} and Planck~\cite{Planck:2018vyg} satellites has complemented the inference of the cosmological parameters in the $\Lambda$CDM model
with statistical uncertainties at the percent level,
giving the best-fit value of the density parameter of the cosmological constant as $\Omega_\Lambda = 0.685\pm 0.007$.\footnote{The density parameter $\Omega_K$
of the spatial curvature is assumed to be negligible. Indeed, the flatness of the Universe is tightly constrained by the latest CMB and BAO measurements to be $\Omega_K = 0.0007 \pm 0.0019$\cite{Planck:2018vyg}.}
LSS observations, 
which can
reveal the growth history of the structure in the Universe over cosmic time, has provided another inspection of the parameters of the $\Lambda$CDM model~\cite{Troster:2019ean,DAmico:2019fhj,HSC:2018mrq,Hamana:2019etx,KiDS:2020suj,DES:2021wwk}. 

The obtained $\Lambda$CDM model, however,  has been concerned with its physics.
The most well-known problem is that
extreme fine-tuning is required for
the cosmological constant to be the observed value
if it originates from the quantum vacuum energy~\cite{Weinberg:1988cp}. The existing data of CMB and LSS have not even determined whether or not the cosmic acceleration is genuinely driven by the cosmological constant.
It has recently been reported that the cosmological parameters inferred from CMB data show a discrepancy with those measured by local distance ladders (see~\cite{Verde:2019ivm,DiValentino:2021izs} for a review), presumably implying internal inconsistency of the $\Lambda$CDM model.
The physics of the late-time evolution of the Universe thus remains unresolved, 
and new observational information and further theoretical developments are awaited. 

In the meantime, gravitational physics in the context of cosmology has drawn attention
over recent years. Most attractively, one of the alternative explanations for cosmic acceleration is modifying Einstein’s general relativity (GR) on cosmological scales.
Exploring the possibilities of modified gravity amounts to examining the underlying hypothesis in the $\Lambda$CDM model that gravity is described by GR.
Small-scale tests of gravity, e.g., the solar-system experiments and decadal observations of the Hulse-Taylor pulsar 
strongly indicate
that GR works quite well on those scales. 
On the contrary, current cosmological tests are less accurate and
hence are insufficient for 
clarifying whether gravity obeys GR or not. 
It is therefore worth probing gravity on cosmological scales to understand the nature of gravity in the late-time Universe. 
%
%
%

Future cosmological observations will provide higher-quality data suitable for probing gravity on large scales.
For instance, as the ground-based CMB experiments, e.g.,
Simons Observatory and CMB Stage 4 (CMB-S4) observatory are planned,
whereas as a space mission, LiteBIRD satellite is expected to be launched
in the late 2020s.
For LSS, several forthcoming observations 
such as Subaru Prime Focus Spectroscopy (PFS), 
Vera C. Rubin Observatory’s Legacy Survey of Space and Time (LSST), 
Nancy Grace Roman Space Telescope, 
Euclid, SPHEREx, and Square Kilometre Array Observatory 
are expected 
to give us the data across wider areas and deeper resolutions
about the evolution of the structure in the Universe.

The purpose of this paper is 
to establish the goal of testing gravity on cosmological scales in order to understand the physics of the late-time Universe.
To properly extract information on gravity from observational data,  it is crucial to prepare well-motivated theories.
In addition, it is in need to select appropriate observables that are able to indicate any sign beyond GR. To this end, it is necessary to link theoretical studies on gravity with observational ones.
Here, it should be emphasised that any expertise should not be separated individually, but 
cooperate 
to handle some technical 
issues, which we aim to clarify in this paper. 

Throughout this paper, we organise studies of the concerned area for the purpose of clarifying the strategy for cosmological tests of gravity. 
After providing all the theoretical knowledge, targeted observables, and specific predictions from theories, we shall create a criterion to qualify the theories with their physical motivations, validity and appealing features, and maturity and calculability 
out of the best knowledge we have. 

The rest of the paper is organised as follows. Sec.~2 provides a collective dictionary of theories of gravity. Physical motivations and features for theories are all given. Sec.~3 provides observables for cosmological probes: CMB and LSS. We particularly focus on how the effects of gravity are captured in observables, introducing commonly-used phenomenological parameters. Sec.~4 provides concrete predictions from theories by analytical computations of cosmological perturbations. We highlight major groups of theories: scalar-tensor theories, vector-tensor theories, and massive gravity and bigravity. Sec.~5 provides computational tools which are indispensable for theoretical predictions. We introduce a Boltzmann solver and tools to analyse non-linear structure formations. Sec.~6 provides an outlook to direct future studies. We clarify which parts of the study would are more critical and highly-prior subjects. Sec.~7 summarises the paper.

\clearpage

\section{Theories of gravity}\label{sec:theories of gravity}


The purpose of this section is to review theoretical aspects of modified gravity.
Let us start by discussing briefly how one can modify
the standard theory of gravity, i.e., GR.
According to Lovelock's theorem~\cite{Lovelock:1971yv,Lovelock:1972vz},
the only possible {\em second-order} Euler-Lagrange equation is
given by the Einstein tensor plus a cosmological term if the action
is {\em diﬀeomorphism invariant} and is constructed from {\em the metric tensor alone}
in {\em four spacetime dimensions}. To modify the left-hand side of the
Einstein equations, one must therefore relax at least one of
these basic postulates of Lovelock's theorem.

Probably the simplest way of modifying gravity is adding
new dynamical degrees of freedom on top of the two tensor modes
that are already present in GR as the two polarisations of gravitational waves.
For example, scalar-tensor theories possess an additional scalar degree of freedom
and have been studied extensively over the past decades.
The scalar-tensor family is of particular importance because,
as will be seen, various theories of modified gravity can be described at least effectively by
scalar-tensor theories.
In a similar way
one can also consider vector-tensor theories and scalar-vector-tensor theories.

The second possibility is to consider higher-derivative generalisations of
GR. This can be done by adding to the Einstein-Hilbert Lagrangian
various terms constructed from the Riemann tensor.
According to Ostrogradsky's theorem~\cite{Ostrogradsky:1850fid,Woodard:2015zca},
this in general results in ghost degrees of freedom (see, e.g., \cite{Stelle:1977ry,Deruelle:2009zk}).
Even though such ghost degrees of freedom could be safe
from the viewpoint of effective field theories,
it is often preferred to avoid their appearance.
It should be noted that if the Lagrangian is a function of the Ricci scalar $R$ only,
then the resultant theory is free of Ostrogradsky ghosts.
This exceptional case is so-called $f(R)$ gravity, which
is in fact equivalent to a certain class of scalar-tensor theories
and will be discussed in some depth below.

The third possibility is (partially) breaking diffeomorphism invariance.
For example, one can consider spatially covariant theories of gravity in which
time diffeomorphism invariance is broken. 
This would be a natural setup in the context of cosmology 
because there exists the preferred time slicing in the universe.
In this case,
however, full diffeomorphism invariance can be recovered by introducing
a St\"{u}ckelberg scalar field, and hence such theories are basically
equivalent to scalar-tensor theories.
Gravity with broken diffeomorphism invariance is also closely related
to massive gravity theories.

The fourth possibility is assuming more than four spacetime dimensions.
Traditional Kaluza-Klein theory is in this family. In the effective four-dimensional
description, it gives rise to Kaluza-Klein scalar and vector modes,
and hence it is essentially a theory with additional dynamical degrees of freedom.
A more non-trivial example is given by the braneworld scenario
such as the Arkani-Hamed-Dimopoulos-Dvali (ADD)~\cite{Arkani-Hamed:1998jmv}
and Randall-Sundrum (RS)~\cite{Randall:1999vf,Randall:1999ee} models,
in which our four-dimensional universe is realised as a brane embedded
in a higher-dimensional spacetime.
The Dvali-Gabadadze-Porrati model~\cite{Dvali:2000hr} is particularly interesting,
because it not only admits a self-accelerating universe that could be
an alternative to dark energy~\cite{Deffayet:2001pu}, but also yields a cubic galileon theory
as an effective theory on the brane~\cite{Nicolis:2004qq}.

The fifth possibility is to change the geometrical interpretation of gravity. 
GR and most modified gravity theories are assumed to be based on Riemannian geometry,
but non-Riemannian geometry can be considered as a foundation of gravity.
Gravity based on the metric-affine geometry is called metric-affine gravity. 
In this setup, the Riemannian geometry is treated as a low-energy effective description of the spacetime.

\begin{table}[t]
\caption{List of theories of modified gravity that we will discuss in this review. We classify the theories into five classes according to their dynamical degrees of freedom (DOFs), and also indicate which assumptions of the Lovelock theorem are not satisfied (only four-dimensional theories are considered in this review). The first column indicates whether Euler-Lagrange equations are second-order differential equations (in the case of vector-tensor theories, after the St\"uckelberg scalar are introduced), the second column expresses whether diffeomorphism invariance is broken, and the third column shows whether a theory does not contain extra (dynamical or auxiliary) degrees of freedom other than the metric $g_{\mu\nu}$.  In the last column, each number corresponds to the number of DOFs of massless tensor field (2), massless/massive scalar field (1), massive tensor field (5), and massive vector field (3), respectively. There are two exceptions: (i) a special case of Lorentz-violating massive gravity only has 2 DOFs although gravitational waves acquire a non-vanishing mass and (ii) the number of DOFs in metric-affine gravity has not been fully understood yet.}
\label{table:theories}
  \centering
  \begin{tabular}{lcccc}
  \hline
   & 2nd-order & Diff-invariance & Metric only & DOFs \\
  \hline \hline 
  General Relativity & \checkmark & \checkmark & \checkmark &  2 \\
  \hline 
  \multicolumn{5}{l}{{\it Modified gravity with a scalar DOF} (Sec.~\ref{sec:Scalar-tensor theories})} \\
  \quad Horndeski & \checkmark & \checkmark & $\times$ & $2+1$ \\
  \quad DHOST & $\times$ & \checkmark & $\times$ & $2+1$ \\
  \quad $f(R)$ & $\times$ & \checkmark & \checkmark & $2+1$ \\
  \hline
  \multicolumn{5}{l}{{\it Modified gravity with a massive graviton} (Sec.~\ref{sec:Massive gravity and bigravity})} \\
  \quad dRGT & \checkmark & $\times$ & \checkmark & $5$ \\
  \quad Mass-varying/quasi-dilaton & \checkmark & $\times$ & $\times$ & $5+1$ \\
  \quad Translation breaking & \checkmark & $\times$ & \checkmark & $5$ \\
  \quad Lorentz-violating & \checkmark & $\times$ & \checkmark & $5$ or $2$ \\
  \quad Bigravity & \checkmark & \checkmark & $\times$ & $2+5$ \\
    \hline
  \multicolumn{5}{l}{{\it Modified gravity with a vectorial DOF} (Sec.~\ref{sec:Vector-tensor theories})} \\
\quad Generaized Proca & \checkmark & \checkmark & $\times$ & $2+3$ \\
\quad Extended vector & $\times$  & \checkmark & $\times$ & $2+3$ \\
    \hline
  \multicolumn{5}{l}{{\it Modified gravity based on non-Riemannian geometry} (Sec.~\ref{sec:Metric-affine gravity})} \\
\quad Metric-affine & \checkmark & \checkmark & $\times$ & N/A \\
    \hline
  \multicolumn{5}{l}{{\it Modified gravity without new DOF} (Sec.~\ref{sec:Cuscuton and minimally modified gravity})}  \\
\quad Cuscuton & \checkmark & \checkmark & $\times$ & 2 \\
\quad Minimally modified & \checkmark & $\times$ & \checkmark & 2 \\
    \hline
  \end{tabular}
\end{table}

In this section, 
we provide a concise dictionary of gravity theories as summarised in Table~\ref{table:theories}.
We will review scalar-tensor theories such as the Horndeski theory and the degenerate higher-order
scalar-tensor (DHOST) theory in Sec.~\ref{sec:Scalar-tensor theories}, massive gravity and bigravity in Sec.~\ref{sec:Massive gravity and bigravity}, and
vector-tensor theories in Sec.~\ref{sec:Vector-tensor theories}.
As yet other possible ways of modifying gravity, metric-affine gravity and cuscuton/minimally modified gravity will be introduced
in Secs.~\ref{sec:Metric-affine gravity} and \ref{sec:Cuscuton and minimally modified gravity}, respectively.
Having reviewed those theories, we will discuss two theoretical aspects of modified gravity concerning small-scale physics. The theories are thought of as infrared modifications of gravity (gravity is modified at cosmological distances), but meanwhile, the predictions of GR should be recovered at small distances to evade the solar-system constraints. The restoration to GR can be achieved by a {\it screening mechanism}. 
In Sec.~\ref{sec:screening}, we will
discuss the Vainshtein, chameleon, and symmetron mechanisms as representative screening mechanisms.
Furthermore, even if GR is recovered, GR is not ultraviolet complete. Therefore, modified gravity, as well as GR, should be regarded as a low-energy effective field theory of an (unknown) ultraviolet completion of gravity. {\it Positivity bounds} provide necessary conditions to have an ultraviolet completion under certain assumptions which we will discuss in Sec.~\ref{sec:Positivity bound}.

\subsection{Scalar-tensor theories}
\label{sec:Scalar-tensor theories}

Scalar-tensor theories are modified gravity theories with one scalar and two tensor degrees of freedom.
This class of modified gravity has been studied extensively in the past decades.
An ancient example is the Jordan-Brans-Dicke theory~\cite{Brans:1961six} whose Lagrangian is given by
\begin{align}
    {\cal L}=\frac{1}{2}\left[\phi R-\frac{\omega}{\phi}(\partial\phi)^2\right],
    \label{Lag:Brans-Dicke}
\end{align}
where $\omega$ is a constant parameter. (In this expression, $\phi$ has dimensions of (mass)$^2$.)
More recently, many efforts have been devoted to exploring general frameworks
of scalar-tensor theories that are free of Ostrogradsky ghosts.
In this subsection, we review the Horndeski and degenerate higher-order scalar-tensor (DHOST)
theories as general frameworks for healthy scalar-tensor theories,
as well as $f(R)$ gravity, which is another well-studied example of modified gravity
and is closely related to a certain scalar-tensor theory.
See also Ref.~\cite{Kobayashi:2019hrl} for a comprehensive review
on the Horndeski theory and its generalisations.

\subsubsection{Horndeski theory}
\label{sec:Horndeski}

Suppose that the Lagrangian for a scalar-tensor theory is given by
the metric, the scalar field, and their derivatives:
\begin{align}
    {\cal L}={\cal L}(g_{\mu\nu},\partial_\lambda g_{\mu\nu},
    \partial_\lambda\partial_\rho g_{\mu\nu},\cdots,
    \phi, \partial_\mu\phi, \partial_\mu\partial_\nu\phi,\cdots).
\end{align}
If the resultant Euler-Lagrange equations for the metric and the scalar field 
are of second order, then the theory is obviously free of Ostrogradsky ghosts.
Already in 1974, Horndeski determined the most general Lagrangian
yielding the second-order equations of motion both for the metric and the scalar field~\cite{Horndeski:1974wa}.
The Lagrangian of the Horndeski theory is given by
\begin{align}
    {\cal L}_{\rm Horndeski}&=G_2(\phi,X)-G_3(\phi,X)\Box\phi + G_4(\phi,X)R
    +G_{4X}  \left[(\Box\phi)^2-\phi^{\mu\nu}\phi_{\mu\nu}\right]
  \notag \\ & \quad
  +G_5(\phi,X) G^{\mu\nu}\phi_{\mu\nu}
  -\frac{G_{5X}}{6}\left[
  (\Box\phi)^3-3\Box\phi\phi^{\mu\nu}\phi_{\mu\nu}
  +2\phi_{\mu\nu}\phi^{\nu\lambda}\phi_\lambda^\mu
  \right]
    ,
    \label{Lag:Horndeski}
\end{align}
where $G_2$, $G_3$, $G_4$, and $G_5$ are arbitrary functions of
$\phi$ and $X:=-g^{\mu\nu}\partial_\mu\phi\partial_\nu\phi$/2, and
we introduced the notation $\phi_{\mu\nu}:=\nabla_\mu\nabla_\nu\phi$ and $g_X := \partial g/ \partial X$ for a function $g$ of $X$.
In fact, the expression~\eqref{Lag:Horndeski},
which is now frequently used in the literature,
is the one found more recently
in a different context
in the course of generalising the galileon theory~\cite{Deffayet:2011gz}.
The original Lagrangian that Horndeski discovered has a different form,
and from the derivation it is not clear at first sight that
the generalised galileon theory~\eqref{Lag:Horndeski} is equivalent to
the Horndeski theory.
The equivalence was first proven in~\cite{Kobayashi:2011nu}.

A number of scalar-tensor theories studied in the literature belong to the Horndeski family.
Choosing the arbitrary functions in Eq.~\eqref{Lag:Horndeski} appropriately, 
one can reproduce any specific second-order scalar-tensor theory.
When $G_4={\rm const.}=M_{\rm Pl}^2/2$ and others are taken to be zero,
the standard Einstein-Hilbert action can be reproduced.
By taking $G_2=Z(\phi)X+U(\phi)$ and $G_4=F(\phi )$, it can be seen as
generalisations of the Jordan-Brans-Dicke theory~\eqref{Lag:Brans-Dicke}.
As will be seen later, the $f(R)$ gravity can be recast as second-order scalar-tensor
and hence is a subclass of the Horndeski theory.
The $G_2$ term gives the well-known action describing the k-inflation~\cite{Armendariz-Picon:1999hyi} and k-essence~\cite{Chiba:1999ka},
and the $G_3$ term is investigated in the context of kinetic gravity braiding~\cite{Deffayet:2010qz} and G-inflation~\cite{Kobayashi:2010cm}.
We note that the effective theory of the DGP braneworld model~\cite{Dvali:2000hr} naturally includes the $G_3$ terms such as $X\Box\phi$ called cubic galileon~\cite{Nicolis:2008in}.
One of the most non-trivial examples is the non-minimal coupling to
the Gauss-Bonnet term,
\begin{align}
  \xi(\phi)\left(R^2-4R_{\mu\nu}R^{\mu\nu}
  +R_{\mu\nu\rho\sigma}R^{\mu\nu\rho\sigma}\right),\label{eq:nonminimal_GB}
\end{align}
which corresponds to the following choice of the Horndeski functions~\cite{Kobayashi:2011nu}:
\begin{align}
  &G_2=8\xi^{(4)}X^2\left(3-\ln X\right),
  \quad  G_3=4\xi^{(3)}X\left(7-3\ln X\right),
  \notag \\
  &G_4=4\xi^{(2)}X\left(2-\ln X\right),
  \quad\;\; G_5=-4\xi^{(1)} \ln X,\label{eq:GB_Galileon_expression}
\end{align}
where $\xi^{(n)}=\partial^n\xi/\partial\phi^n$.

\subsubsection{Degenerate higher-order scalar-tensor theories}
\label{sec:DHOST}

Since the equations of motion in the Horndeski theory are of second order,
it evades the Ostrogradsky instability by construction. However, having
second-order equations is not a necessary condition for a theory to be
free of Ostrogradsky ghosts. The trick is that if the equations of motion
are degenerate, the system contains less number of dynamical degrees of freedom
than is expected from the derivative order of the equations of motion~\cite{Motohashi:2016ftl,Klein:2016aiq,Motohashi:2017eya,Motohashi:2018pxg,Crisostomi:2017aim}.
This idea leads us to consider degenerate higher-order scalar-tensor (DHOST) theories
beyond Horndeski
that propagate one scalar and two tensor degrees of freedom
and hence are Ostrogradsky-stable
even though the Euler-Lagrange equations are of higher order.

The first example of DHOST theories was obtained by means of
a disformal transformation~\cite{Bekenstein:1992pj},
\begin{align}
    g_{\mu\nu}\to \Omega(\phi,X)g_{\mu\nu}+\Gamma(\phi,X)\phi_\mu\phi_\nu,
    \label{dttrs}
\end{align}
from the Horndeski theory~\cite{Zumalacarregui:2013pma}, with $\phi_\mu :=\partial_\mu\phi$.
This transformation in general yields higher-derivative terms in the equations of motion,
but the number of dynamical degrees of freedom remains the same
as long as the transformation is invertible~\cite{Takahashi:2017zgr}.
A further extension of the DHOST theories along this direction was made in \cite{Babichev:2019twf,Babichev:2021bim,Minamitsuji:2021dkf,Takahashi:2021ttd,Naruko:2022vuh,Takahashi:2022mew} with a higher-derivative generalisation of the disformal transformation.

Systematic constructions of DHOST theories have been developed in~\cite{Langlois:2015cwa,Langlois:2015skt,Crisostomi:2016czh,BenAchour:2016cay,BenAchour:2016fzp},
assuming that the Lagrangian depends quadratically and cubically on the second derivatives of the scalar field $\phi_{\mu\nu}$.
A physically and phenomenologically interesting class is given by
the following subset of the quadratic DHOST theories:
\begin{align}
{\cal L}_{\rm qDHOST}=G_2(\phi,X)-G_3(\phi,X)\Box\phi+
f(\phi,X)R+\sum_{I=1}^5A_I(\phi,X)L_I,\label{qDHOSTL1}
\end{align}
where
\begin{align}
&L_1=\phi_{\mu\nu}\phi^{\mu\nu},\quad L_2=(\Box\phi)^2,
\quad L_3=\Box\phi \phi^\mu\phi^\nu\phi_{\mu\nu}
\notag \\ &
L_4=\phi^\mu\phi_{\mu\alpha}\phi^{\alpha \nu}\phi_\nu,
\quad
L_5=(\phi^\mu\phi^\nu\phi_{\mu\nu})^2.
\end{align}
The above Lagrangian represents all the possible contractions of the second-order derivatives $\phi_{\mu\nu}$ with
the metric $g_{\mu\nu}$ and the scalar field gradient $\phi_\mu$.
Although the Lagrangian \eqref{qDHOSTL1} contains $8$ arbitrary functions of $\phi$ and $X$, the functions cannot be chosen arbitrarily.
In order that the resultant theory contains a single scalar degree of freedom, the functions except for $G_2$ and $G_3$ must satisfy
the degeneracy conditions.
The degeneracy conditions read
\begin{align}
    A_2&=-A_1,\label{degcond1}
    \\
    A_4&=\frac{1}{2(f+2XA_1)}\Bigl[
			8XA_1^3+\left( 3f+16Xf_X\right) A_1^2-X^2fA_3^2+2X\left( 4Xf_X-3f\right) A_1A_3
    \notag\\
	&       +2f_X\left( 3f+4Xf_X\right) A_1+2\left( Xf_X-f\right) A_3+3ff_X^2
		\Bigr],
    \\
    A_5&=-\frac{\left( f_X+A_1+XA_3\right)\left( 2fA_3-f_XA_1-A_1^2+3XA_1A_3\right)}{2\left( f+2XA_1\right)^2},\label{degcond3}
\end{align}
with $f+2XA_1\neq 0$. The arbitrary functions are taken to be $f$, $A_1$, and $A_3$,
and then $A_2$, $A_4$, and $A_5$ are determined accordingly from
the degeneracy conditions.
The conditions~\eqref{degcond1}--\eqref{degcond3} ensure
that the metric and scalar sectors are degenerate.
This class is called class Ia in the terminology of~\cite{Langlois:2015cwa,Crisostomi:2016czh,BenAchour:2016cay}.

The Horndeski theory is obtained as the special case
with $A_1=-A_2=-f_X$ and $A_3=A_4=A_5=0$.
The so-called Gleyzes-Langlois-Piazza-Vernizzi (GLPV) theory~\cite{Gleyzes:2014dya,Gleyzes:2014qga}, which is also studied frequently in the literature,
corresponds to the case $A_1=-A_2=f_X+XA_3$, which is equivalent to $A_4=-A_3$ and $A_5=0$.
More importantly, all class Ia DHOST theories can be generated
through the disformal transformation~\eqref{dttrs} from
the Horndeski Lagrangian with $G_5=0$~\cite{BenAchour:2016cay,Crisostomi:2016czh}.
This, however, does not mean that class Ia DHOST theories are simply equivalent to
the Horndeski theory because the story here is entirely about vacuum spacetime.
In the presence of matter, DHOST and Horndeski theories are clearly inequivalent.

The DHOST theory can be further extended to the so-called U-DHOST theory~\cite{DeFelice:2018ewo,DeFelice:2021hps,DeFelice:2022xvq}. In the DHOST theory, the degeneracy conditions hold for arbitrary gauge choices. However, when the theory is regarded as an EFT, the validity of the EFT generically depends on the background configuration of the field, and some scalar-tensor theories may be valid only when the gradient of the scalar field is timelike. In such a theory, it would be sufficient to impose the degeneracy conditions only in the unitarity gauge $\phi=\phi(t)$, leading to the U-DHOST theory. Away from the unitarity gauge (while keeping to assume a timelike gradient $\partial_{\mu}\phi$), the degeneracy conditions do not hold and there apparently exists an extra mode, called the shadowy mode in~\cite{DeFelice:2018ewo,DeFelice:2021hps,DeFelice:2022xvq}. However, the shadowy mode is non-propagating as it satisfies an elliptic differential equation, so the U-DHOST theory is Ostrogradsky-stable.

Finally, we should add a comment on the constraints from the detection of gravitational waves from the binary neutron star merger  GW170817~\cite{LIGOScientific:2017vwq}, observed by the LIGO/Virgo collaboration, 
and its optical counterpart gamma-ray burst (GRB) 170817A~\cite{LIGOScientific:2017zic,LIGOScientific:2017ync}.
The combination of these observations  gives
a remarkably precise measurement of the propagation speed of gravitational waves: it is compatible with the speed of light with deviations
smaller than $\text{a few} \times 10^{-15}$. 
Since the speed of gravitational waves with respect to a cosmological background can be computed for all DHOST theories,
it is a straightforward exercise to identify the DHOST theory that survives after GW170817.
In particular, the propagation speed of gravitational waves in the class Ia DHOST theory is given by
\begin{align}
	c_{\rm GW}^2=\frac{f}{f+2XA_1}
	\,.
\end{align}
Hence, the requirement $c_{\rm GW}=c$ for any background imposes the simple condition $A_1=0$~\cite{Langlois:2017dyl}.\footnote{
In many dark energy models, the typical strong coupling scale is given by $ (M_{\rm Pl} H_0^2)^{1/3} \sim 260$~Hz which is close to the LIGO frequency scale.
Hence, the LIGO frequency may be beyond the regime of validity of the theory and more careful investigations are necessary to impose $A_1=0$ for such a theory~\cite{deRham:2018red}.}
The Lagrangian of the viable subclass in the class Ia DHOST theories reduces to
\begin{align}
	{\cal L}_{{\rm qDHOST}}^{c_{\rm GW}=c}
		=&G_2(\phi,X)-G_3(\phi,X)\Box\phi+f(\phi,X)R
			+A_3(\phi,X)\Box\phi \phi^\mu\phi^\nu\phi_{\mu\nu}
	\notag\\
		&+\frac{1}{2f}\Bigl[3f_X^2+2\left( Xf_X-f\right) A_3-X^2A_3^2\Bigr]\phi^\mu\phi_{\mu\alpha}\phi^{\alpha \nu}\phi_\nu
	\notag\\
		&-\frac{A_3\left(f_X+XA_3\right)}{f}(\phi^\mu\phi^\nu\phi_{\mu\nu})^2
	\,.\label{eq:LqDHOST cGW=c}
\end{align}
Moreover, \cite{Creminelli:2018xsv} has pointed out that 
the stability of graviton against decay into scalar field constrains
the structure of the DHOST theory.
The Lagrangian of the DHOST theory satisfying these two constraints obtained from
the gravitational waves is given by
\begin{align}
    {\cal L}_{\rm qDHOST}^{c_{\rm GW}=c,\text{no-decay}}
        =G_2(\phi ,X)-G_3(\phi ,X)\Box\phi
        +f(\phi ,X)R+\frac{3f_X^2}{2f}\phi^\mu\phi_{\mu\alpha}\phi^{\alpha\nu}\phi_\nu
    \,.\label{eq:LqDHOST cGW=c nodecay}
\end{align}
One can also specialise the above results to the Horndeski theories, characterised by $f=G_4$, $A_1=-A_2=2G_{4X}$, $A_3=A_4=A_5=0$.
Combining this with the condition $A_1=0$, we then obtain the reduced Lagrangian as
\begin{align}
	{\cal L}_{{\rm Horndeski}}^{c_{\rm GW}=c}
		=G_2(\phi,X)-G_3(\phi,X)\Box\phi+f(\phi)R
	\,.
\end{align}

\subsubsection{\texorpdfstring{$f(R)$}{fR} gravity}\label{subsec:fr}

The Lagrangian for {$f(R)$} gravity is given by an arbitrary function of
the Ricci scalar~\cite{Buchdahl:1970ynr,Starobinsky:1980te}:
\begin{align}
    S=\int {\rm d}^4x\sqrt{-g}f(R).\label{fr01}
\end{align}
By introducing an auxiliary field $\phi$,
this can be written equivalently as
\begin{align}
    S=\int {\rm d}^4x\sqrt{-g}\left[f(\phi)+f_\phi(R-\phi)\right],\label{fr02}
\end{align}
where $f_\phi = {\rm d} f / {\rm d} \phi$.
The equation of motion for $\phi$ reads $f_{\phi\phi}(R-\phi)=0$,
and hence one has $\phi=R$ provided that $f_{\phi\phi}\neq 0$.
Substituting $\phi=R$ back to Eq.~\eqref{fr02}, one obtains the original action~\eqref{fr01}.
Therefore, the two expressions are indeed equivalent.
Now, it can be seen that the action~\eqref{fr02}
is that of a particular scalar-tensor theory. In terms of the Horndeski functions,
this corresponds to $G_2=f-\phi f_\phi$ and $G_4=f_\phi$ with $G_3=G_5=0$.

Performing a conformal transformation provides further insight into
$f(R)$ gravity. In terms of the conformally transformed metric
$\widetilde g_{\mu\nu}:=f_\phi g_{\mu\nu}$, the action~\eqref{fr01}
can be written as
\begin{align}
    S=\int {\rm d}^4x\sqrt{-\widetilde g}\left[ 
    \widetilde R-\frac{3}{2}\widetilde g^{\mu\nu}\partial_\mu\ln f_\phi \partial_\nu\ln f_\phi 
    -\frac{1}{f_\phi^2}\left(\phi f_\phi-f\right)
    \right],\label{fr03}
\end{align}
where $\widetilde R$ is the Ricci tensor constructed from $\widetilde g_{\mu\nu}$.
(Here we assumed that $f_\phi>0$ for simplicity.)
A further field redefinition $\varphi=\ln f_\phi$ recasts the action~\eqref{fr03}
into a more standard form with the Einstein-Hilbert term plus a scalar field
with the linear kinetic term and the potential.
This is called the Einstein frame action.
Note that if the matter is coupled only to $g_{\mu\nu}$ in the original frame,
then it is non-minimally coupled to $\phi$ in the Einstein frame.

\subsection{Massive gravity and bigravity}
\label{sec:Massive gravity and bigravity}

\subsubsection{dRGT massive gravity}
Although a graviton in GR is massless as a consequence of general covariance, 
whether the graviton is massless or massive is an interesting question in both theoretical and experimental points of view.
Starting from the pioneering work by Fierz and Pauli in 1939~\cite{Fierz:1939ix}, the theoretical construction of consistent gravitational theories of
a massive spin-2 field has attracted considerable attention in theoretical physics. The Fierz-Pauli (FP) theory which describes a massive spin-2 field is given by
\begin{align}
	S_{\rm FP}=\int {\rm d}^4x 
	\biggl[
	-\frac{1}{4}h^{\mu\nu}{\cal E}^{\alpha\beta}_{~~~\mu\nu}h_{\alpha\beta}
	-\frac{m^2}{4} (h_{\mu\nu}h^{\mu\nu}-h^2)
	+\frac{1}{M_{\rm Pl}}h_{\mu\nu}T^{\mu\nu}
	\biggr],
	\label{FPaction}
\end{align}
where $h_{\mu\nu}$ is a rank-2 symmetric tensor field, which can be thought as the fluctuation tensor around the Minkowski metric $g_{\mu\nu}=\eta_{\mu\nu}+h_{\mu\nu}/M_{\rm Pl}$, ${\cal E}^{\mu\nu\alpha\beta}$ is the linearised Einstein-Hilbert kinetic operator defined as 
\begin{align}
{\cal E}^{\mu\nu}_{~~\alpha\beta}
= 
\left[ \eta^{(\mu}_{~\alpha}\eta^{\nu)}_{~\beta}-\eta^{\mu\nu}\eta_{\alpha\beta}\right] \square
-2 \partial^{(\mu}\partial_{(\alpha}\eta^{\nu)}_{~\beta)}
+\partial^\mu \partial^\nu \eta_{\alpha\beta} + \partial_\alpha \partial_\beta \eta^{\mu\nu} \,,
\label{HR_lin_kin_op}
\end{align}
and $m$ is the mass of the graviton. Due to the presence of the mass term in Eq.~\eqref{FPaction}, the gauge symmetry, $h_{\mu\nu} \to h_{\mu\nu} + \partial_\mu \xi_\nu + \partial_\nu \xi_\mu$, which is present in the case of linearised GR carrying two degrees of freedom, is explicitly broken, and four additional degrees of freedom.
In total, general massive gravity has six degrees of freedom. On the other hand, a massive spin-2 particle should have five degrees of freedom corresponding to helicity $\pm 2, \pm 1, 0$ modes. The additional sixth mode is a ghost.
However, the FP
tuning in the mass term, $h_{\mu\nu}h^{\mu\nu}-h^2$, guarantees that there are no more than five degrees of freedom propagating, and thus this dangerous degree of freedom is absent on flat space. 
One would naively expect that it recovers GR as the mass of the graviton goes to zero. However, one finds order-one deviations from GR in the linear theory~\eqref{FPaction}, which is known as the van Dam-Veltman-Zakharov (vDVZ) discontinuity~\cite{Zakharov:1970cc,V1}. 
Meanwhile, this discontinuity turned out to be an artifact of the truncation at linear order, and it has the continuous massless limit when taking into account non-linearities by embedding the FP mass term into GR as pointed out by Vainshtein~\cite{Vainshtein}. In contrast with this successful mechanism, non-linearities reintroduce the sixth degree of freedom called the Boulware-Deser (BD) ghost~\cite{PhysRevD.6.3368}. In the decoupling limit description, the scalar mode of a massive graviton contains higher derivative self-interactions, and this is the origin of the BD ghost leading to Ostrogradsky's instabilities. It had been long thought that Lorentz-invariant massive gravity theories are plagued by ghost instabilities (e.g.,~\cite{Creminelli:2005qk}). Surprisingly, in 2010, de Rham and Gabadadze performed a systematic construction of the Lagrangian which is free from the BD ghost by carefully choosing the mass terms consisting of an infinite series of interactions, and all higher derivative interactions vanish thanks to the total derivative properties in the decoupling limit~\cite{deRham:2010ik}. The infinite series of potential terms can be resumed into a compact form by introducing the square root of a rank-2 tensor~\cite{deRham:2010kj} (see~\cite{deRham:2014zqa,Hinterbichler:2011tt} for a review). The absence of the BD ghost in a full theory has been proven in the Hamiltonian formalism~\cite{Hassan:2011hr,Mirbabayi:2011aa,Hassan:2012qv}.  This ghost-free non-linear massive gravity, referred to as the de Rham-Gabadadze-Tolley (dRGT) theory,
is described by the following action,\footnote{The original action of the dRGT potential is written in terms of ${\cal K}^{\mu}_{~\nu}=\delta^{\mu}_{~\nu}-(\sqrt{g^{-1}f})^\mu_{~\nu}$ as $\sum_{n=0}^4\,{\alpha}_n \,e_n({\cal K})$ where $\alpha_n$ are constant parameters related with $\beta_n$. We adopt the form in Eq.~\eqref{action:dRGT} throughout this paper. } 
\begin{align}
S_{\rm dRGT} = \int {\rm d}^4x\,\sqrt{-{g}}\frac{M_{\rm Pl}^2}{2}\left[{R} -2m^2\sum_{n=0}^3\,{\beta}_n \,e_n(\sqrt{{g}^{-1}f})\right] + S_{\rm m}  [g, \psi] \,,
\label{action:dRGT}
\end{align}
where $\beta_n$ are constant parameters, $f_{\mu\nu}$ is called the fiducial metric, and $e_n ({\cal Q})$ are the symmetric polynomials given by
\begin{align}
e_0 ({\cal Q})
=& 1\,, \\
e_1 ({\cal Q})=& [{\cal Q}]\,, \\
e_2 ({\cal Q})=& \frac{1}{2!} \left([{\cal Q}]^2-[{\cal Q}^2]\right)\,, \\
e_3 ({\cal Q})=& \frac{1}{3!} \left([{\cal Q}]^3-3[{\cal Q}][{\cal Q}^2]
+2[{\cal Q}^3] ) \right)\,, \\
e_4 ({\cal Q})=& \frac{1}{4!} \left(
[{\cal Q}]^4-6[{\cal Q}]^2[{\cal Q}^2]
+3[{\cal Q}^2]^2+8[{\cal Q}][{\cal Q}^3]-6[{\cal Q}^4]\right)\,.
\label{eq:elementarypoly}
\end{align}
Here, $[{\cal Q}]$ denotes the trace of the matrix ${\cal Q}$ and 
a square root of the matrix represents the matrix that satisfies 
$\sqrt{{\cal Q}}^\mu_{~\rho}\sqrt{{\cal Q}}^\rho_{~\nu} = {\cal Q}^{\mu}_{~\nu}$.
Although the simplest choice of the fiducial metric would be the Minkowski metric, 
it can be generalised to an arbitrary metric~\cite{Hassan:2011tf}. In this paper, we mainly focus on the case where $f_{\mu\nu}=\eta_{\mu\nu}$. Note that in this case the tensor inside a square root can be written as $(g^{-1}f)^\mu_{~\nu}=\delta^\mu_{~\nu}-h^\mu_{~\nu}$, and the mass term up to quadratic order after expanding in terms of $h_{\mu\nu}$ gives the FP mass term.

Although the general covariance is broken in massive gravity, one can restore the gauge symmetry by introducing a set of four scalar fields called the St\"uckelberg scalars $\phi^a$ via~\cite{Arkani-Hamed:2002bjr}
\begin{align}
f_{\mu\nu}=\eta_{ab}\partial_{\mu}\phi^a \partial_{\nu}\phi^b\,.
\label{fiducial_metric}
\end{align}
Note that the fiducial metric~\eqref{fiducial_metric} is manifestly invariant under the Poincar{\'e} symmetry in the internal St\"uckelberg field space. 
Choosing the gauge $\phi^a=\delta^a_{~\mu}x^\mu$ called {\it the unitary gauge}, the fiducial metric reduces to the Minkowski metric.
To see the connection with the galileon theories, let us consider the decoupling limit which captures physics at distances in the range $\Lambda_3^{-1} := (M_{\rm Pl}m^2)^{-1/3} \lesssim r \ll m^{-1}$. We then introduce the small fluctuation of the St\"uckelberg field around the unitary gauge and the metric around the Minkowski spacetime as
$\phi^a =\delta^a_{~\mu}x^\mu -\eta^{a\mu}A_\mu / M_{\rm Pl} m-\eta^{a\mu}\partial_\mu \pi/ M_{\rm Pl} m^2$ and $g_{\mu\nu}=\eta_{\mu\nu} +h_{\mu\nu}/M_{\rm Pl}$, where $\pi$, $A_\mu$, and $h_{\mu\nu}$ respectively represent the helicity 0, 1, 2 components of the massive graviton. 
The decoupling limit is defined in the following way:
\begin{align}
	m \to 0, \quad M_{\rm Pl} \to \infty, \quad T \to \infty, \quad \Lambda_3~{\rm and}~\frac{T}{M_{\rm Pl}}~{\rm are~fixed}.
\end{align}
Then, the scalar-tensor parts of an effective field theory with a strong coupling scale $\Lambda_3$ are given by
\begin{align}
	&S_{\rm DL}=\int {\rm d}^4x \biggl[-\frac{1}{4} \tilde h_{\mu\nu}{\cal E}^{\mu\nu,\alpha\beta} \tilde h_{\alpha\beta}
	+\sum_{n=2}^{5}\frac{\tilde \beta_n}{(\Lambda_3^3)^{n-2}}{\cal L}^{(n)}_{\rm galileon}
	+\frac{\tilde \beta_X}{\Lambda_3^6} \tilde h^{\mu\nu}  X^{(3)}_{\mu\nu}\nonumber\\
	&~~~~~~~~~~~~~~~~~
	+\frac{1}{M_{\rm Pl}} \tilde h_{\mu\nu}T^{\mu\nu}
	+\frac{1}{M_{\rm Pl}} \pi T-\frac{\tilde \beta_T}{2\Lambda_3^3M_{\rm Pl}}\partial_\mu \pi\partial_\nu \pi T^{\mu\nu}
	\biggr],
	\label{action:DL}
\end{align}
where $\Pi_{\mu\nu}:=\partial_{\mu}\partial_{\nu}\pi$, $\tilde \beta_i$ and  $\tilde \beta_{X, T}$ are functions of the dRGT parameters $\beta_i$, 
\begin{align}
	{\cal L}_{\rm galileon}^{(2)}=&-\frac{1}{2}(\partial\pi)^2,\\
	{\cal L}_{\rm galileon}^{(3)}=&-\frac{1}{2}(\partial\pi)^2[\Pi],\\
	{\cal L}_{\rm galileon}^{(4)}=&-\frac{1}{2}(\partial\pi)^2\left([\Pi]^2-[\Pi^2]\right),\\
	{\cal L}_{\rm galileon}^{(5)}=&-\frac{1}{2}(\partial\pi)^2\left([\Pi]^3-3[\Pi][\Pi^2]+2[\Pi^3]\right),\\
	X^{(3)}_{\mu\nu} =& \left([\Pi]^3-3[\Pi][\Pi^2]+2[\Pi^3]\right)\eta_{\mu\nu}-3\left([\Pi]^2-[\Pi^2]\right)\Pi_{\mu\nu}+6[\Pi]\Pi_{\mu\nu}^2-6\Pi_{\mu\nu}^3\,,
\end{align}
and 
\begin{align}
\tilde h_{\mu\nu}= h_{\mu\nu} + \pi\eta_{\mu\nu} -\frac{\tilde\beta_T}{\Lambda_3^3} \pi \Pi_{\mu\nu}\,.
\end{align}
The purely scalar derivative interactions, ${\cal L}_{\rm galileon}^{(n)}$, are the covariant galileon Lagrangian in a flat spacetime~\cite{Nicolis:2008in} and the action \eqref{action:DL} belongs to the decoupling limit in the Horndeski theory~\cite{Koyama:2013paa}. 
This galileon structure prevents the BD ghost from appearing in massive gravity, and the massless limit smoothly connects to GR through the Vainshtein mechanism as will be explained later. 

\subsubsection{Extensions of dRGT massive gravity}
As will be explained in Sec.~\ref{sec:Perturbations in massive gravity theories}, the dRGT theory cannot unfortunately accommodate stable cosmological solutions. 
For this reason, many works have been done in the literature to seek extended theories of massive gravity which allows healthy homogeneous and isotropic cosmological background. Here, we outline the simplest modification of the dRGT theory involving a scalar degree of freedom. The basic idea of modifications is based on promoting the dRGT constant parameters to be functions of a scalar field $\sigma$ in a way such that the BD ghost is absent. 

The simplest example is the mass-varying dRGT theory in which mass parameter $m$ is a function of an scalar field $\sigma$~\cite{DAmico:2011eto,Huang:2012pe}, 
\begin{align}
	S_\text{mass-varying} &= \int {\rm d}^4x\,\sqrt{-{g}}\frac{M_{\rm Pl}^2}{2}\left[{R} -2m^2(\sigma)\sum_{n=0}^4\,{\beta}_n \,e_n(\sqrt{{g}^{-1}f})
	-\frac{1}{2} (\partial\sigma)^2 -V(\sigma)\right] 
 \notag \\ &\quad 
 + S_{\rm m}  [g, \psi] \,. 
	\label{action:mass-varying}
\end{align}
Mass-varying massive gravity can be further extended by introducing multiple scalar fields and non-minimal coupling with the Einstein-Hilbert action, which is intensively investigated in~\cite{Huang:2013mha}.

There is
another example called quasi-dilaton massive gravity, in which a new global symmetry for a scalar field, 
$\sigma \to \sigma -\alpha M_{\rm Pl}$ and $\phi^a \to e^\alpha \phi^a $ where $\alpha$ is a constant parameter, is imposed~\cite{DAmico:2012hia}. This restricts the fiducial metric as
\begin{align}
	f_{\mu\nu}^{(\rm QD)}=e^{2\sigma / M_{\rm Pl}}\eta_{ab}\partial_{\mu}\phi^a \partial_{\nu}\phi^b \,.
\end{align}
Inspired from the dRGT theory, the action which respects the new global symmetry is given by 
\begin{align}
	S_{\rm quasi-dilaton} = \int {\rm d}^4x\,\sqrt{-{g}}\frac{M_{\rm Pl}^2}{2}\left[{R} -2m^2\sum_{n=0}^4\,{\beta}_n \,e_n(\sqrt{{g}^{-1}f^{(\rm QD)}})
	-\frac{1}{2} (\partial\sigma)^2\right] + S_{\rm m}  [g, \psi] \,. 
	\label{action:quasi-dilaton}
\end{align}
The quasi-dilaton theory can be interpreted that the parameter $\beta_n$ is promoted to a function of the scalar field while the modification of the mass term only appears in the mass parameter $m$ in the mass-varying theory. 
The quasi-dilaton theory can be further extended by redefining the new fiducial metric $\bar f_{\mu\nu}^{(\rm QD)} :=  f_{\mu\nu}^{(\rm QD)} -(\alpha_\sigma /M_{\rm Pl}^2m^2)e^{-2\sigma / M_{\rm Pl}} \partial_\mu \sigma \partial_\nu \sigma$ where $\alpha_\sigma$ is a constant parameter, which still satisfies the global symmetry~\cite{DeFelice:2013tsa}. Furthermore, these theories~\eqref{action:mass-varying} and~\eqref{action:quasi-dilaton} can accommodate the Horndeski action for the new scalar field $\sigma$, which is investigated in the context of stable Vainshtein solutions in the quasi-dilaton theory~\cite{Gabadadze:2014gba}. 

\subsubsection{Translation breaking theories}
Although the previous examples~\eqref{action:mass-varying} and~\eqref{action:quasi-dilaton} carry six degrees of freedom, a number of attempts to seek massive gravity theories with five degrees of freedom beyond the dRGT theory have been also performed in the literature \cite{Gao:2014jja,deRham:2013tfa,Hinterbichler:2013eza,Folkerts:2011ev,Kimura:2013ika}. The ghost-free examples have been investigated in~\cite{Gumrukcuoglu:2020utx,deRham:2014lqa}. The idea to extend the dRGT theory is abandoning the global translation invariance  $\phi^a \to \phi^a + c$ while keeping the global Lorentz invariance. This relaxation allows to include the new Lorentz-invariant function of the 
St\"uckelberg fields, e.g., $X:= \phi^a\phi_a$ and $Y^\mu_{~\nu} := g^{\mu\alpha} \phi_a  \phi_b \partial_\alpha \phi^a\partial_\nu\phi^b$. In~\cite{Gumrukcuoglu:2020utx}, two novel classes of massive gravity theories with five degrees of freedom have been proposed. 

The first class is an extension of generalised massive gravity (GMG)~\cite{deRham:2014lqa} with a non-minimal coupling to the St\"uckelberg fields, which is given by
\begin{align}
	S_{\rm GMG} = \int {\rm d}^4x\,\sqrt{-{g}}\frac{M_{\rm Pl}^2}{2} \left[ 
	GR+ 
	\frac{6 G_X^2}{G} [Y]
	-2m^2\sum_{n=0}^3\,{\beta}_n(X) \,e_n\left(\sqrt{{g}^{-1}\tilde{f}}\right)\right] + S_{\rm m}  [g, \psi] \,. 
	\label{action:GMG}
\end{align}
where $G=G(X)$, $G_X=\partial G/\partial X$, and ${\tilde f}_{\mu\nu} = (\eta_{ab} + D(X)\,\phi_a\phi_b)\,\partial_\mu \phi^a\,\partial_\nu \phi^b$
is the disformally distorted fiducial metric. This reduces to the dRGT theory when $G=1$ and $D=0$.

The second class can be constructed using the projected fiducial metric ${\bar f}_{\mu\nu} = P_{ab}\,\partial_\mu \phi^a\,\partial_\nu \phi^b$, where $P_{ab} = \eta_{ab} - {\phi_a\phi_b /X}$ is the projection tensor, which manifestly eliminates one of the St\"uckelberg~fields along $\phi^a$. The action for the projected theory is given by
\begin{align}
S_{\rm PMG}=
\int {\rm d}^4 x \sqrt{-g} \frac{M_{\rm Pl}^2}{2}\Biggl[ G \,R  + \frac{6 G_X^2}{G} [Y] 
+ m^2  \,{\cal U} \bigl(X, [Z], [Z^2], [Z^3] \bigr)
\Biggr]+ S_{\rm m}  [g, \psi]\,,
\label{action:PMG}
\end{align}
where $Z^\mu_{~\nu}= (g^{-1}{\bar f})^\mu_{~\nu}$.
The crucial difference between this theory and the dRGT theory~\eqref{action:dRGT}/GMG theory~\eqref{action:GMG} is that
in the projected theory
the potential for a massive graviton is no longer a function of the square-root tensor, but is an arbitrary function of $X$ and $[Z^n]$.
In the GMG theory and the projected massive gravity theory,
a non-minimal coupling to the Einstein-Hilbert term, which was absent in the dRGT case,
is allowed as a consequence of the translation breaking of the St\"uckelberg fields.

\subsubsection{Lorentz-violating massive gravity}
\label{Sec.th.LVMG}
The theoretical difficulty for avoiding the appearance of the BD ghost can be easily solved by Lorentz-violating graviton masses~\cite{Rubakov:2004eb,Dubovsky:2004sg}. One of the phenomenologically interesting and intensively studied Lorentz-violating massive gravity is the minimal theory of massive gravity (MTMG).\footnote{The action for the MTMG theory is too long to display here. See~\cite{DeFelice:2015moy} for the explicit expression for the action.} This theory shares the same background equations for an FLRW background of the dRGT theory, and the gravitational waves acquire a non-vanishing mass while it has only two gravitational degrees of freedom at a fully non-linear level, i.e., the scalar and vector modes are absent~\cite{DeFelice:2015hla,DeFelice:2015moy}. 
 The similar Lorentz-violating extensions for quasi-dilation and bigravity can be found in the literature~\cite{DeFelice:2020ecp, Minamitsuji:2022vfv,DeFelice:2017wel,Gumrukcuoglu:2017ioy,DeFelice:2017rli,DeFelice:2018vkc}. 
Another class of Lorentz-violating massive gravity is obtained when the spacetime is filled with a solid-type material~\cite{Endlich:2012pz,Aoki:2022ipw}. In this case, the temporal diffeomorphism invariance is preserved in a different way from the other massive gravity theories.

\subsubsection{Massive bigravity theory}
\label{Sec:bigravity}
Other interesting extension of massive gravity theories is to promote the fiducial metric in the dRGT theory to be a dynamical variable by adding an extra Einstein-Hilbert action for $f_{\mu\nu}$ metric. This is called the Hassan-Rosen ghost-free bigravity theory, whose action is given by~\cite{Hassan:2011zd}
\begin{align}
S_{\rm HR}=&
\frac{M_{\rm Pl}^2}{2}\int {\rm d}^4x\sqrt{-g} R[g]
+\frac{\kappa M_{\rm Pl}^2}{2}\int d^4x\sqrt{-f} R[f]
 \nonumber\\ 
&+m^2 M_{\rm Pl}^2\int d^4x\sqrt{-g} \sum_{n=0}^4\,{\beta}_n \,e_n(\sqrt{{g}^{-1}f})+S_{\rm m}[g,\psi_g]+S_{\rm m}[f,\psi_f],
%
	\label{action:HR}
\end{align}
where $R[g]$ and $R[f]$ respectively denotes the scalar curvature for $g_{\mu\nu}$ and $f_{\mu\nu}$, $\kappa$ represents the ratio of the squared Planck masses for $g_{\mu\nu}$ and $f_{\mu\nu}$, and the matter field $S_{\rm m}$ can be coupled to either or both metric fields. 
When we linearise both metrics around Minkowski spacetime as  $g_{\mu\nu}= \eta_{\mu\nu} + h_{\mu\nu}$ and $f_{\mu\nu}= \eta_{\mu\nu} + l_{\mu\nu}$, it becomes manifest that the physical degrees of freedom can be decomposed into five from the massive spin-2 field and two from the massless spin-2 field.
This has been also confirmed in the Hamiltonian analysis based on the $3+1$ decomposition~\cite{Hassan:2011zd,Hassan:2011ea}.  

\subsection{Vector-tensor theories}
\label{sec:Vector-tensor theories}

In analogous to the attempts to construct gravitational theories beyond GR 
by introducing a scalar field as explained in Sec.~\ref{sec:Scalar-tensor theories},
there are also attempts to 
construct modified gravitational theories by introducing a vector field 
instead of a scalar field. The most general action in the presence of an Abelian 
vector field with a non-minimal coupling to gravity is derived by Horndeski 
in 1976~\cite{Horndeski:1976gi}. 
This action is constructed under the conditions that the equations of motion are kept up to 
second order for derivatives and the standard Maxwell equations are recovered on Minkowski 
spacetime. In a manner analogous to Galileon gravity in scalar-tensor theories,  
there are attempts to generalise Abelian vector field theories by introducing derivative 
self-interactions~\cite{Deffayet:2010zh,Tasinato:2013oja}, while it is shown that 
the Galileon-type interactions are not  allowed for a single vector field as long as 
both the $U(1)$ gauge invariance and the Lorentz invariance are preserved 
on the flat space-time~\cite{Deffayet:2013tca}. However, this no-go theorem 
cannot be applied to a vector field with a mass term, i.e. Proca field, 
due to the fact that the existence of the mass breaks the $U(1)$ gauge invariance 
which is one of the assumptions of the no-go theorem. 
In Proca theories, there is a longitudinal propagating degree of freedom originating from 
the broken gauge symmetry in addition to the two transverse polarisations. 
In Ref.~\cite{Heisenberg:2014rta}, the derivative self-interactions are introduced 
into Proca theories by keeping the number of propagating degrees of freedom, 
and the resultant theory is dubbed the generalised Proca (GP) theory (see also 
Refs.~\cite{Allys:2015sht,BeltranJimenez:2016rff,Allys:2016jaq,Heisenberg:2018vsk}). 
On the curved spacetime with a dynamical metric, 
GP theories are described by the action, 
\begin{equation}
S=\int\,{\rm d}^4x\sqrt{-g}\sum^6_{i=2}{\cal L}_i\,,
\label{GPact}
\end{equation}
where 
\begin{align}
{\cal L}_2 =&\,\, 
G_2(X,F,U,Y)\,,
\label{L2}\\
{\cal L}_3 =&\,\,  G_3(X) \nabla_{\mu}A^{\mu}\,,
\label{L3}\\
{\cal L}_4 =&\,\, 
G_4(X)R+
G_{4,X}(X) \left[ (\nabla_{\mu} A^{\mu})^2
-\nabla_{\rho}A_{\sigma}
\nabla^{\sigma}A^{\rho} \right]\,,\label{L4} \\
{\cal L}_5 =&\,\, 
G_{5}(X) G_{\mu \nu} \nabla^{\mu} A^{\nu}
-\frac16 G_{5,X}(X) [ (\nabla_{\mu} A^{\mu})^3
-3\nabla_{\mu} A^{\mu}
\nabla_{\rho}A_{\sigma} \nabla^{\sigma}A^{\rho}
+2\nabla_{\rho}A_{\sigma} \nabla^{\gamma}
A^{\rho} \nabla^{\sigma}A_{\gamma}] \nonumber \\
& -g_5(X) \tilde{F}^{\alpha \mu}
{\tilde{F^{\beta}}}_{\mu} \nabla_{\alpha} A_{\beta}\,,
\label{L5}\\
{\cal L}_6 =&\,\,  G_6(X) L^{\mu \nu \alpha \beta}
\nabla_{\mu}A_{\nu} \nabla_{\alpha}A_{\beta}
+\frac12 G_{6,X}(X) \tilde{F}^{\alpha \beta} \tilde{F}^{\mu \nu}
\nabla_{\alpha}A_{\mu} \nabla_{\beta}A_{\nu}\,.
\label{L6}
\end{align}
Here, 
$A_{\mu}$ is a vector field with the strength 
$F_{\mu \nu}=\nabla_{\mu}A_{\nu}-\nabla_{\nu}A_{\mu}$. 
The function $G_2$ depends on the following four quantities, 
\begin{equation}
X =-\frac12 A_{\mu} A^{\mu}\,,\qquad
F = -\frac14 F_{\mu \nu} F^{\mu \nu}\,,\qquad
U = -\frac14 F_{\mu \nu} \tilde{F}^{\mu \nu}\,,\qquad
Y = A^{\mu}A^{\nu} {F_{\mu}}^{\alpha}
F_{\nu \alpha}\,,
\end{equation}
where $\tilde{F}^{\mu \nu}$ is the dual strength tensor defined by 
\begin{equation}
\tilde{F}^{\mu \nu}=\frac12 \epsilon^{\mu \nu \alpha \beta}
F_{\alpha \beta}\,,
\end{equation}
with the Levi-Civita tensor $\epsilon^{\mu \nu \rho \sigma}$. 
The other functions, $G_{3,4,5,6}$ and $g_5$, are arbitrary functions of $X$ alone 
with the notation $G_{i,X} := \partial G_{i}/\partial X$. 
The Lagrangians ${\cal L}_{4,5,6}$ contain the non-minimal couplings 
where the Ricci scaƒlar $R$, the Einstein tensor $G_{\mu\nu}$, 
and the double dual Riemann tensor $L^{\mu \nu \alpha \beta}$ defined by 
\begin{equation}
L^{\mu \nu \alpha \beta}=\frac14 \epsilon^{\mu \nu \rho \sigma}
\epsilon^{\alpha \beta \gamma \delta} R_{\rho \sigma \gamma \delta}\,,
\end{equation}
with the Riemann tensor $R_{\rho \sigma \gamma \delta}$, 
are coupled to the vector field, respectively. 
These non-minimal couplings originate from 
the requirement for keeping the three propagating degrees of freedom 
of Proca field as well as second-order equations of motion. 
We note that the sixth order Lagrangian ${\cal L}_6$ reduces to the $U(1)$ invariant 
interaction derived by Horndeski~\cite{Horndeski:1976gi} for the constant $G_6$.

Taking the scalar limit, $A_{\mu}\to\nabla_{\mu}\phi$, the action (\ref{GPact}) reduces to 
that of shift-symmetric Horndeski theories. 
The late-time cosmic acceleration can be realised in GP theories since 
a temporal component of the vector field can play a similar role to the scalar field 
in shift-symmetric Horndeski theories on the cosmological background~\cite{Tasinato:2014eka,Tasinato:2014mia,DeFelice:2016yws,Heisenberg:2016wtr}. 
In contrast, the existence of intrinsic vector modes in GP theories gives rise to 
several different properties as compared to Horndeski theories, which we will discuss in Sec.~\ref{sec:pert_VT}.

It is a natural follow-up question to ask whether the extension of 
GP theories is possible. In the case of scalar-tensor theories, 
Horndeski theories are 
the most general scalar-tensor theories 
with second-order equations of motion. Since the second-order property 
guarantees the absence of additional degrees of freedom associated with 
Ostrogradski instabilities, there is only one scalar degree of freedom 
in these theories. In contrast to the case of scalar-tensor theories,
no rigorous construction of \textit{the most general}
second-order vector-tensor theory (\`{a} la Horndeski)
with one scalar, two transverse vector, and two tensor modes has been known so far. 
It is therefore fair to say that we do not even know whether or not
the GP theories are the most general second-order vector-tensor theories.
Having said that, 
there are some attempts to go beyond the GP theories by allowing for
higher derivative equations of motion.
The existence of higher-order derivatives in equations of motion does not immediately mean that the number of 
propagating degrees of freedom increases, as we have already learned
from the study of scalar-tensor theories.
Indeed, in 
GLPV theories~\cite{Gleyzes:2014dya}, 
which are more general than Horndeski theories
and hence have higher-order equations of motion,
the Hamiltonian analysis shows that 
the number of propagating degrees of freedom is the same as that in Horndeski 
theories~\cite{Lin:2014jga,Gleyzes:2014qga}. In an analogous way, the extension of 
second-order GP theories to the domain of beyond-generalised Proca (BGP) theories 
is performed in Ref.~\cite{Heisenberg:2016eld}. 
In addition to the GP Lagrangians (\ref{L2})--(\ref{L6}), the BGP action consists of 
the following new derivative interactions: 
\begin{align}
&
{\cal L}_4^{\rm N}
=f_4 \hat{\delta}_{\alpha_1 \alpha_2 \alpha_3 \gamma_4}^{\beta_1 \beta_2\beta_3\gamma_4}
A^{\alpha_1}A_{\beta_1}
\nabla^{\alpha_2}A_{\beta_2} 
\nabla^{\alpha_3}A_{\beta_3}\,, \label{L4N}\\
&
{\cal L}_5^{\rm N}
=
f_5 \hat{\delta}_{\alpha_1 \alpha_2 \alpha_3 \alpha_4}^{\beta_1 \beta_2\beta_3\beta_4}
A^{\alpha_1}A_{\beta_1} \nabla^{\alpha_2} 
A_{\beta_2} \nabla^{\alpha_3} A_{\beta_3}
\nabla^{\alpha_4} A_{\beta_4}\,,\label{L5N} \\
&
\tilde{{\cal L}}_5^{\rm N}
=
\tilde{f}_{5}
\hat{\delta}_{\alpha_1 \alpha_2 \alpha_3 \alpha_4}^{\beta_1 \beta_2\beta_3\beta_4}
A^{\alpha_1}A_{\beta_1} \nabla^{\alpha_2} 
A^{\alpha_3} \nabla_{\beta_2} A_{\beta_3}
\nabla^{\alpha_4} A_{\beta_4}\,,
\label{L5Nd}\\
&
{\cal L}_6^{\rm N}
=\tilde{f}_{6}
 \hat{\delta}_{\alpha_1 \alpha_2 \alpha_3 \alpha_4}^{\beta_1 \beta_2\beta_3\beta_4}
\nabla_{\beta_1} A_{\beta_2} \nabla^{\alpha_1}A^{\alpha_2}
\nabla_{\beta_3} A^{\alpha_3} \nabla_{\beta_4} A^{\alpha_4}\,,
\label{L6N}
\end{align}
where $f_4$, $f_5$, $\tilde{f}_5$, $\tilde{f}_6$ are arbitrary functions depending on $X$ alone, 
and we introduced the operator 
$\hat{\delta}_{\alpha_1 \alpha_2 \alpha_3 \alpha_4}^{\beta_1 \beta_2\beta_3\beta_4}
=\epsilon_{\alpha_1 \alpha_2 \alpha_3 \alpha_4} \epsilon^{\beta_1 \beta_2\beta_3\beta_4}$.
Although these new self-interaction terms generate derivatives higher than second order 
in equations of motion after taking the scalar limit, $A_{\mu}\to\nabla_{\mu}\phi$, the number of propagating degrees of freedom in BGP theories 
is the same as that in GP theories (two tensor polarisations, two transverse vector modes, 
and one scalar mode) on the cosmological background~\cite{Heisenberg:2016eld}. 
The application of BGP theories to the late-time cosmic acceleration has been studied in~\cite{Nakamura:2017dnf}, 

One can further perform a healthy extension of GP theories by keeping the number of 
propagating degrees of freedom analogously to DHOST theories~\cite{Langlois:2015cwa,Crisostomi:2016czh,BenAchour:2016fzp} which
are the further extension of the GLPV theories. The key ingredient is the so-called degeneracy conditions 
that can eliminate an unwanted propagating degree of freedom even though 
the equations of motion contain derivatives higher than second order. In Ref.~\cite{Kimura:2016rzw}, 
the authors applied this idea to vector-tensor theories and derived the most general vector-tensor Lagrangian built out of
quadratic (and lower) terms in the first derivatives of the vector field
supplemented with the corresponding degeneracy conditions in curved spacetime. 
This is the vector-tensor extension of DHOST theories dubbed as extended vector-tensor (EVT) theories described by the following action, 
\begin{equation}
S= \frac{1}{2}\int {\rm d}^4 x \sqrt{-g} \Bigl[ f(X) R + C^{\mu\nu\rho\sigma}\nabla_{\mu}A_{\nu}\nabla_{\rho}A_{\sigma} + G_3(X)\nabla_{\mu}A^{\mu} + G_2(X,F,U,Y)   \Bigr] \,,
\label{EVTact}
\end{equation}
where 
$C^{\mu\nu\rho\sigma}$ is defined by
\begin{align}
C^{\mu\nu\rho\sigma} =&
\alpha_1(X) g^{\mu(\rho}g^{\sigma)\nu} + \alpha_2(X) g^{\mu\nu}g^{\rho\sigma}
+ \frac{1}{2}\alpha_3(X) (A^\mu A^\nu g^{\rho\sigma} + A^\rho A^\sigma g^{\mu\nu}) 
\notag\\
&
+ \frac{1}{2} \alpha_4(X) (A^\mu A^{(\rho} g^{\sigma)\nu}
+ A^\nu A^{(\rho} g^{\sigma)\mu}) 
+ \alpha_5(X) A^{\mu} A^{\nu} A^{\rho} A^{\sigma} + \alpha_6(Y) g^{\mu[\rho}g^{\sigma]\nu}
\notag\\
&
+ \frac{1}{2}\alpha_7(X) (A^\mu A^{[\rho} g^{\sigma]\nu} - A^\nu A^{[\rho} g^{\sigma]\mu})
+ \frac{1}{4}\alpha_8(X) (A^\mu A^\rho g^{\nu\sigma} - A^{\nu}A^{\sigma} g^{\mu\rho})
+\frac{1}{2}\alpha_9(X) \epsilon^{\mu\nu\rho\sigma} \,.
\end{align}
Imposing to the action~(\ref{EVTact}) the degeneracy conditions that eliminate the would-be Ostrogradski mode, a new class of degenerate vector-tensor theories that 
cannot be included in GP and BGP theories has been found. The cosmological application 
of EVT theories is studied in Ref.~\cite{Kase:2018tsb}. In sharp contrast to the case of 
DHOST theories, there exist healthy cosmological solutions in non-trivial degenerate 
domain isolated from GP and BGP theories. 

\subsection{Metric-affine gravity}
\label{sec:Metric-affine gravity}
Although the Riemannian geometry is a foundation of GR and most modified gravity theories, there have been many attempts to use non-Riemannian geometries as a foundation of gravity. The history is almost as old as the usual Riemannian formulation of gravity. Here, we provide a brief review of their underlying ideas and some recent developments. A
comprehensive review on this subject can be found in~\cite{Hehl:1994ue,Gronwald:1997bx,Hehl:1999sb,Blagojevic:2002du,Blagojevic:2012bc,Blagojevic:2013xpa}.

It is often said that spacetime has the structure of a differentiable manifold, but the concept of the manifold itself is too general. It might be plausible to assume that we can perform two operations in spacetime: we can compare tensors at different points and we can measure lengths of curves and angles between vectors. Then, the spacetime is required to have additional structures, namely the parallel transport and the inner product. Let a manifold be equipped with the parallel transport characterised by the linear connection $\Gamma^{\mu}_{~\nu\rho}$, which has no (anti-)symmetric indices in general. 
Using $\Gamma^{\mu}_{~\nu\rho}$, the covariant derivatives for a vector is defined as $\overset{\scriptscriptstyle  \Gamma}{\nabla } {}_{\alpha}A^{\mu}=\partial_{\alpha}A^{\mu}+\Gamma^{\mu}_{~\beta\alpha}A^{\beta}$,
where we put $\Gamma$ to clarify that we are using a general connection. 
Using the connection, we can introduce two geometrical objects, called the curvature and the torsion, defined respectively as
\begin{align}
\overset{\scriptscriptstyle  \Gamma}{R } {}^{ \mu}{}_{\nu\alpha\beta}&:=\partial_{\alpha}\Gamma^{\mu}_{~\nu\beta}-\partial_{\beta}\Gamma^{\mu}_{~\nu\alpha}
+\Gamma^{\mu}_{~\sigma\alpha } \Gamma^{\sigma}_{~\nu\beta }-\Gamma^{\mu}_{~\sigma \beta} \Gamma^{\sigma}_{~\nu\alpha } 
\,, 
\label{def_curvature} \\
T^{\mu}{}_{\alpha\beta}&:=\Gamma^{\mu}_{~\beta\alpha}-\Gamma^{\mu}_{~\alpha\beta}
\,.
\label{def_torsion}
\end{align}
Note that the notion of the inner product, i.e. the metric, has not been introduced so far, implying that the curvature and the torsion are independent concepts from the metric. With the help of the metric, $g_{\mu\nu}$, we can define the non-metricity tensor
\begin{align}
Q_{\mu}{}^{\alpha\beta}&:=\overset{\scriptscriptstyle  \Gamma}{\nabla } {}_{\mu}g^{\alpha\beta}\,.
\label{def_non-metricity}
\end{align}
The three independent tensors,~\eqref{def_curvature}--\eqref{def_non-metricity}, specify the properties of the most general geometry equipped with the inner product and the parallel transport, namely the metric-affine geometry. The Riemannian geometry corresponds to the subclass satisfying $T^{\mu}{}_{\alpha\beta}=0$ and $Q_{\mu}{}^{\alpha\beta}=0$, where the connection is uniquely determined by the metric due to the additional conditions.

The metric-affine geometry appears as a geometric interpretation of gauge theories of gravity, initiated by~\cite{Utiyama:1956sy,Kibble:1961ba,sciama1962analogy}. For instance, one may think of gravity as a gauge force associated with the Poincar\'{e} group, similarly to other fundamental forces. Since the Poincar\'{e} group is the semi-direct product of the translation and rotation (Lorentz) groups, we may have two corresponding gauge fields, $e^a=e^a_{\mu}dx^{\mu}$ and $\omega^{ab}=\omega^{ab}{}_{\mu}dx^{\mu}$, where $a,b,\dots$ are the Lorentz indices and the Lorentz connection implies $\omega^{ab}=\omega^{[ab]}$, where the parenthesis around the indices denotes anti-symmetrisation of a tensor. As long as ${\rm det} \,e^a_{\mu}\neq 0$, the gauge fields $e^a$ and $\omega^{ab}$ can be regarded as the tetrad and the spin connection of the spacetime, and their field strengths are identified with the curvature and the torsion, respectively. In addition, the Lorentz connection, $\omega^{ab}=\omega^{[ab]}$, concludes that $Q_{\mu}{}^{\alpha\beta}=0$. Therefore, the Poincar\'{e} gauge theories (PGT) of gravity lead to the Riemann-Cartan geometry, where the curvature and the torsion are independent while the non-metricity tensor vanishes. Note that the precise treatment would require a spontaneous symmetry breaking down to the Lorentz group and PGT are generically theories in the symmetry-breaking phase~\cite{MacDowell:1977jt,Stelle:1979aj,Percacci:2009ij}. If one considers the general affine group, the semi-direct product of the translation group, and the general linear group as the gauge group of gravity, the connection no longer has the anti-symmetric indices~\cite{Hehl:1976my,Lord:1978qz}. Then, the metric-affine geometry is the basis of gravitational theories, called metric-affine gravity (MAG). In MAG, the energy-momentum tensor induces the curvature as usual, while the hyper-momentum tensor, which represents spin, dilation, and shear currents of matter, induces the torsion and the non-metricity.

The action of PGT/MAG is given by
\begin{align}
S=\int {\rm d}^4x \sqrt{-g}\mathcal{L}
\,, \qquad
\mathcal{L}=\mathcal{L}_G (g,\Gamma) + \mathcal{L}_{\phi}(g,\Gamma,\phi)
\,,
\end{align}
where $\mathcal{L}_G$ is the purely gravitational part and $\mathcal{L}_{\phi}$ is the Lagrangian for a matter field $\phi$. The first term may be expanded as
\begin{align}
\mathcal{L}_G = \frac{M_{\rm Pl}^2}{2}\mathcal{L}_2 + \frac{M_{\rm Pl}^2}{M_*^2 }\mathcal{L}_4 + \sum_{n>4} \frac{M_{\rm Pl}^2}{\Lambda^{n-2} }\mathcal{L}_n
\,,
\label{MAG_L}
\end{align}
with the mass scales $M_*$ and $\Lambda$, where $\mathcal{L}_n$ are operators with scaling dimension $n$. Assuming that $[g_{\mu\nu}]=0$ and $[\Gamma^{\mu}_{~\nu\rho}]=1$, the lower-dimensional operators are schematically given by
\begin{align}
\mathcal{L}_2 = \overset{\scriptscriptstyle  \Gamma}{R } {} + T^2 +Q^2 + TQ
\,, \quad 
\mathcal{L}_4 = \overset{\scriptscriptstyle  \Gamma}{R } {}^2 + \cdots,
\label{qMAG}
\end{align}
with indices omitted where the ellipsis stands for terms containing either non-linear order in or derivatives of $T$ and $Q$. In the absence of these operators, the gravitational theories are called the quadratic Poincar\'{e} gauge theories (qPGT) when $Q_{\mu}{}^{\alpha\beta}=0$ and the quadratic metric-affine gravity (qMAG) when $Q_{\mu}{}^{\alpha\beta}\neq 0$, respectively. Roughly speaking, the dimension-4 operators lead to the ``kinetic terms'' of the connection, $(\partial \Gamma)^2$, while the dimension-2 operators involve the ``mass terms'', $\Gamma^2$. Therefore, the additional degrees of freedom associated with the independent connection generically represent massive states. The linear particle spectrum and the stability conditions around the Minkowski background have been well-investigated in qPGT~\cite{Sezgin:1979zf,Hayashi:1980qp,Sezgin:1981xs,Karananas:2014pxa,Lin:2018awc,Blagojevic:2018dpz} while there are less studies in qMAG~\cite{Percacci:2020ddy} and theories including derivatives of the torsion and/or non-metricity~\cite{Aoki:2019snr}. Furthermore, the non-linear analysis of \eqref{MAG_L} has not been thoroughly discussed. Although the general PGT (and MAG) yield a non-linear ghost even if it is free from ghost at the linear level~\cite{Yo:1999ex,Yo:2001sy,Jimenez:2019qjc,BeltranJimenez:2019acz,BeltranJimenez:2020sqf}, there are special classes in which the ghost is absent at fully non-linear orders. The known ghost-free PGT can admit the massive spin-0 state(s)~\cite{Yo:1999ex,Jimenez:2019qjc} and the massive spin-2 state~\cite{Aoki:2020rae}, in addition to the massless graviton. Note that the appearance of the non-linear ghost implies that the EFT is valid only below the mass of the ghost. One can study phenomenology well below the cutoff scale even in the presence of the ghost. Phenomenological studies on PGT/MAG mainly focus on the early universe since the higher curvature terms would be expected to represent a UV modification of gravity. Nonetheless, similarly to $f(R)$ dark energy, it would be also interesting to explore the possibilities that the modification is active in the late-time universe as investigated in \cite{Nikiforova:2016ngy,Nikiforova:2018pdk,Barker:2020gcp}.

When one is interested in low-energy phenomena where only the lowest-dimensional operator $\mathcal{L}_2$ is important, the connection (the massive degrees of freedom of \eqref{MAG_L}) can be integrated out. Then, the metric (the massless graviton) is the only dynamical variable of the geometry at low energies, implying that the Riemannian geometry arises as a low-energy effective description of spacetime. Nonetheless, similarly to the relations between the weak interaction and Fermi's interaction, integrating out the connection modifies interactions of fields that couple to the connection at the microscopic level. Furthermore, the hyper-fluid, a hypothetical fluid having a non-vanishing hyper-momentum tensor, produces the torsion and the non-metricity at the macroscopic level. Recent studies on cosmology with the hyper-fluid can be found in \cite{Iosifidis:2020gth,Iosifidis:2020upr,Iosifidis:2021fnq}.

There have been a lot of studies about fermions since the minimal coupling already predicts the coupling between the fermions and the connection through the covariant derivative, and recent papers study non-minimal couplings of the fermions~\cite{Shaposhnikov:2020aen,Karananas:2021zkl}. 
On the other hand, non-minimal couplings between gravity and a scalar field have been widely discussed in cosmology and modified gravity, so we will detail them below. As for vector fields, the minimal coupling depends on whether the vector is a spacetime vector or a vector of the internal Lorentz (or general linear) group: for instance, a gauge field $A=A_{\mu}dx^{\mu}$ is a spacetime vector and then the field strength $F:=dA$ does not lead to the coupling to the connection, while a vector of the internal gauge group $A^a$ could give a coupling through the covariant derivative $(\delta^a{}_b\partial_{\mu}+\omega^a{}_{b\mu})A^b$. 

The discovery of cosmological mysteries has motivated us to study a general framework of a scalar field with non-minimal coupling to gravity. One of the simplest non-minimal couplings is the Brans–Dicke type, $f(\phi) \overset{\scriptscriptstyle  \Gamma}{R } {}(g,\Gamma)$. If one thinks of $\phi$ as dark energy, the field $\phi$ has a non-vanishing expectation value on cosmological scales, which yields a background torsion and/or non-metricity through the coupling to the connection. As a result, phenomenological predictions are different from the metric counterpart $f(\phi)R(g)$, which was noted in \cite{Bauer:2008zj} in the context of inflation.\footnote{In the context of inflation, the literature often assumes the Palatini formalism of gravity where the torsion vanishes while the non-metricity does not vanish. See~\cite{Tenkanen:2020dge} for a review on the inflationary cosmology in Palatini gravity.} Recall that the connection is non-dynamical at low energies and can be integrated out. One can translate scalar-tensor theories with the general connection into those in the Riemannian formalism~\cite{Aoki:2018lwx,Aoki:2019rvi,Helpin:2019kcq} and can compute phenomenological predictions in the equivalent Riemannian formalism~\cite{Kubota:2020ehu}.

We finally comment on scale transformations and a relation to the degenerate scalar-tensor theories. In the metric-affine geometry, there are three different scale transformations~\cite{Iosifidis:2018zwo}:
\begin{align}
&\text{1. The Weyl transformation: }
g_{\mu\nu} \to e^{2\Omega(x)}g_{\mu\nu}\,, \quad \Gamma^{\mu}_{~\nu\rho}\to \Gamma^{\mu}_{~\nu\rho}
\,, \label{Weyl_trans}
\\
&\text{2. The frame rescaling: }
g_{\mu\nu} \to e^{2\Omega(x)}g_{\mu\nu}\,, \quad \Gamma^{\mu}_{~\nu\rho}\to \Gamma^{\mu}_{~\nu\rho}+\delta^{\mu}_{\nu}\partial_{\rho}\Omega(x)
\,, \label{frame_trans}
\\
&\text{3. The projective transformation: }
g_{\mu\nu} \to g_{\mu\nu}\,, \quad \Gamma^{\mu}_{~\nu\rho}\to \Gamma^{\mu}_{~\nu\rho}+\delta^{\mu}_{\nu}\xi_{\rho}(x)
\,, \label{proj_trans}
\end{align}
where the frame rescaling \eqref{frame_trans} is equivalent to $e^a_{\mu} \to e^{\Omega(x)} e^a_{\mu} ,~ \omega^{ab}{}_{\mu} \to \omega^{ab}{}_{\mu}$. Conformal invariant theories, in the sense that the action is invariant under \eqref{Weyl_trans} (or \eqref{frame_trans}), can be systematically obtained by introducing the Weyl covariant derivative, following the idea of the Weyl gauge theory. The projective transformation is a kind of scale transformation associated with the parallel transport (and also a transformation that preserves the auto-parallel equation, hence the name, ``projective''). The U-DHOST theory~\cite{DeFelice:2018ewo} arises as the most general projective invariant scalar-tensor theory containing up to quadratic terms in the connection~\cite{Aoki:2019rvi}, where the degeneracy condition is automatically satisfied by the projective invariance.

\subsection{Cuscuton and minimally modified gravity}
\label{sec:Cuscuton and minimally modified gravity}

According to the Lovelock theorem, GR plus a cosmological constant is the only metric theory with general covariance and having second-order Euler-Lagrange equations in four dimensions.
This implies that any modified gravity theory, at least effectively, contains additional 
DOFs on top of the metric.
Nevertheless, it is possible to construct non-trivial models without introducing a new dynamical DOF.
Such models yield only two physical DOFs and they are identified as the two polarisation modes of gravitational waves as in GR, and hence can be regarded as {\it minimal modifications of gravity}.
The simplest example within scalar-tensor theories is the cuscuton model~\cite{Afshordi:2006ad}, whose action is given by
    \begin{align}
    S=\int {\rm d}^{4}x\sqrt{-g}\left[\frac{M_{\rm Pl}^{2}}{2}R+\mu^{2}\sqrt{2|X|}-V(\phi)\right]\,,
    \end{align}
with $\mu$ being a non-vanishing constant and $V$ being an arbitrary function of $\phi$.
In terms of the Arnowitt-Deser-Misner~(ADM) variables, the above action can be written as
    \begin{align}
    S=\int {\rm d}t{\rm d}^{3}x\sqrt{\gamma}\left[\frac{M_{\rm Pl}^{2}}{2}N\left(\mathcal{R}+K_{ij}K^{ij}-K^{2}\right)+\mu^{2}-NV(t)\right]\,. \label{eq2.3.3.1}
    \end{align}
Here, $N$ is the lapse function, $\mathcal{R}$ is the scalar curvature associated with the spatial metric~$\gamma_{ij}$, $K_{ij}$ is the extrinsic curvature with its trace denoted by $K=K^{i}{}_{i}$, and we chose the unitary gauge where $\phi$ is taken to coincide with the time~$t$.
The point is that the modification from the Einstein-Hilbert term is at most linear in $N$, which makes the Hamiltonian structure of this model the same as that of GR.
More generally, the authors of Ref.~\cite{Lin:2017oow} performed a Hamiltonian analysis for theories of the form
    \begin{align}
    S=\int {\rm d}t{\rm d}^{3}x\sqrt{\gamma}\left[NF(\gamma_{ij},K_{ij},\mathcal{R}_{ij},D_{i},t)+G(\gamma_{ij},\mathcal{R}_{ij},D_{i},t)\right]\,,
    \end{align}
with $\mathcal{R}_{ij}$ being the spatial Ricci tensor and $D_i$ being the covariant derivative associated with $\gamma_{ij}$, and derived the necessary and sufficient conditions for the Lagrangian to yield two (or less) physical DOFs.

Interestingly, there exist more general classes of minimally modified gravity, which can be systematically constructed by means of Hamiltonian analysis~\cite{Mukohyama:2019unx,Gao:2019twq,Lin:2020nro}.
In particular, a general class within DHOST theories was specified in Ref.~\cite{Iyonaga:2018vnu}.
In the ADM language, the action is written as
    \begin{align}
    S=\int {\rm d}t{\rm d}^{3}xN\sqrt{\gamma}\,\bigg[&\left(u_{0}+\frac{u_{1}}{N}\right)\mathcal{R}+\frac{u_{2}N}{N+u_{3}}\left(K_{ij}K^{ij}-K^{2}\right) \nonumber \\
    &+\frac{u_{4}}{N+u_{3}}K+u_{5}+\frac{u_{6}}{N}-\frac{3u_{4}^{2}}{8u_{2}N(N+u_{3})}\bigg]\,,
    \end{align}
where $u$'s are arbitrary functions of $t$.
Here, we omitted terms cubic in $K_{ij}$ for simplicity.
Note that the cuscuton model~\eqref{eq2.3.3.1} amounts to the choice~$u_{0}(t)=u_{2}(t)=M_{\rm Pl}^{2}/2$, $u_{5}(t)=-V(t)$, $u_{6}(t)=\mu^2$, $u_{1}(t)=u_{3}(t)=u_{4}(t)=0$.
Cosmology in this model was studied in Ref.~\cite{Iyonaga:2020bmm}, where it was shown that this class of models serves as a viable dark-energy model.
Further generalisations (in the U-DHOST family) can be found in Ref.~\cite{Gao:2019twq}.
It is intriguing to note that the authors of~\cite{Iyonaga:2021yfv} proposed a model
within the class of theories of~\cite{Gao:2019twq}
that gives the same predictions as those in GR for weak gravitational fields and the propagation speed of gravitational waves.
It is also possible to choose the model parameters so that the homogeneous and isotropic cosmological dynamics is close to or even identical to that of the $\Lambda$CDM model.

It should be noted that there is a subtlety in the two-DOF nature of the theories of this type for an inhomogeneous configuration of $\phi$~\cite{Iyonaga:2018vnu}.
Indeed, when $\partial_{\mu}\phi$ is spacelike, there is a third propagating DOF.
However, so long as $\partial_{\mu}\phi$ is timelike, this unwanted DOF does not propagate if an appropriate boundary condition is imposed.
This means that there is a preferred slicing in the cuscuton-like theories.
The situation is reminiscent of the shadowy mode in U-DHOST theories~\cite{DeFelice:2018ewo,DeFelice:2021hps,DeFelice:2022xvq}.
Thus, $\partial_{\mu}\phi$ has to be timelike to keep two-DOF nature. Then, we can always adopt the gauge $\phi=t$ and write the action in the ADM form by using the spacetime metric only. Owing to the existence of the preferred slicing, cuscuton and minimally modified gravity are Lorentz-violating theories of gravity, and they do not contradict the uniqueness of GR in the sense of the Lovelock theorem. On the other hand, the scattering amplitude arguments indicate that GR would be the unique theory of the massless spin-2 field (i.e., the theory having two tensorial DOFs with the relativistic dispersion relation) under certain assumptions of unitarity and locality~\cite{Benincasa:2007qj, Arkani-Hamed:2008bsc, Carballo-Rubio:2018bmu, Pajer:2020wnj} even when the Lorentz invariance is omitted from the assumptions~\cite{Pajer:2020wnj}. Cuscuton and minimally modified gravity are consistent with the latter uniqueness results as well since the locality condition does not hold thanks to the shadowy mode, giving rise to non-trivial models of modified gravity without a new dynamical DOF.

Another possible way to construct minimally modified gravity theories is to perform a canonical transformation on GR~\cite{Aoki:2018zcv,Aoki:2018brq}, as a canonical transformation (or any invertible field transformation) does not change the number of physical DOFs~\cite{Domenech:2015tca,Takahashi:2017zgr,Babichev:2019twf,Babichev:2021bim}.
Although the theory obtained after the transformation is mathematically equivalent to GR, this equivalence no longer holds when matter fields are taken into account:
If matter fields are minimally coupled to gravity in the new frame, then there are non-minimal interactions in the original frame where gravity is described by GR.
In this sense, we obtain a new class of theories via canonical transformation, which offers a rich phenomenology~\cite{Aoki:2018brq}.
Likewise, performing an invertible disformal transformation~\cite{Bekenstein:1992pj} (or its generalisation proposed in \cite{Takahashi:2021ttd}) on known minimally modified gravity models yields a novel class of minimally modified gravity theories in general.

Let us make a comment on the classification of modified gravity theories.
According to Ref.~\cite{Aoki:2018brq}, a theory is called type-I if there exists an Einstein frame, or otherwise called type-\mbox{I\hspace{-1pt}I}.
For instance, the class of theories studied in Refs.~\cite{Aoki:2018zcv,Aoki:2018brq} belongs to type-I, as there exists an Einstein frame by construction.
An obvious example of type-\mbox{I\hspace{-1pt}I} minimally modified gravity is the minimal theory of massive gravity~\cite{DeFelice:2015hla,DeFelice:2015moy}, which we have discussed in Sec.~\ref{Sec.th.LVMG}, where the dispersion relation of the graviton is different from that in GR.
Another example is the VCDM model~\cite{DeFelice:2020eju,DeFelice:2020cpt}, which is obtained by adding a cosmological constant in a canonically transformed frame of GR and then performing the inverse canonical transformation.

\subsection{Evading solar-system tests}
\label{sec:screening}
Basically,
additional 
DOFs appearing in modified gravity theories mediate an additional gravitational force at all scales. Although a modification of the gravitational law on cosmological scales is necessary to account for the present cosmic acceleration, the ``fifth force'' on small scales is strongly constrained by the precision tests of gravity on the Earth and
in the solar system. Modified gravity theories are therefore required to
have a screening mechanism to reproduce GR on small scales
for their phenomenological success as a dark energy alternative.
Here, we briefly overview three major screening mechanisms: the Vainshtein, chameleon, and symmetron mechanisms. Although the details of screening depend on the mechanisms, they are essentially classified into two classes: the fifth force is weakened by a suppression of the effective coupling to matter (the Vainshtein and symmetron mechanisms) or the effective mass of the field carrying the fifth force increases (the chameleon mechanism). 
 For a comprehensive review
 of screening mechanisms in modified gravity, see~\cite{Babichev:2013usa,deRham:2014zqa,Burrage:2017qrf}.
\subsubsection{Vainshtein screening}
\label{sec:Vainshtein screening}


The Vainshtein mechanism was first noticed in the context of vDVZ discontinuity in massive gravity~\cite{Vainshtein:1972sx} as explained in Sec.~\ref{sec:Massive gravity and bigravity}. This mechanism relies on kinetic interactions involving higher derivatives
of the scalar degree of freedom, and their non-linearities 
become dominant within a certain length scale, leading to a strong suppression of the fifth force. Such higher derivative interactions appear not only in the decoupling limit theory of massive gravity given in~\eqref{action:DL}, but also in the Horndeski and DHOST theories.  
One of the simplest examples is given by the cubic galileon model
coupled to matter:
\begin{align}
	{\cal L}_{\rm cG}&=
	 -\frac{1}{2}(\partial\phi)^2\left(1 +\frac{\Box\phi}{2\Lambda^3}\right) +\frac{1}{M_{\rm Pl}}\phi T^{\mu}_{~\mu},
	\label{eq: cG}
\end{align}
where $T^{\mu}_{~\mu}$ is the trace of the energy-momentum tensor for matter and $\Lambda$ is a mass scale. This example arises as a subset of the Horndeski theory
as well as the decoupling limit theory with the particular parameter choice ${\tilde \beta_4}={\tilde \beta_5}={\tilde \beta_X}={\tilde \beta_T}=0$ in
Eq.~\eqref{action:DL}. 

To see how the Vainshtein screening operates, let us consider a point source
represented by $T^{\mu}{}_{\nu}=-M\delta^{(3)}(r)\delta^{\mu}_0\delta^0_{\nu}$, where $M$ is the mass of the source. This static and spherically symmetric setup allows us to consider  $\phi=\varphi(r)$ (where $r$ is the radial coordinate),
and then the equation of motion for the scalar field can be integrated once, 
yielding
\begin{align}
	\frac{\varphi^\prime}{r}+\frac{1}{\Lambda^3}\left(\frac{\varphi^\prime}{r}\right)^2
	=\frac{M}{4\pi M_{\rm Pl}r^3}
	\,,
\end{align}
where a prime stands for the derivative with respect to $r$.
This algebraic equation determines $\varphi'$, i.e.
the force due to the scalar field acting on the point mass.
In the two limiting cases the solution is given by
\begin{align}
	\frac{\varphi^\prime}{M_{\rm Pl}} &\approx
	\begin{cases}
		\displaystyle{\frac{2G_{\rm N}M}{r^{2}}\left(\frac{r}{r_{\rm V}}\right)^{3/2}~(r \ll r_{\rm V}),}\vspace{2mm}\\
		\displaystyle{\frac{2G_{\rm N}M}{r^{2}}~~~~~~~~~~~~\, ( r_{\rm V}\ll r),}
	\end{cases}
	\label{eq: soln}
\end{align}
where $G_{\rm N}=1/(8\pi M_{\textrm{Pl}}^2)$ and
we defined the so-called Vainshtein radius
\begin{align}
r_{\rm V} := \frac{1}{\Lambda}\left(\frac{M}{M_{\rm Pl}}\right)^{1/3}\,.
\end{align}
The Vainshtein radius gives the boundary between the linear and non-linear regimes. The fifth force should be compared with the usual Newtonian force
($\Phi'\sim G_{\textrm{N}}M/r^2$), which can be read off from
the 00 component of the metric, $g_{00}=-(1+2\Phi)$.
Inside the Vainshtein radius, the fifth force due to the scalar field is suppressed by the factor $(r/r_{\rm V})^{3/2}$ while it is comparable to the Newtonian force outside the Vainshtein radius. It is natural to assume that $\Lambda\sim (M_{\rm Pl}H_0^2)^{1/3}$ in the models accounting for the present accelerated expansion of the Universe.
In this case, the Vainshtein radius is given by $r_{\rm V}\sim {\cal O}(100)$ pc for the Sun. This implies that the fifth force contribution can be well screened at small scales and the solar-system constraints can be evaded
thanks to the Vainshtein mechanism.

The basic structure is the same even in the presence of higher-order galileon interactions in a spherically symmetric setup~\cite{Nicolis:2008in,DeFelice:2011th}. Although the Vainshtein screening mechanism can in principle be successful in the case of a homogeneous and isotropic cosmological background in the Horndeski theory~\cite{Kimura:2011dc},
there are several caveats. If $G_4$ is a function of the scalar field, the gravitational coupling even inside the Vainshtein radius,  denoted by $G_{\rm local}$, is a time-independent function,  $G_{\rm local}=G_{\rm local}(t)$. As we will discuss in Sec.~\ref{sec:Density perturbations in scalar-tensor theories}, $G_{\rm local}$ is equal to the gravitational coupling appearing in the modified Friedmann equation given in the form $3H^2 = 8\pi G_{\rm local}(t) \rho_m +\,$(up to linear in $H$) where the last term may be interpreted as the effective dark energy component of the Horndeski theory.
In addition, the deviation of the parameterised post-Newtonian (PPN) parameter from the GR value, $1-\gamma_{\rm PPN}$, where $\gamma_{\rm PPN} :=\Psi / \Phi$ and $\Psi$ is another metric potential (the spatial curvature)
read off from the spatial metric, $g_{ij}=(1-2\Psi)\delta_{ij}$,
is always suppressed by the factor given by the ratio between $r$ and $r_{\rm V}$ after imposing $c_{\rm GW}=c$. 
Furthermore, in the case of $G_{5X}\neq 0$, which is also unacceptable from the requirement $c_{\rm GW}=c$, 
the inverse-square law of gravity cannot be even reproduced  well inside the Vainshtein radius. 
In addition, there are several conditions in order that the Vainshtein solution exists.
These conditions include successful matching of the screened solution
and the outer solution in the linear regime~\cite{Kimura:2011dc} and
the perturbative stability conditions~\cite{Koyama:2013paa}. 
There are also studies of
more general setups beyond weak gravity/spherical symmetry:
low-density stars with slow
rotation~\cite{Chagoya:2014fza},
static relativistic stars~\cite{Ogawa:2019gjc},
a two-body system~\cite{Hiramatsu:2012xj}, and a disk with a hole at its centre~\cite{Ogawa:2018srw}. 
In the case of massive gravity theories and their extensions,
it has been confirmed e.g. in Refs.~\cite{Renaux-Petel:2014pja,Babichev:2009us,Babichev:2010jd,Gabadadze:2014gba,Aoki:2016eov,Aoki:2016vcz,Gumrukcuoglu:2021gua}
that the Vainshtein mechanism successfully works.
The Vainshtein screening successfully works in the 
GP theory as well, and the corrections to gravitational potentials become small 
enough to satisfy the solar-system tests of gravity~\cite{DeFelice:2016cri,Nakamura:2017lsf}. 

In the quadratic DHOST theory, non-linear derivative interactions involving higher derivatives again play an important role in order to screen the fifth force~\cite{Crisostomi:2017lbg,Langlois:2017dyl,Dima:2017pwp} (the earlier studies of the Vainshtein screening beyond Horndeski were done in Refs.~\cite{Kobayashi:2014ida,Koyama:2015oma}).
After integrating out the scalar field,
the crucial difference between the Horndeski and DHOST theories can be seen in the gravitational potentials 
well inside the Vainshtein radius: 
\begin{align}
    \frac{{\rm d} \Phi}{{\rm d} r} &= \frac{G_{\rm N}M(r)}{r^2} +\Xi_1G_{\rm N}M''(r),\label{eq: mod Phi}
\\
    \frac{{\rm d} \Psi}{{\rm d} r} &= \frac{G_{\rm N}M(r)}{r^2} 
    +\Xi_2\frac{G_{\rm N}M'(r)}{r}
    +\Xi_3G_{\rm N}M''(r),
\end{align}
where $M(r)$ is the enclosed mass inside a radius $r$,
$G_{\rm N}$ is the effective Newton's constant, and $\Xi_i$ are the dimensionless background-dependent coefficients.
Here, $G_{\rm N}$ and $\Xi_i$ are written in terms of the functions
in the Lagrangian of the quadratic DHOST theory, and their explicit expressions can be found in~\cite{Langlois:2017dyl}.
Outside the matter source, $M$ is constant and the standard behaviour of gravity
is recovered. In contrast, the derivatives of $M(r)$ do not vanish inside the matter source, leading to  ``partial breaking of Vainshtein screening"~\cite{Kobayashi:2014ida}.

Since 
breaking of the screening mechanism would change the structure 
of matter distributions, 
one obtains new predictions specific to the quadratic DHOST theory and
can thereby place observational/experimental constraints on the theory~\cite{Koyama:2015oma,Saito:2015fza,Sakstein:2015zoa,Sakstein:2015aac,Jain:2015edg,Sakstein:2016ggl,Babichev:2016jom,Sakstein:2016lyj,Salzano:2017qac,Sakstein:2017xjx,Saltas:2018mxc,Chagoya:2018lmv,Kobayashi:2018xvr,Babichev:2018rfj,Cermeno:2018qed,Saltas:2019ius,Anson:2020fum,Banerjee:2020rrd,Chowdhury:2020wfr,Boumaza:2021hzr,Ikeda:2021skk}.
For example, from Eq.~(\ref{eq: mod Phi}), the modified hydrostatic and Lane-Emden equations for a polytropic fluid have been derived~\cite{Saito:2015fza}.
The universal bound on $\Xi_1$ has been obtained from the requirement that gravity is attractive at the stellar centre, and the analytic/numerical solutions of the modified Lane-Emden equation have been calculated for some particular polytropic indices.
The most stringent bound on $\Xi_1$ has been obtained from the helioseismic observations as $-7.2\times10^{-3}\leq\Xi_1\leq 4.8\times10^{-3}$ (at $2\sigma$)~\cite{Saltas:2019ius}.

Let us mention the Vainshtein mechanism in the particular
DHOST theory respecting $c_{\rm GW}=c$ and the absence of graviton decay
described by the Lagrangian~\eqref{eq:LqDHOST cGW=c nodecay},
which turns out to break in a different way from that in
generic DHOST theories~\cite{Hirano:2019scf} (see also Ref.~\cite{Crisostomi:2019yfo}).
Technically,
this particular theory corresponds to the special case where
the denominators of $\Xi_{1}$ and $\Xi_3$ vanish, and
one thus needs to perform a separate analysis of this case. It is found that, even in the exterior region of the matter distribution inside the Vainshtein radius,
the recovery of the standard behaviour of gravity
is not automatic but requires
the fine-tuning of the theory parameters. 
Inside the matter distribution, the two metric potentials $\Phi$ and $\Psi$ do not coincide, and the effective Newton's constant in the matter interior is different from its exterior value.
The difference between $\Phi$ and $\Psi$ can be measured by comparing the X-ray and lensing profiles of galaxy clusters~\cite{Terukina:2013eqa,Wilcox:2015kna,Sakstein:2016ggl}.
The difference between 
the interior and exterior Newton's constants could potentially be tested by using the sound speed and neutrino fluxes in the Sun.
The 
tightest constraint on this particular DHOST theory 
comes from the orbital decay of the Hulse-Taylar pulsar~\cite{BeltranJimenez:2015sgd,Dima:2017pwp}.


We finally note that there is another screening mechanism called  ``kinetic screening/k-Mouflage gravity"~\cite{Babichev:2009ee}. Even if higher derivative interactions responsible for the Vainshtein mechanism is absent, non-linear kinetic terms
such as $X^2$
in k-essence theory can suppress the fifth force on small scales
by enhancing effectively the kinetic term for the fluctuation of the scalar field.
%
For the detail of the kinetic screening mechanism, one may refer to~\cite{Gabadadze:2012sm,Brax:2012jr,Joyce:2014kja}.

\subsubsection{Chameleon and symmetron}
\label{sec:Chameleon and symmetron}

%

The Vainshtein and kinetic screening mechanisms
introduced in the previous subsection
rely on non-linear derivative interactions of the scalar
degree of freedom.
In this subsection, we review other two screening mechanisms
called the chameleon and symmetron~\cite{Brax:2012bsa,Burrage:2017qrf},
which are implemented through an interplay between the
potential term of the scalar degree of freedom and
a conformal coupling to matter.
It is known that
some $f(R)$ models and corresponding scalar-tensor theories
exploit the Chameleon mechanism.

To see how the chameleon mechanism operates,
let us consider the scalar-tensor theory described by the action
\begin{align}
    S = \int{\rm d}^4x \sqrt{-\tilde{g}} \left[ \frac{M_{\mathrm{Pl}}^2}{2} \tilde{R} 
    - \frac{1}{2}\tilde g^{\mu\nu} \partial_{\mu} \phi \partial_{\nu} \phi - V(\phi) \right]
    + S_{\mathrm{m}}(g_{\mu \nu}, \psi_{i})
    \label{Eq: chameleon action}
    \, ,
\end{align}
where $\psi_{i}$ stands for the matter field(s).
The Jordan frame metric $g_{\mu \nu}$ and the Einstein frame metric $\tilde{g}_{\mu \nu}$ are related through the conformal transformation as follows\footnote{One may find in the literature a different convention
in which the metrics in the two frames are related through
$g_{\mu \nu} = A^{2}(\phi) \tilde{g}_{\mu \nu}$ with $A(\phi)=e^{-\beta \phi/ M_{\mathrm{Pl}}}$.}:
\begin{align}
    \tilde{g}_{\mu \nu} = A^{2}(\phi) g_{\mu \nu}
    \label{Eq: weyl transformation}
    \, .
\end{align}
Couplings between the scalar field $\phi$ and
the matter fields $\psi_{i}$ are introduced by 
the conformal factor $A(\phi)$.
As an illustration, let us consider the dilatonic coupling
of the form $A(\phi)=e^{\beta \phi/ M_{\mathrm{Pl}}}$,
where $\beta$ is a constant parameter.
The equation of motion for the scalar field $\phi$ is given by
\begin{align}
    \tilde{\Box} \phi = \frac{\partial V(\phi)}{\partial \phi} 
    + \frac{\beta}{M_{\mathrm{pl}}} T^{\mu}_{\ \mu} A^{-4}(\phi)
    \label{Eq: scalaron eom}
    \, ,
\end{align}
where $T^{\mu}_{\ \mu}:=T^{\mu\nu}g_{\mu\nu}$
is the trace of the energy-momentum tensor $T_{\mu \nu}$ in the Jordan frame
defined by $T^{\mu\nu}:=(2/\sqrt{-g})\delta S_{\textrm{m}}/\delta g_{\mu\nu}$.
(This energy-momentum tensor is conserved
in the Jordan frame, $\nabla_\mu T^{\mu\nu}=0$.)
The right-hand side may be regarded as an effective potential slope,
and the conformal coupling to matter
thus affects the scalar field dynamics in an environment-dependent way.
In the case where the scalar field plays the role of
dark energy, the bare mass of the field is assumed to be given by the dark energy scale, which is light. 
Due to the coupling to the trace of the matter energy-momentum tensor,
the \textit{effective} mass of the scalar field can
be significantly larger than its bare mass,
reflecting the hierarchy between the dark energy scale and other physical scales.
Here, it is convenient to define the energy-momentum tensor with a tilde as
$\tilde T^\mu_{\ \nu}=A^{-3}T^\mu_{\ \nu}$
and its trace as $\tilde T^\mu_{\ \mu}=A^{-3}T^\mu_{\ \mu}$.
For nonrelativistic matter with $\tilde T^\mu_{\ \nu}= \text{diag}(-\tilde\rho,0,0,0)$,
this energy-momentum tensor is conserved in the Einstein frame,
$\tilde \nabla_\mu \tilde T^\mu_{\ 0}=0$.\footnote{The
energy-momentum tensor with a tilde here is different from
the energy-momentum tensor in the Einstein frame,
$T^{\mu\nu}_{\textrm{(E)}}:=(2/\sqrt{-\tilde g})\delta S_{\textrm{m}}/\delta \tilde g_{\mu\nu}=A^{-6}T^{\mu\nu}$, which is not conserved. For a more general
fluid with the equation of state $w=p/\rho=\,$const, one would define
$\tilde T^\mu_{\ \nu}:=A^{-3(1+w)}T^\mu_{\ \nu}$ for a conserved energy-momentum
tensor in the Einsiten frame.}
In terms of $\tilde T^\mu_{\ \mu}$, 
the effective potential is written as
\begin{align}
    V_{\mathrm{eff}} (\phi) =  V(\phi) - \tilde{T}^{\mu}_{\ \mu} A^{-1}(\phi)
    \label{Eq: scalaron effective potential}
    \, .
\end{align}
Suppose that $\beta>0$ $(\beta<0)$ and the bare potential $V(\phi)$ is a monotonically
increasing (decreasing) function of $\phi$.
The effective potential then has
a minimum at some $\phi= \phi_{\min}$,
and one can evaluate the effective mass at the minimum as
\begin{align}
    m_{\phi\, \mathrm{eff}}^{2} := \left. \frac{\partial^{2} V_{\mathrm{eff}}(\phi)}{\partial \phi^{2}} \right|_{\phi= \phi_{\min}}
    \label{Eq: scalaron effective mass}
    \, .
\end{align}
The effective mass is larger for larger $\tilde \rho \,(=-\tilde T^\mu_{\ \mu})$
(here we simply assume that $\tilde T^\mu_{\ \nu}$ is given by
that of non-relativistic matter), and
the typical energy density in the local environment is
much
larger than the dark energy density.
The propagation of the scalar field is consequently suppressed, 
which enables the scalar-tensor theory to evade the constraints from local gravitational experiments. 
Since this screening mechanism relies on the environment-dependent
effective mass,
we call it the chameleon mechanism~\cite{Khoury:2003aq,Khoury:2003rn}.

Despite this efficient screening effect, the coupling parameter can be constrained e.g.
by the E\"{o}t-Wash experiment~\cite{Adelberger:2003zx,Kapner:2006si,Lambrecht:2005km}, Casimir force tests~\cite{Lamoreaux:2005zza,Lambrecht:2011qm}, atom interferometry experiments~\cite{Burrage:2014oza,Burrage:2015lya,Elder:2016yxm}, and precision atomic measurements~\cite{Jaeckel:2010xx,Schwob:1999zz,Simon:1980hu}. The constraints on the specific model
with the potential $V(\phi) = \Lambda^{n+4} / \phi^n$ as well as $f(R)$ gravity are provided in the literature \cite{Burrage:2017qrf}.
In cosmology, the cosmic history makes non-trivial effects on the scalar field dynamics through the chameleon mechanism~\cite{Erickcek:2013oma,Motohashi:2012tt,Nishizawa:2014zra,Yashiki:2020naf,Katsuragawa:2016yir,Katsuragawa:2017wge,Chen:2019kcu,Chen:2022zkc}.


Let us give several comments on the conformal
coupling to matter.
In general, one may assign different coupling parameters $\beta_{i}$ to
different matter species $\psi_{i}$.  
As conformal symmetry suggests a vanishing trace of the energy-momentum tensor,
the scalar field does not couple to the radiation components (massless gauge bosons)
at the classical level.
However, at the quantum level,
the trace anomaly generates a non-vanishing trace~\cite{Fujikawa:1980vr,Ferreira:2016kxi,Katsuragawa:2021wmw}, 
and thus the coupling between the scalar field and radiation shows up.
One can find the non-trivial scalar-field dependence in the matter Lagrangian, such as the kinetic term of the Standard-Model Higgs.
Although scalar-field dependence is obscure in the perfect-fluid description and hidden in the energy density and pressure,
a field-theoretical approach reveals that quasi-static configuration of matter can allow us to ignore such a scalar-field dependence~\cite{Katsuragawa:2018wbe}, which would justify the conventional definition Eq.~\eqref{Eq: scalaron effective potential}.


It should be noted that $f(R)$ gravity can be recast into
a particular scalar-tensor theory whose action is of the form
of Eq.~\eqref{Eq: chameleon action} with the dilatonic coupling function
$A=e^{\phi/(\sqrt{6}M_{\textrm{Pl}})}$,
i.e. $\beta=1/\sqrt{6}$ (see Sec.~\ref{subsec:fr}).
The shape of the potential $V(\phi)$ depends on the concrete form of the
function $f(R)$.
Thus, $f(R)$ gravity 
can exhibit the chameleon mechanism with an appropriate choice of the function $f(R)$.
Well-studied models of such chameleon $f(R)$ gravity
accounting for dark energy include~\cite{Starobinsky:2007hu,Hu:2007nk}.

Before closing this subsection, let us discuss briefly 
the symmetron mechanism~\cite{Hinterbichler:2010es},
which also relies on the interplay between the potential of
the scalar field and the conformal coupling to matter.
In the symmetron model,
the coupling function is given by
$A^{-1}(\phi) = 1 + \beta\phi^2/(2M_{\mathrm{Pl}}^{2})$,
while
the potential is chosen to be a Mexican-hat form,
\begin{align}
    V(\phi) = - \frac{\mu^{2}}{2} \phi^{2}+ \frac{\lambda}{4} \phi^{4}
    \, .
\end{align}
Now, in the presence of
non-relativistic matter with $\tilde{T}^{\mu}_{\ \mu} = -\tilde{\rho}$,
the effective potential for the scalar field takes the following form:
\begin{align}
    V_{\mathrm{eff}} (\phi) =  \frac{1}{2} \left( \frac{\beta \tilde{\rho}}{M_{\mathrm{Pl}}^{2}} - \mu^{2} \right) \phi^{2}
    + \frac{\lambda}{4} \phi^{4} 
    \, .
\end{align}
From this, we see that the energy density $\tilde\rho$
controls the vacuum structure.
One can define the critical density $\rho_*:=\mu^2 M_{\textrm{Pl}}^2/\beta$,
and the scalar field acquires a non-zero VEV, $\phi_{\textrm{VEV}}\sim \pm\mu /\sqrt{\lambda}$,
in the low-density region ($\rho<\rho_*$) where the $Z_{2}$ symmetry
is spontaneously broken, while
$\phi$ has the vanishing VEV
in the high-density region ($\rho>\rho_*$) where the symmetry is restored. Since fluctuations $\delta\phi$ around $\phi=\phi_{\textrm{VEV}}$ couple to
matter as $\sim (\beta/M_{\textrm{Pl}}^2)\phi_{\textrm{VEV}}\delta\phi\tilde\rho$,
the coupling between the scalar field and matter is switched off
in the high-density region where $\phi_{\textrm{VEV}}=0$.
This is how the symmetry mechanism operates.



Although the chameleon and symmetron mechanisms are both
controlled by the matter field,
they screen the scalar-mediated force in different ways:
the chameleon mechanism increases the effective mass and suppresses the propagation of the fifth force, 
while the symmetron mechanism suppresses the coupling to matter.
Therefore, the way screening works in the symmetron mechanism is
in some sense similar to that in the Vainshtein mechanism.

\subsection{Positivity bound}
\label{sec:Positivity bound}

Despite the diversity of modified gravity models and their predictions, almost all of them share the same property: most of them are described by non-renormalisable Lagrangian and hence would be understood as low-energy effective field theories (EFTs) of some unknown ultraviolet (UV) completion of gravity. From this viewpoint, it is interesting to ask which modified gravity models can arise as low-energy EFTs from a model of quantum gravity. Once this question is answered, one can get some insight into quantum gravity indirectly from observational constraints on modified gravity models.

Recently, it has been recognised that EFTs which can be embedded into standard UV completions have to satisfy several consistency conditions, so-called positivity bounds. 
Roughly speaking, the EFT interactions arise by integrating out UV degrees of freedom and the interactions between UV degrees of freedom and EFT degrees of freedom are subject to unitarity constraints; accordingly, the coupling constants of the EFT should be also subject to the unitarity constraints.
The bounds are originally formulated in the context of non-gravitational theories~\cite{Pham:1985cr,Ananthanarayan:1994hf,Adams:2006sv}. Their extensions to gravitational theories have been discussed recently (see e.g.,~\cite{Cheung:2014ega,Hamada:2018dde,Andriolo:2018lvp,Bellazzini:2019xts,Chen:2019qvr,Alberte:2020jsk,Tokuda:2020mlf,Loges:2020trf,Herrero-Valea:2020wxz, Caron-Huot:2021rmr,Alberte:2020bdz,Aoki:2021ckh,Noumi:2021uuv,Alberte:2021dnj,Chiang:2022jep,Caron-Huot:2022ugt} and references therein). In particular, constraints on lowest-order coefficients are recently extended in a way applicable to gravitational theories with some additional working assumptions~\cite{Hamada:2018dde, Tokuda:2020mlf, Caron-Huot:2021rmr}. Below, we review positivity bounds by following~\cite{Adams:2006sv,Tokuda:2020mlf} and then discuss their implications. 

\subsubsection{Non-gravitational positivity bound}
\begin{figure}
\centering
\includegraphics[width=100mm, trim=130 200 100 70]{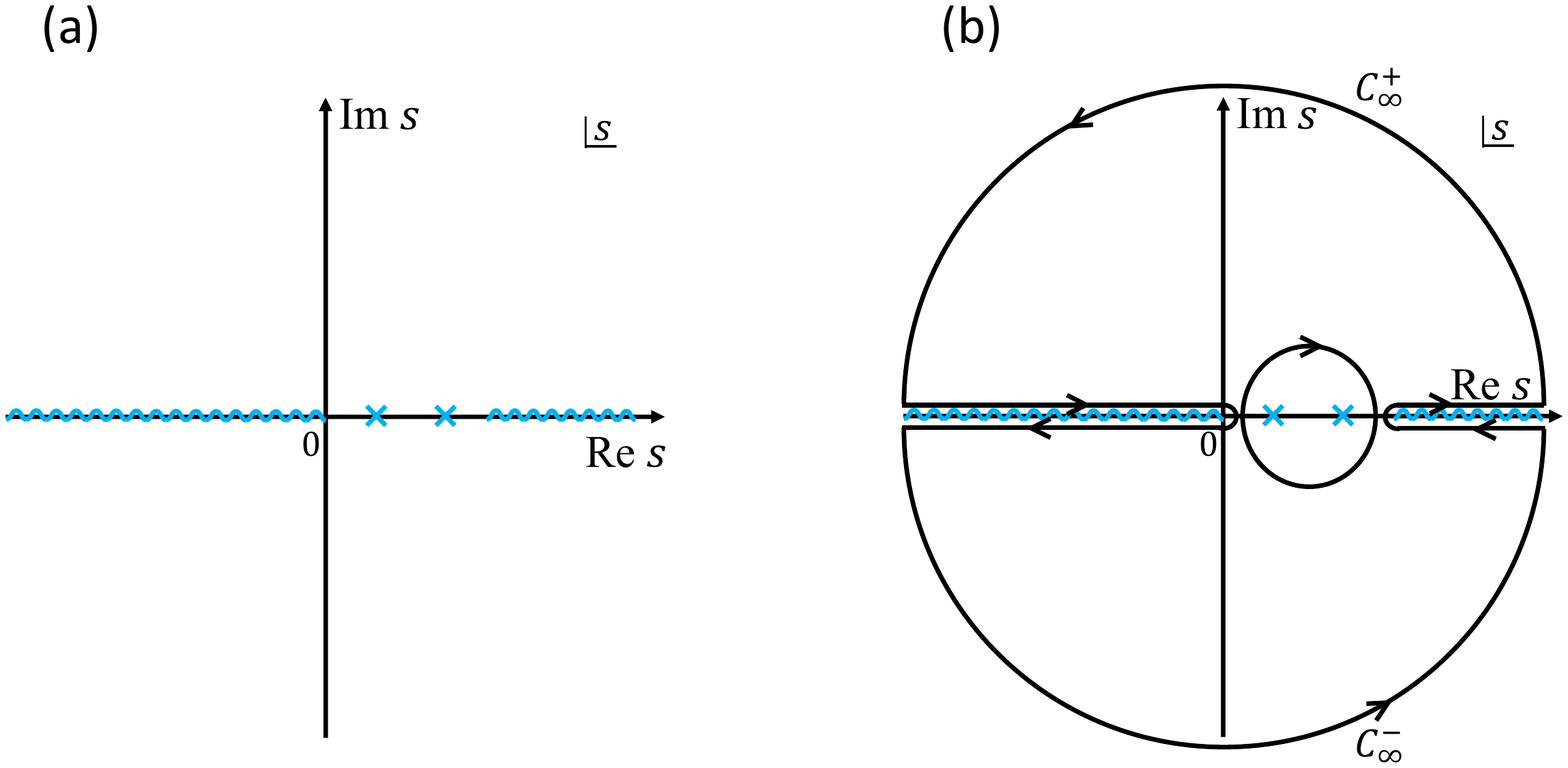}
\caption{Fig.(a): the analytic structure of $\mathcal{M}(s,0)$ in the complex $s$-plane. The light-blue "$\times$" and the wavy lines are poles and branch cuts, respectively. Fig.(b): the integration contour to derive Eq.~\eqref{eq:disp1}. }
\label{fig:cont}
\end{figure}

Positivity bounds are formulated in terms of the 2 to 2 scattering of light fields which are contained in EFT spectra. As the simplest example, we focus on a scalar EFT in four dimensions {\it without} gravity,
\begin{align}
\mathcal{L}=-\frac{1}{2}(\partial\phi)^2-\frac{m^2}{2}\phi^2-V(\phi)+\frac{\alpha}{\Lambda^4}(\partial\phi)^4+\cdots\,,\label{eq:lag1}
\end{align}
and consider the scattering amplitude $\mathcal{M}(s,t,u)$ of the process $\phi(k_1)\phi(k_2)\to\phi(k_3)\phi(k_4)$ (all momenta ingoing). Here, $s$, $t$, and $u$ are usual Mandelstam variables defined by $s:=-(k_1+k_2)^2$, $t:=-(k_1+k_3)^2$, and $u:=-(k_1+k_4)^2$. These variables satisfy an equality $s+t+u=4m^2$ and we regard $\mathcal{M}$ as a function of $(s,t)$. We write $\mathcal{M}$ as $\mathcal{M}(s,t)$ below. Note that we do not distinguish between $m$ and the physical pole mass for simplicity. Let us consider the series expansion around the $s\leftrightarrow u$ crossing symmetric point $s=2m^2-t/2$,
\begin{align}
\mathcal{M}(s,t)=(s,t,u\text{-channel poles})+\sum_{n=0}^\infty c_n(t)\left(s-2m^2+\frac{t}{2}\right)^n\,.
\end{align}
Suppose that the EFT cutoff scale $\Lambda$ satisfies $\Lambda\gg m$. The low-energy part of the amplitude is calculable within the EFT region $\{(s,t,u):|s|,|t|, |u|<\Lambda^2\}$, and then the expansion coefficients $\{c_n(t)\}_{\forall n}$ are expressed in terms of the coupling constants of the EFT Lagrangian.
In contrast, we cannot compute $\mathcal{M}(s,t)$ outside the EFT region without specifying the UV completion.
Nonetheless, it has been known that $\mathcal{M}(s,t)$ must possess several properties (even outside the EFT region) if UV completion is  Lorentz invariant, unitary, local, and causal. Causality implies analyticity: $\mathcal{M}(s,0)$ is analytic except for usual poles and cuts associated with single particle exchanges and multi-particle exchanges in the complex $s$-plane, see Fig.~\ref{fig:cont}. 
Analyticity, unitarity, and locality in the form of polynomial-boundedness assumptions on $\mathcal{M}$ give the $s^2$-boundedness $\lim_{|s|\to\infty}|\mathcal{M}(s,0)/s^2|=0$ thanks to the Phragm{\'{e}}n-Lindel{\"{o}}f theorem in the absence of massless $t$-channel pole.\footnote{For the derivation of positivity bounds in the absence of polynomial boundedness assumptions and its relation to non-localisability of theories, see \cite{Keltner:2015xda,Tokuda:2019nqb}.
} This is an extension of the Froissart-Martin bound~\cite{Froissart:1961ux, Martin:1962rt}  
to the complex $s$-plane. These properties admit a dispersive sum rule for $c_2$ by integrating $\mathcal{M}(s,0)/(s-2m^2)^3$ along the contour shown in Fig.~\ref{fig:cont}~\cite{Adams:2006sv}:
\begin{align}
&c_2(0)=\int^\infty_{4m^2}\frac{\mathrm d s}{\pi}\,\frac{\mathrm{Disc}_s\mathcal{M}(s,0)/i}{(s-2m^2)^3}>0\,,\label{eq:disp1}
\end{align}
where $\text{Disc}_s\mathcal{M}(s,t):=\mathcal{M}(s+i\epsilon,t)-\mathcal{M}(s-i\epsilon,t)$. Unitarity ensures $\text{Disc}_s\,\mathcal{M}(s,0)/i>0$, giving the strict inequality for $c_2(0)$. Separating EFT and UV quantities in the above equation by introducing $B(\Lambda):=c_2(0)-\int^{\Lambda^2}_{4m^2}\frac{\mathrm d s}{\pi}\,\frac{\mathrm{Disc}_s\mathcal{M}(s,0)/i}{(s-2m^2)^3}$, we have~\cite{Bellazzini:2016xrt, deRham:2017imi} 
\begin{align}
&B(\Lambda)
=\int^\infty_{\Lambda^2}\frac{\mathrm d s}{\pi}\,\frac{\mathrm{Disc}_s\mathcal{M}(s,0)/i}{(s-2m^2)^3}>0\,.\label{eq:disp2}
\end{align}
Eqs.~\eqref{eq:disp1} and \eqref{eq:disp2} are called positivity bounds. (Eq.~\eqref{eq:disp2} is sometimes called improved positivity bounds because the bounds are strengthened by subtracting the calculable part of the integral.) For instance, we have $c_2(0)=16\alpha$ at the tree-level approximation, meaning that an EFT \eqref{eq:lag1} with $\alpha\leq0$ is inconsistent with the standard UV completion. 
Note that one can derive dispersive sum rules for higher-order coefficients including $\{c_n\}_{n>2}$. It is also possible to go away from the forward limit~\cite{deRham:2017avq}. 
Also, one can derive new bounds on Wilson coefficients by using crossing symmetries~\cite{Bellazzini:2020cot, Tolley:2020gtv, Caron-Huot:2020cmc, Arkani-Hamed:2020blm} which lead to two-sided bounds; that is, the Wilson coefficients are bounded from above and below.

\subsubsection{Gravitational positivity bound}
It is very non-trivial to extend the above analysis in a way applicable to gravitational theories. The proof of $s^2$-bound $\lim_{|s|\to\infty}|\mathcal{M}(s,t<0)/s^2|=0$ is not known due to the presence of massless graviton $t$-channel pole. The $s^2$-bound is satisfied in known examples of UV-complete gravitational amplitudes such as string amplitudes, however.\footnote{
For recent discussions on the high-energy behaviour of gravitational scatterings (see e.g.,~\cite{Haring:2022cyf}). Also, the $s^2$-bound is consistent with the result of \cite{Caron-Huot:2021enk}.}
We assume the $s^2$ bound and discuss its implications. Under this assumption, we can derive a dispersive sum rule for $B(\Lambda)$ as we did in non-gravitational setups. Importantly, because graviton $t$-channel exchange gives a singular term $\mathcal{M}\ni M_\text{pl}^{-2}t^{-1}su$ which grows as fast as $s^2$, a sum rule for $B(\Lambda)$ now reads
\begin{align}
&B(\Lambda)=\lim_{t\to0^-}\left\{\int^\infty_{\Lambda^2}\frac{\mathrm d s}{\pi}\,\frac{\mathrm{Disc}_s\mathcal{M}(s,t)/i}{(s-2m^2+(t/2))^3}+\frac{1}{M_\text{pl}^2t}\right\}\,.\label{eq:disp3}
\end{align}
The right-hand side is ``$\infty-\infty$'' whose sign is unclear. To realise the cancellation of $\mathcal{O}(t^{-1})$ singular term, it will be required that the $\text{Disc}_s\,\mathcal{M}(s,t)$ grows as fast as $s^2$ at $t=0$ while slower than $s^2$ for $t<0$. This suggests us to parameterise the growth rate as $\text{Disc}_s\mathcal{M
}(s,t)\sim s^{\alpha(t)}$ , called Regge behaviour, in the UV region. Physically, the Regge behaviour implies that the UV theory has a tower of higher spin states and this behaviour is indeed realised in examples such as Virasoro-Shapiro amplitude. In~\cite{Tokuda:2020mlf}, the cancellation of $\mathcal{O}(t^{-1})$ term is demonstrated under the assumption of the Regge behaviour $\text{Disc}_s\mathcal{M}(s,t)/i=f(t)\left(s/M_\text{regge}^2\right)^{2+\alpha't+\alpha''t^2+\cdots}+(\text{sub-leading terms})$ with $\alpha'>0$. Here, $M_\text{regge}$ denotes the lightest mass scale of the heavy physics which Reggeizes the amplitude. In the tree-level superstring example, we have $\alpha' \sim M_\text{regge}^{-2}$ with $\alpha''=0$.
An explicit computation of finite terms of the right-hand side of \eqref{eq:disp3} shows the gravitational positivity bound~\cite{Tokuda:2020mlf}~(see also \cite{Hamada:2018dde, Loges:2020trf, Herrero-Valea:2020wxz}): 
\begin{align}
B(\Lambda)>-\mathcal{O}(M_\text{pl}^{-2}M_\text{regge}^{-2})\,,\label{eq:gravposi}
\end{align} 
under the single scaling assumption $\alpha'\sim|\alpha''/\alpha'|\sim|\partial_tf/f|_{t=0}\sim\mathcal{O}(M_\text{regge}^{-2})$. As long as the scale of Reggeization $M_\text{regge}$ is high enough, the bound \eqref{eq:gravposi} can be regarded as an approximate positivity bound. The bound \eqref{eq:gravposi} is distinct from those conjectured in \cite{Alberte:2020jsk, Alberte:2020bdz} which depends on EFT cutoff scales or mass scales of fields in EFTs. 

Note that the subtleties discussed above are absent if the graviton is massive. Then, positivity bounds on massive gravity are discussed in the literature (see e.g., \cite{Bellazzini:2017fep,deRham:2017xox,deRham:2018qqo,Alberte:2019lnd}).

\subsubsection{Implications of gravitational positivity bound}
Gravitational positivity bounds have been applied to various models: for cosmological models, bounds on Horndeski theory are obtained in \cite{Melville:2019wyy,deRham:2021fpu}, and the bounds on scalar-tensor EFT and DHOST theory are obtained in \cite{Tokuda:2020mlf}. Such bounds, however, constrain Wilson coefficients of models expanded around the Minkowski solution, not those expanded around the cosmological background. Phenomenologically, we are interested in the latter. Because Lorentz invariance is broken due to the non-trivial background, it is not fully understood how one can extend positivity bounds in a way applicable to such situations (for recent developments, see e.g., \cite{Baumann:2015nta, Grall:2021xxm, Aoki:2021ffc,Melville:2022ykg,Creminelli:2022onn}). 

Having said that, it is possible to {\it guess} several implications for the models of modified gravity. The Lagrangian \eqref{eq:lag1} is known as the k-essence theory, one of the simplest theories within the Horndeski family. For simplicity, let us assume the shift symmetry $(m=V(\phi)=0)$ and consider perturbations around a ``cosmological'' background $\phi=\epsilon \Lambda^2 t + \pi(t,{\bf x})$ with a constant parameter $\epsilon$. The quadratic action is given by
\begin{align}
\mathcal{L}^{(2)}= \frac{1}{2}(1 + 12  \epsilon^2 \alpha) ( \dot{\pi}^2 - c_s^2 \partial_i \pi \partial^i \pi ) + \cdots
\,,
\end{align}
with
\begin{align}
c_s^2=\frac{1+4\epsilon^2 \alpha}{1+12 \epsilon^2 \alpha},
\end{align} 
where gravitational effects are ignored for simplicity. In general, higher-order terms such as $(\partial \phi)^6$ also contribute, but they can be ignored when $\epsilon$ is sufficiently small. One can see that the positivity bound $\alpha>0$ states that the sound speed $c_s$ has to be subluminal, $c_s<1$. In this sense, the positivity bound could be interpreted as a causality constraint on the EFT~\cite{Adams:2006sv} (see also \cite{CarrilloGonzalez:2022fwg} for a recent study). Even though the validity of this relation is not trivial in generic situations, especially in the presence of gravity~\cite{Camanho:2014apa,Hollowood:2015elj,deRham:2019ctd,deRham:2020zyh,Bellazzini:2021shn,Chen:2021bvg,deRham:2021bll,Caron-Huot:2022ugt}, prohibiting a superluminal propagation provides important implications for modified gravity. First of all, the sound speed of perturbations around a cosmological background generically deviates from the speed of light, and a large parameter space might be excluded if a subluminal propagation speed is imposed. The situation is more stringent in the screening mechanisms. A superluminal propagation generically appears around solutions exhibiting the kinetic screening or the Vainshtein screening~(see e.g.,~\cite{Joyce:2014kja}). In the case of the k-essence theory, the sign for the superluminal propagation and the sign for suppressing the local fifth force precisely agree with each other even considering a generic functional form~\cite{Joyce:2014kja, Aoki:2021ffc}. The superluminal propagation also appears around the Vainshtein screening, and it is recently shown that the (weakly-broken) galileon theory around the flat spacetime is incompatible with the two-sided bounds obtained by using full crossing symmetry~\cite{Bellazzini:2020cot, Tolley:2020gtv, Caron-Huot:2020cmc, Arkani-Hamed:2020blm}. It would be important to understand the precise relations between the assumptions on the UV completion and the screening mechanisms latter of which can be tested observationally.

The above arguments are essentially based on the knowledge from non-gravitational positivity bounds, while gravitational positivity bounds can get one step further.
Interestingly, once renormalisable theories are coupled to gravity, models are often constrained by the gravitational positivity bounds \eqref{eq:gravposi}. When considering the system consisting of massless $U(1)$ gauge bosons and charged fermions/scalars, gravitational positivity bounds on scatterings of $U(1)$ gauge bosons often lead to the Weak-Gravity-Conjecture-like bounds~\cite{Cheung:2014ega,Andriolo:2018lvp,Chen:2019qvr}, and could be even stronger~\cite{Alberte:2020bdz}. In particular, when applying these bounds to QED, the cutoff scale reads $\Lambda\sim10^8$~GeV~\cite{Alberte:2020bdz}. This scale is much lower than the GUT scale or the expected quantum gravity scales. It is found that the cutoff is raised to around $10^{16}$~GeV in the Standard Model, which is comparable to the expected GUT scale~\cite{Aoki:2021ckh}. 
One can also derive non-trivial constraint on scalar theories~\cite{Alberte:2020jsk, Noumi:2021uuv}: for example, precise bounds on scalar potentials are obtained in~\cite{Noumi:2021uuv} which may be understood as swampland conditions for scalar potentials. Also, a recent work~\cite{Noumi:2022zht} discusses the implications of \eqref{eq:gravposi} to the physics of dark sector. See e.g., \cite{Andriolo:2018lvp,Alberte:2021dnj} for earlier discussions along this line of considerations. Ref.~\cite{Noumi:2022zht} discusses dark photon models as an illustrative example and computes the photon-dark photon amplitudes. Interestingly, it is found that the bounds are so stringent that can be tested against ongoing experiments. These results clarify the phenomenological importance of gravitational positivity bounds.

One should keep in mind, however, that the assumptions which lead to the bound~\eqref{eq:gravposi} are not yet fully justified. For instance, once the single scaling assumption mentioned below \eqref{eq:gravposi} is violated, then negative value of $B(\Lambda)$ may be allowed to alter the implications discussed above. A recent interesting paper \cite{Alberte:2021dnj} pointed out that for graviton-photon amplitudes the single scaling assumption should be violated for consistency of the sum rule. It is important to understand how generic this feature is beyond the graviton-photon scattering. It is desired to derive positivity bounds with fewer assumptions.\footnote{Ref.~\cite{Caron-Huot:2021rmr} achieved the derivation of positivity bounds in dimensions higher than four by assuming the $s^2$-bound and analyticity only, but the obtained bound is much weaker than \eqref{eq:gravposi} once loop corrections are included. The loop corrections are essential for obtaining non-trivial consequences of gravitational positivity bounds.}

\section{Observables for testing gravity}\label{sec:observables for testing gravity}

This section describes cosmological observables to test modified gravity. 
The principal observables are the statistical correlations of the fluctuations of CMB temperature/polarisations and the spatial distribution of galaxies at large scales. As secondary effects, gravitational lensing/Sunyaev-Zel'dovich effects give detailed information on photon scattering via gravitational potentials and environments. These observables trace the physics of gravitational interactions on cosmological scales. 
%

To detect signatures of modified gravity, it is necessary to signify deviations from the standard predictions {of the $\Lambda$CDM model} in the observables.
Although the detailed predictions depend on models of modified gravity, the major changes are caused by the change of the background dynamics and the modifications of the Poisson and lensing equations at the linear order in perturbations. As the first step for testing possible deviations from the $\Lambda$CDM model, we shall introduce phenomenological parameters to characterise these effects of modified gravity. The phenomenological parameters enable us to discuss general constraints on the modification of gravity regardless of the details of the model.
Although there are many possible ways of introducing such parameters, here we introduce some of the most commonly used parameters in current analyses in Sec.~\ref{sec:Signatures of modified gravity against LCDM model}. 

We then give comprehensive information about the observations and the observables of CMB and LSS in Secs.~\ref{sec:Cosmic Microwave Background} and \ref{subsec:LSS}, respectively.
By utilising the phenomenological parameters (or using predictions of a concrete theory), we can test the modifications of gravity with observational data.
Forthcoming surveys can overcome systematic uncertainties and gain statistical power thanks to wider and deeper probes over the cosmic volume. 
However, we stress that the actual observables are not phenomenological parameters, and the phenomenological parameters are used to translate the observational data into the constraints on the modifications of gravitational physics. It is, therefore, important to enumerate the {\it observables} in a coverall manner and understand observational strategies for testing gravity.

\subsection{Basic equations for testing modified gravity against \texorpdfstring{$\Lambda$CDM}{LCDM} model}
\label{sec:Signatures of modified gravity against LCDM model}

This subsection briefly introduces phenomenological parametrisations commonly used for testing gravity on cosmological scales, based on the standard formulae for the evolution of the background universe and the growth of the matter density fluctuations. The signatures of modified gravity arise in observables as deviations from the predictions of the standard $\Lambda$CDM model and GR. Appropriate phenomenological parametrisations help observationally verify to what extent the 
GR is consistent, as well as constraining theoretical information therein. 
To clarify the connection to the observables and of keeping the spirit of minimalism of introduced parameters, we admit a minimum set of phenomenological parameters here. 
\\


We start from the flat Friedmann-Lem\^aitre-Robertson-Walker (FLRW) metric in the form
\begin{align}
	{\rm d}s^2=-{\rm d}t^2+a^2(t)\left({\rm d}\chi^2+\chi^2{\rm d}\Omega^2\right)
	\,,
\end{align}
where $\chi$ is the comoving radial distance and $a$ is the scale factor.
In modified gravity, it is often the case that an additional dynamical degree of freedom contributes as a dark energy component,
and such a component generally provides the time-varying energy density and pressure.
To describe the effect of the dynamical dark energy component on the background evolution of the universe, the effective equation-of-state parameter $w(a)$ is commonly used,
and such a model is called $w$CDM model instead of the standard $\Lambda$CDM one. 
The definition of $w(a)$ is given as
\begin{align}
	w(a)=\frac{P_{\rm DE}(a)}{\rho_{\rm DE}(a)}
	\,,
\end{align}
where $\rho_{\rm DE}$ and $P_{\rm DE}$ denote the background energy density and pressure of the dark energy component, respectively.
Assuming that the dark energy component minimally couples
to other matter sectors\footnote{Once the non-minimal coupling between the dark energy and matter components is introduced,
each component no longer satisfies the conservation law individually~\cite{Chibana:2019jrf,Kimura:2017fnq,Koivisto:2012za,Zumalacarregui:2012us}.
In this case, the formula of the expansion rate in the $w$CDM should be modified. 
} and obeys the usual energy-conservation law, the expansion rate for the $w$CDM model in the late-time universe can be written as
\begin{align}
	H^2(a)=H_0^2\Biggl[\frac{\Omega_{\rm m}}{a^3}+\left(1-\Omega_{\rm m}\right)\exp\left(-3\int_0^{\ln a}[1+w(a^\prime )]\,{\rm d}\ln a^\prime \right)\Biggr]
	\,,
\end{align}
where the present scale factor is set to $a=1$. We have neglected the radiation energy density.
The basic parameters for the background are the Hubble-Lem\^aitre constant $H_0$ and the present matter density parameter $\Omega_{\rm m}$.
In general, 
the way the equation-of-state parameter $w(a)$ varies with time depends on the model of modified gravity under consideration.
As a simple example of the functional form for the model-independent analysis,
an expanded form of function with $1-a$ is usually employed, defined as~\cite{Linder:2002et,Chevallier:2000qy}
\begin{align}
	w(a)=w_0+\left(1-a\right)w_a
	\,,\label{eq:CPL model}
\end{align}
where $w_0$ and $w_a$ denote the two phenomenological constant parameters.
In this model, the expansion rate can be rewritten as
\begin{align}
	H^2(a)=H_0^2\biggl[\frac{\Omega_{\rm m}}{a^3}+\left(1-\Omega_{\rm m}\right) a^{-3(1+w_0+w_a)}e^{-3w_a(1-a)}\biggr]
	\,.  
\end{align}
This $(w_0,w_a)$ parametrsation is known to reproduce the expansion history of the late-time universe relatively well
for a wide class of modified gravity theories up to a sufficiently high redshift and has been used widely for
the constraints on dynamical dark energy and modified gravity model.
It should be noted that there are several subtleties in this approach. In the general class of modified gravity theories, 
it may be non-trivial to identify the energy density and the pressure of the dark energy component, $\rho_{\rm DE}$ and $P_{\rm DE}$, in the system of the background Friedmann and Raychaudhuri equations. In that case,
the equation-of-state parameter is not well defined. This issue will be discussed in Sec.~\ref{sec:Density perturbations in scalar-tensor theories}.
\\

Not only the background dynamics but also the spatial matter density fluctuations
in the FLRW universe should be affected by the modification of gravity theory,
and they can be scrutinised from the precise observations of the evolution of the LSS.
Then, let us introduce the phenomenological parameters to characterise the matter density fluctuations.
For the current purpose, we will adopt the metric perturbations in the Newtonian gauge around
the background FLRW spacetime,
\begin{align}
	{\rm d}s^2=-\Bigl[ 1+2\Phi (t,{\bf x})\Bigr]{\rm d}t^2+a^2(t)\Bigl[1-2\Psi (t,{\bf x})\Bigr]\delta_{ij}{\rm d}x^i{\rm d}x^j
	\,.\label{eq:conformal Newton metric}
\end{align}
For the matter sector, we consider
the density fluctuation, $\delta$, which is defined by
\begin{align}
\rho_{\rm m} (t, {\bf x}) =
\rho_0 (t) \left[ 1+ \delta (t, {\bf x}) \right],
\end{align}
where $\rho_0$ denotes the background matter density.
Though various parametrisations
describing the effect of modified gravity on the growth of structure
have been considered in the literature,
one of the extensively-discussed parametrisations
is $(\mu, \Sigma$) parametrisations in which
%
the Poisson and lensing equations 
take the following form in the Fourier space~\cite{Silvestri:2013ne,Zhao:2010dz,Zhao:2008bn,Pogosian:2010tj}:
\begin{align}
	&k^2\Phi (t,{\bf k})=-4\pi G_{\rm N}\mu (t,k)\rho_0 (t)\delta (t,{\bf k})
	\,,\label{eq:Poisson eq}\\
	&k^2\Bigl[\Phi  (t,{\bf k})+\Psi (t,{\bf k})\Bigr] =-8\pi G_{\rm N}\Sigma (t,k)\rho_0 (t)\delta (t,{\bf k})
	\,,\label{eq:lensing eq}
\end{align}
by introducing
the phenomenological functions $\mu (t,k)$ and $\Sigma (t,k)$ which represent the deviations from GR ($\mu=\Sigma=1$ in GR).
We should note that the Newton constant $G_{\rm N}=1/(8\pi M_{\rm Pl}^2)$ introduced in Eqs.~\eqref{eq:Poisson eq} and \eqref{eq:lensing eq} 
is now defined not through the cosmological equations but through the local test of Newton's law.
Instead of $\Sigma$, the gravitational slip parameter $\eta (a,k)$ is also often used, which is defined as\footnote{The definition of the gravitational slip parameter is the same as for the PPN parameter, $\gamma_{\rm PPN}$, which was introduced in Sec.~\ref{sec:screening}. Conventionally, $\gamma_{\rm PPN}$ is used as a phenomenological parameter for solar-system tests, while the gravitational slip parameter is used for cosmological observations.}
\begin{align}
	\eta (t,k):=\frac{\Psi}{\Phi}
	\,.
\end{align}

The recent observational constraints based on this parametrisation by
using CMB measurements and weak lensing data from
LSS surveys have been reported in~\cite{DES:2022ygi,DES:2018ufa,Joudaki:2016kym,Planck:2015bue,Simpson:2012ra}.
To practically obtain the observational constraints, a functional form of these parameters as a function
of time and scale should be specified as in the case of the equation-of-state parameter $w(a)$.
As a specific choice for large-scale observations, the scale-independent functional form is widely employed:
$\mu (a)=1+\mu_0\,\Omega_{\rm DE}(a)/\Omega_{\rm DE,0}$ and $\Sigma (a)=1+\Sigma_0\,\Omega_{\rm DE}(a)/\Omega_{\rm DE,0}$,
where $\Omega_{\rm DE}(a)$ is the density parameter of the dark energy component at cosmic scale factor $a$ ($\Omega_{\rm DE} (a) = 1 - \Omega_{\rm m} (a)$ in the late-time flat Universe, where $\Omega_{\rm m}(a)$ represents the density parameter of the matter at cosmic scale factor $a$,
which is given by $\Omega_{\rm m}(a) =H_0^2\Omega_{\rm m}/a^3H^2(a)$), and $\Omega_{\rm DE,0}$
is its present value.
It is simply motivated by the natural thought that the deviations from GR have to be associated with cosmic acceleration, and hence
in the matter-dominated Universe, GR should be recovered,
that is, $\mu = \Sigma = 1$.
We should note that the mapping of these phenomenological parameters to specific theories of gravity is not fully understood and 
the effective gravitational couplings have difficulty in choosing the theoretical prior distributions of parameters.
As shown in Sec.~\ref{sec:Perturbative predictions in modified gravity}, $\mu$ and $\Sigma$ for specific models of modified gravity have complicated dependence on
arbitrary functions of a parent theory. 
Although there are these subtleties, this parametrisation is still useful to understand how signals in observables deviate from
the standard scenario.
\\

While the above parametrisation focused on the effects of modified gravity that appear in the equations relating to matter and gravity, now let us introduce an extensively-studied effective parameter that describes the growth of matter fluctuations.
Assuming that the matter sector is minimally coupled to the gravity sector\footnote{When one breaks the assumption of the minimal coupling to the gravity sector,
both the continuity and Euler equations get modified. 
In particular, the physical interpretation of observational data obtained
by measurements of redshift-space distortions changes drastically~\cite{Chibana:2019jrf,Kimura:2017fnq,Koivisto:2012za,Zumalacarregui:2012us}.} and focusing on the evolution on sub-horizon scales,
the matter density fluctuation $\delta (t,{\bf x})$ and the velocity field $v^i(t,{\bf x})$ 
obey the standard fluid equations in the Newtonian limit.
The continuity and Euler equations yield
\begin{align}
	&\frac{\partial\delta}{\partial t}+\frac{1}{a}\partial_i\bigl[(1+\delta )v^i\bigr] =0
	\,,\label{eq:continuity eq}\\
	&\frac{\partial v^i}{\partial t}+Hv^i+\frac{1}{a}v^j\partial_j v^i=-\frac{1}{a}\partial^i\Phi
	\,.
\end{align}
To linear order, these equations are combined to give
\begin{align}
	\frac{\partial^2\delta}{\partial t^2}+2H\frac{\partial\delta}{\partial t}-\frac{\nabla^2}{a^2}\Phi =0
	\,.
	\label{eq:delta eq}
\end{align}
If the (effective) Poisson equation is properly given, the equation for the linear density fluctuation can be solved.
The effects of modified gravity come into play only through the gravitational potential $\Phi$.
Substituting Eq.~\eqref{eq:Poisson eq} into the above equation, we obtain the evolution equation for the density fluctuation:
\begin{align}
	\frac{\partial^2\delta}{\partial t^2}+2H\frac{\partial\delta}{\partial t}-4\pi G_{\rm N}\mu\rho_0 \delta =0
	\,.
\end{align}
When the phenomenological function in the Poisson equation, $\mu$, depends only on time,
the time dependence of the linear density field can be expressed independently of the wave number. In particular, on large scales, the density fluctuations can be simply expressed as
$\delta (a,{\bf k})=D(a)\delta_{\rm ini}({\bf k})$, where $D(a)$ and $\delta_{\rm ini} ({\bf k})$ denote the linear growth and
the initial density fluctuation.\footnote{On smaller scales, matter density fluctuations that enter the horizon in the radiation-dominated era should have scale dependence in terms of when they enter the horizon due to the growth suppression of fluctuations during the radiation-dominated era. In addition, so-called baryon acoustic oscillations also appear scale-dependent, and these scale dependencies are often expressed as a transfer function.}
Once the solution of $\delta$ is obtained, from the continuity equation \eqref{eq:continuity eq},
the linear velocity divergence field defined as $\theta (a,{\bf x}):= -\partial_i v^i(a,{\bf x})/aH(a)$ is written as
\begin{align}
	\theta (a,{\bf x})= f(a)\delta (a,{\bf x})
	\,,
\end{align}
where $f$ represents the linear growth rate of structure, which is defined as
\begin{align}
	f(a):=\frac{{\rm d}\ln D(a)}{{\rm d}\ln a}
	\,.
 \label{eq:LGR}
\end{align}
Among varieties of current cosmological observational data, measuring the growth rate $f(a)$ is believed to be a powerful
tool for the test of the modified gravity responsible for the current cosmic acceleration, which would not be possible just by looking at the evolution of the homogeneous and isotropic universe.
To compare the observational data and theoretical predictions efficiently, it should be useful to introduce a constant phenomenological
parameter, which depends on the theory. 
A minimal approach that has been conventionally used is to introduce the gravitational growth index $\gamma$, defined through
\begin{align}
	f(a)=[\Omega_{\rm m}(a)]^\gamma
	\,.\label{eq:f parametrisation}
\end{align}
This parametrisation was originally developed in \cite{1980lssu.book.....P}, and 
the potential to distinguish dark energy model 
and modified gravity
has been discussed in \cite{Linder:2005in,Linder:2007hg,Wang:1998gt}.
Within this formalism, $\gamma$ is a parameter that is different for different cosmological models:
in the $\Lambda$CDM model with GR, we expect the growth index to be approximately constant with $\gamma\approx 6/11$,
while $\gamma\approx 0.69$ for the self-accelerating DGP model~\cite{Linder:2005in}, $0.40$--$0.43$ for $f(R)$ model~\cite{Gannouji:2008wt,Tsujikawa:2009ku,Narikawa:2009ux,Motohashi:2009qn},
 $3n(16n-5)/(110n^2-47n+5)$ for the $n$-th kinetic braiding model~\cite{Kimura:2010di}.
See also \cite{Yamauchi:2017ibz} for the Horndeski theory and \cite{Hirano:2019nkz,Yamauchi:2021nxw} for the DHOST theory.
\\

So far we have looked at linear growth, but we can consider higher order.
Even if the initial density fluctuation is well described by linear theory, 
the continuity and Euler equations do indeed have non-linear terms.
The non-linearity of
the gravitational dynamics eventually dominates and we must correctly take into account the non-linear 
growth of fluctuations.
For the quasi-nonlinear regime, 
the density fluctuations in the Fourier space can be formally expanded in terms of
the linear density field, $\delta_{\rm L}$, as
\begin{align}
	\delta ({\bf k},a)=\delta_{\rm L}({\bf k},a)+\int\frac{{\rm d}^3p_1{\rm d}^3p_2}{(2\pi)^3}\delta_{\rm D}^3({\bf k}-{\bf p}_1-{\bf p}_2)F_2({\bf p}_1,{\bf p}_2;a)\delta_{\rm L}({\bf p}_1,a)\delta_{\rm L}({\bf p}_2,a)+\cdots
	\,.\label{eq:F2}
\end{align}
In the case of the Einstein-de Sitter universe in 
GR, the symmetric kernel function $F_2$ 
can be solved analytically and found to be time-independent~\cite{Bernardeau:2001qr}.
However, in addition to the non-linearity in the continuity and Euler equations, since the modification of gravity theory alters 
the non-linear structures of gravitational interactions,
the form of the kernel function should depend on the underlying theory of gravity.
In particular, when we focus on the specific type of scalar-tensor theory discussed in this paper, namely Horndeski theory and DHOST theory, 
the explicit form of the second-order kernel function can be derived as~\cite{Takushima:2013foa, Takushima:2015iha, Crisostomi:2019vhj, Lewandowski:2019txi, Hirano:2020dom}
\begin{align}
	F_2({\bf k}_1,{\bf k}_2,a)=\kappa (a)\biggl[ 1+\frac{1}{2}(\widehat{\bf k}_1\cdot\widehat{\bf k}_2)\left(\frac{k_1}{k_2}+\frac{k_2}{k_1}\right)\biggr]
				-\frac{2}{7}\lambda (a)\biggl[1-(\widehat{\bf k}_1\cdot\widehat{\bf k}_2)^2\biggr]
	\,,
\end{align}
with $\widehat{\bf k}:={\bf k}/k$.
This expression reproduces the well-known analytic result for the case of the Einstein-de Sitter universe in GR 
when $\kappa =\lambda =1$, while in the case of $\Lambda$CDM model, $\lambda$ slightly deviates from unity~\cite{Bernardeau:2001qr,Bouchet:1994xp, Fasiello:2022lff}. 
The effect of the modification of gravity theories can be captured in these functions and the time evolution of them drastically changes depending on the theory~\cite{Takushima:2013foa, Takushima:2015iha, Yamauchi:2017ibz, Hirano:2018uar,Hirano:2020dom, Yamauchi:2021nxw, Namikawa:2018erh}.
Thus, the late-time evolution of the coefficients that trace the non-linear growth of structure 
can deliver new information on the modification of gravity theories that would not be imprinted in the background and the linear growth~\cite{Yamauchi:2017ibz}.
\\

In this subsection, we have introduced the phenomenological parametrisations commonly utilised.
Both background and perturbative features beyond the concordance $\Lambda$CDM model 
play an important role in constraining the range of viable gravity theories.
In the following subsections, based on these phenomenological parametrisations, we will give comprehensive information about
cosmological observations such as CMB and LSS.

\subsection{Cosmic Microwave Background}
\label{sec:Cosmic Microwave Background}
\begin{figure*}
\centering
\includegraphics[width=0.85\textwidth]{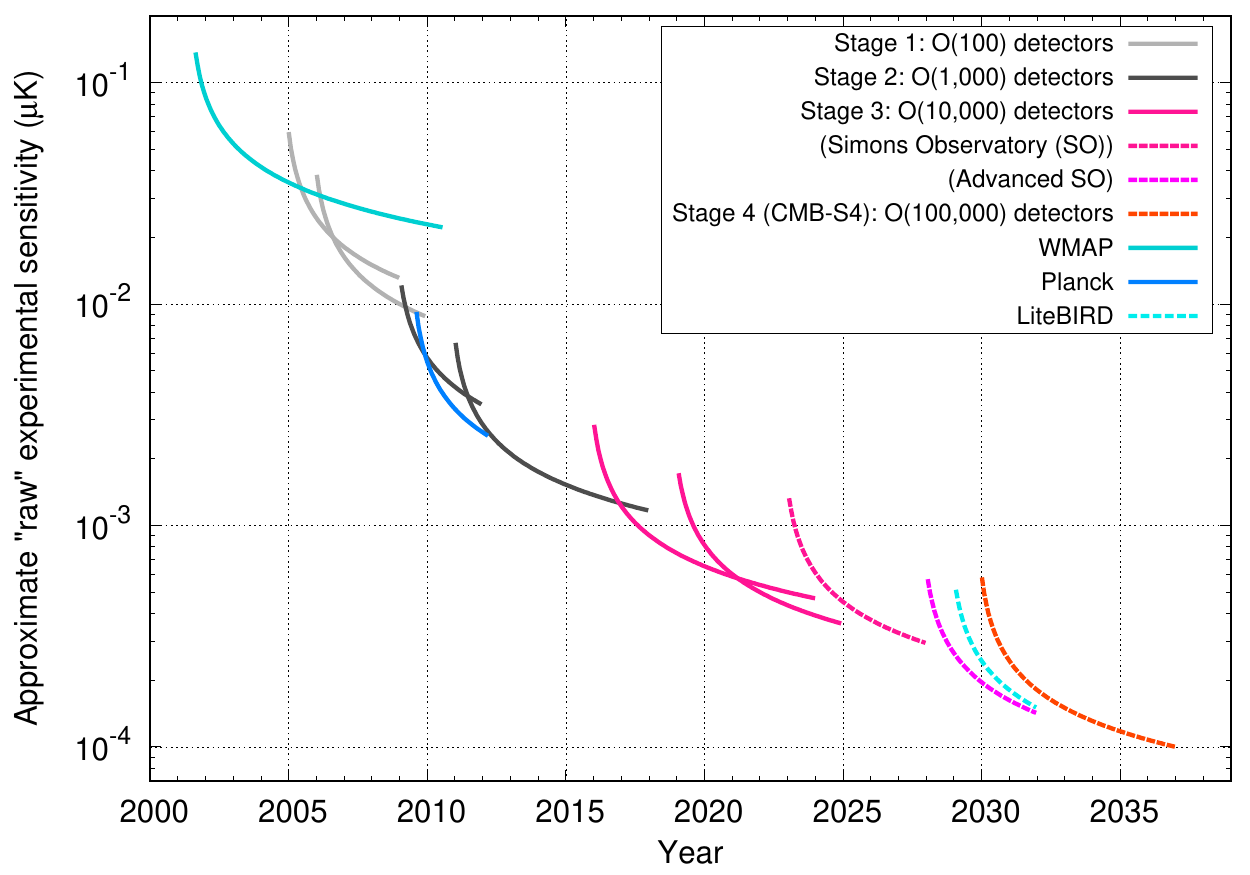}
\caption{The evolution of the ``raw'' sensitivity of ground-based experiments and satellite missions over time showing different stages.
The ``raw'' sensitivity is scaled as the combination of per-detector sensitivity, the number of detectors, and the length of observations.
The figure is adopted from the original figure by \cite{ABAZAJIAN201566}.}
\label{fig:forecastCMB}
\end{figure*}

%
The cosmic microwave background (CMB) is the relic photons of the Big Bang,
which is observed as isotropic radiation with the temperature $T=2.7255~$K.
There are tiny anisotropies in the CMB of the order of $10^{-5}$ K
originating from the quantum fluctuations during inflation (there are many review articles\footnote{See also \url{http://background.uchicago.edu/}} and textbooks on the CMB such as~\cite{Dodelson:2003ft,Lyth:2009zz}).
Thus, CMB 
is usually available to determine the cosmology and fundamental physics at early times,
while it can probe late-time Universe in combination with additional cosmological data, constraining how modified gravity or dark energy evolves at lower redshifts~\cite{Planck:2015bue}.
In this subsection,
as preparation for experimental investigations of modified gravity in the future,
first, we summarise the status and schedule of ongoing and future CMB experiments. The current and forecast sensitivities are shown in Fig.~\ref{fig:forecastCMB}.
Then, we also give a brief review of the CMB observables potentially relevant to constrain the modified gravity.
Details of the formulation for observables will be presented in Sec.~\ref{subsec:Boltzmann}.

\subsubsection{Summary of the status and schedule of ongoing and future CMB experiments}

\paragraph{ACT/ACTPol/AdvACT}

The Atacama Cosmology Telescope~(ACT) project~\cite{Thornton_2016} has been observing the CMB sky with arcmin resolution thanks to a 6~m aperture telescope on Cerro Toco in the Atacama Desert in Chile.
The first generation of receivers was the Millimeter Bolometric Array Camera~(MBAC),
which made CMB observations at 150, 220, and 280~GHz from 2007 to 2010.
The second was ACTPol, which was polarisation-sensitive and made CMB observations at 90 and 150~GHz from 2013 to 2015.
The final camera is AdvACT, which stands for Advanced ACTPol,
and is being made CMB observations at 30/40, 90, 150, and 220~GHz since 2016.

The ACT project made large data releases, called DR4 and DR5,
including temperature and polarisation maps at arcminute resolution
that cover ${\sim} 18{,}000\,{\rm deg^2}$ of the low-foreground sky
having a declination between $-60^\circ \leq \delta \leq 22^\circ$.
The survey has a large overlap with optical surveys like BOSS~\cite{Bautista:2020ahg},
HSC~\cite{Aihara:2018a}, DES~\cite{10.1093/mnras/stw641}, DESI~\cite{2016arXiv161100036D} and LSST~\cite{Ivezi__2019}
for cross-correlation studies.

\paragraph{{\sc Polarbear}/Simons Array}

The Simons Array~\cite{2016JLTP..184..805S} is an upgraded experiment on Cerro Toco in the Atacama Desert in Chile
as the successor to the {\sc Polarbear} experiment~\cite{Kermish_SPIE2012,Arnold_SPIE2012},
which had achieved the first measurement of the $B$-mode power spectrum in 2014~\cite{P_A_R_Ade_2014,Ade_2017}.
The Simons Array will cover 95, 150, and 220/270~GHz frequency bands with three telescopes.
Each telescope and receiver combination is named PB2-A, PB2-B, and PB2-C.
PB2-A and PB2-B will have nearly identical receivers operating at 95~GHz and 150~GHz.
PB2-A was deployed and achieved the first light in January 2019 and PB-2B is being deployed in 2021/2022.
PB2-C will house a receiver operating at 220/270~GHz.

The selection of the observation fields is driven by the science goals, i.e., $B$-mode measurements
and cross-correlation studies with other surveys.
The Simons Array observes a patch of the southern galactic sky
with low galactic foreground emissions~(so-called ``the southern hole''),
which is being examined by multiple CMB polarisation experiments and several galaxy surveys like LSST~\cite{Ivezi__2019},
and a north patch, overlapping the Subaru HSC WIDE surveys~\cite{POLARBEAR:2019:cross,Aihara:2018a}.

The Simons Array will cover the wide multipole ranges $50\leq\ell\leq 3000$,
thanks to novel demodulation technique by a continuously rotating half-wave plate~\cite{2016SPIE.9914E..2UH} and a 2.5~m aperture telescope,
and yield an upper limit of the tensor-to-scalar ratio $r<0.01$.

\paragraph{Simons Observatory}
The Simons Observatory~(SO)~\cite{Ade_2019} is a new CMB experiment being built on Cerro Toco in the Atacama Desert in Chile.
The SO collaboration is the first attempt at merging two CMB teams,
from the ACT and the {\sc Polarbear}/Simons Array experiments.
The site is also one of those planned for the future CMB-S4 experiment.

SO will measure the temperature and polarisation anisotropy of the CMB in six frequency bands centred
at 27, 39, 93, 145, 225, and 280~GHz.
The initial configuration of SO during the first five years of science observations from 2022 through 2027
will have three small-aperture 0.5-m telescopes~(SATs)
and one large-aperture 6-m telescope~(LAT), with a total of 60,000 cryogenic bolometers.

The SAT targets large angular scale polarisation covering the multipole ranges $30 \leq \ell \leq 300$,
mapping ${\approx} 10\%$ of the sky to a white noise level of 2\,$\mu$K-arcmin in combined 93 and 145~GHz bands,
to measure inflationary $B$-mode signals at a target level of $\sigma(r)= 0.003$.
The SAT integration has begun in 2020 and aims to deploy the first SAT in Chile in 2021.
The LAT maps ${\approx} 40\%$ of the sky at arcminute angular resolution
to a white noise level of 6\,$\mu$K-arcmin in combined 93 and 145~GHz bands,
overlapping with DESI~\cite{2016arXiv161100036D} and LSST~\cite{Ivezi__2019} for cross-correlation studies.
The LAT integration has begun in 2019 and aims to begin full scientific observation in 2023. SO is planned to be upgraded to the Advanced SO, which could achieve twice the sensitivity of SO.

\paragraph{SPT/{\sc SPTpol}/SPT-3G}
The South Pole Telescope~(SPT)~\cite{2014SPIE.9153E..1PB} is a 10~m telescope with arcminutes scale resolution
located at the Amundsen–Scott South Pole Station, Antarctica.
The first major survey and instruments~(SPT-SZ) were designed
to find massive clusters of galaxies through their interaction with the CMB
as well as to observe CMB anisotropies.
The SPT-SZ completed a 2500~${\rm deg^2}$ survey at 95, 150, and 220~GHz.

A new camera~({\sc SPTpol}) was installed and operated on the SPT between 2012 and 2016
with the capability to measure the polarisation.
One of the {\sc SPTpol} wide survey overlaps the ``BICEP'' patch
and the BICEP/Keck and {\sc SPTpol} collaborations have demonstrated
an improved constraint on inflationary $B$-mode signals with delensing.

The third-generation camera~(SPT-3G) was installed in 2017 on the telescope
with an order-of-magnitude better sensitivity over {\sc SPTpol}.

\paragraph{BICEP2/Keck Array/BICEP3/BICEP Array}
The BICEP/Keck experiment~\cite{2020SPIE11453E..14M} has been observing the CMB from the South Pole since 2010.
The instruments are optimised for detecting inflationary $B$-mode signals at degree scales
with a small aperture telescope~(${\sim} 0.3\text{--}0.5~{\rm m}$).
The experiment has developed and deployed the BICEP2~(150~GHz), Keck Array~(95, 150, 220, and 270~GHz),
BICEP3~(95~GHz), BICEP Array~(30/40~GHz) instruments.
The experiment has been mapping so-called the ``BICEP'' patch, which is a part of the ``Southern Hole'' patch.
The polarisation maps now reach depths of 2.8, 2.8, and 8.8\,$\mu$K-arcmin at 95, 150, and 220 GHz, respectively
over an effective area of ${\sim} 600\,{\rm deg^2}$ at 95~GHz and ${\sim} 400\,{\rm deg^2}$ at 150 and 220~GHz.

The BICEP Array is based on the BICEP3 instrument and additional receivers will be deployed
at 95, 150, and 220/270~GHz with 32,000+ total detectors.
The BICEP/Keck experiment will tighten to $\sigma(r)$ between 0.002 and 0.004 at the end of the experiment.

\paragraph{South Pole Observatory}
The SPT and BICEP/Keck programs increase the coordination of the South Pole CMB program
through the formation of a South Pole Observatory~(SPO).
They establish a formal entity to guide the development and observation of the combined South Pole CMB programs,
which will continue to benefit CMB-S4.

\paragraph{CMB-S4}
The CMB Stage 4 observatory~(CMB-S4)~\cite{2019arXiv190704473A} is the next-generation ground-based CMB experiment, anticipated to start first observations by 2030.
CMB-S4 is intended to be the definitive ground-based CMB temperature and polarisation experiment,
located in both Chile and the South Pole, and will consist of a deep survey covering a few percent of the sky
and high-resolution legacy survey covering over half the sky.
The frequency range will be 30--280~GHz.
One of the main science targets is to measure inflationary $B$-mode signals at an upper limit of $r < 10^{-3}$.
CMB-S4 is supported by the United States Department of Energy~(DOE) Office of Science and the National Science Foundation~(NSF).
DOE selects Lawrence Berkeley National Laboratory~(LBNL) as the lead lab for CMB-S4.

\paragraph{LiteBIRD}
LiteBIRD is a space mission to understand primordial cosmology and fundamental physics.
JAXA selected LiteBIRD as a strategic large-class~(L-class) mission in May 2019,
with its expected launch in the late 2020s using JAXA’s H3 rocket.
Full-sky surveys for three years at a Lagrangian point L2
will be carried out for 15 frequency bands centred at 40, 50, 60, 68, 78, 89, 100, 119, 140, 166, 195, 235, 280, 337, and 402~GHz
with three telescopes to achieve a total sensitivity better than 3\,$\mu$K-arcmin
with a typical angular resolution of 0.5~degree at 140~GHz,
covering the multipole ranges $2 \leq \ell \leq 200$~\cite{2020SPIE11443E..2FH}.

Its main scientific objective is to carry out a definitive search for inflationary $B$-mode signals as good as $\sigma(r)<10^{-3}$
either making a discovery or ruling out well-motivated inflationary models.

\paragraph{PIXIE}
The Primordial Inflation Explorer~(PIXIE)~\cite{2011JCAP...07..025K} is a proposed concept for a NASA mission to measure primordial gravitational waves via CMB.
The instrument is designed to achieve $r<10^{-3}$ at multipole ranges of $\ell < 100$ limited by the $2.6^\circ$ diameter beam.
PIXIE consists of a polarising Michelson interferometer configured as a nulling polarimeter
to measure the difference spectrum between orthogonal linear polarisations from two co-aligned beams,
which is conceptually different from the other CMB experiments listed above.
This unique design makes it possible to observe the absolute intensity
and linear polarisation in 400 spectral channels from 30~GHz to 6~THz using only four detectors.
The wide and many channels are powerful to measure distortions in the CMB black-body spectrum~\cite{2021ExA....51.1515C}
as well as to separate contaminations from the foreground emissions.



\subsubsection{CMB Polyspectra and Inflation}
\label{sec:CMB_angular}

Primordial metric perturbations, i.e., the curvature perturbation and the gravitational waves~(GWs) for the scalar and tensor modes, respectively, stretched by the inflationary accelerated expansion are converted to observed CMB temperature and polarisation anisotropies. Since the conversion process is swiftly and rigorously computed (at least, up to linear order), the physical information on the primordial metric perturbations is precisely extractable from observed CMB maps.
Indeed the main target in this review paper is modified gravity as the cause of the late-time accelerated expansion of the Universe. However, modification of the theory of gravity has been extensively discussed as a means of explaining inflation in the very early Universe. Thus, here we briefly mention the basics of the CMB polyspectra (power spectrum, bispectrum, and so on) and their significance
in terms of testing the gravity in the inflationary era.
How constraints from CMB observations can be obtained on the modification of gravity at the late time will be presented in the Boltzmann code Sec.~\ref{subsec:Boltzmann} that follows, together with constraints on specific models.

The $N$-point correlators of temperature ($X = T$) and $E$/$B$-mode polarisation ($X = E, B$) fields are powerful diagnostic tools for statistical properties (e.g., Gaussianity, isotropy, parity, and scale dependence) of the primordial metric perturbations. Since the CMB fields are distributed on the 2-sphere, the analysis is conventionally processed in spherical harmonic space. The correlators are formed by spherical harmonic coefficients:
\begin{align}
  a_{\ell m}^X := \int {\rm d}^2 \hat{n} X(\hat{n}) Y_{\ell m}^*(\hat{n}).
\end{align}
If the primordial metric perturbations are Gaussian, because of Wick's theorem, the odd-point correlators vanish and the even-point ones are given as products of the two-point one (or power spectrum) $\langle a_{\ell_1 m_1}^{X_1} a_{\ell_2 m_2}^{X_2} \rangle$. Therefore, the three-point one (or bispectrum) $\langle a_{\ell_1 m_1}^{X_1} a_{\ell_2 m_2}^{X_2} a_{\ell_3 m_3}^{X_3} \rangle$ can be utilised in testing non-Gaussianity.

  Supposing that the primordial metric perturbations are statistically isotropic, the CMB power spectrum and bispectrum take the forms respectively
\begin{align}
  \left\langle a_{\ell_1 m_1}^{X_1} a_{\ell_2 m_2}^{X_2} \right\rangle
  &= C_{\ell_1}^{X_1 X_2} (-1)^{m_1} \delta_{\ell_1, \ell_2} \delta_{m_1, -m_2} , \\
  \left\langle a_{\ell_1 m_1}^{X_1} a_{\ell_2 m_2}^{X_2} a_{\ell_3 m_3}^{X_3} \right\rangle
  &= B_{\ell_1 \ell_2 \ell_3}^{X_1 X_2 X_3}
  \left( \begin{matrix}
    \ell_1 & \ell_2 & \ell_3  \\
    m_1 & m_2 & m_3
  \end{matrix} \right), 
\end{align}
where the Kronecker's delta and the Wigner $3j$ symbol restrict non-vanishing power spectrum and bispectrum multipoles, respectively, to
\begin{align}
  \ell_1 = \ell_2 ~\text{ for power spectrum}, \ \ \
  |\ell_1 - \ell_2| \leq \ell_3 \leq \ell_1 + \ell_2 ~\text{ for bispectrum}. \label{eq:CMB_iso}
\end{align} 
On the other hand, multipoles allowed under parity conservation obey
\begin{align}
  \begin{aligned}
 & \begin{cases}
    \ell_1 + \ell_2 = {\rm even} &: X_1 X_2 \in TT, \, TE, \, EE, \, BB, \\
    \ell_1 + \ell_2 = {\rm odd} &: X_1 X_2 \in TB, \, EB ,
   \end{cases}
    \\
  & \begin{cases}
      \ell_1 + \ell_2 + \ell_3 = {\rm even} &: X_1 X_2 X_3 \in TTT, \, TTE, \, TEE, \, TBB, \, EEE, \, EBB, \\
      \ell_1 + \ell_2 + \ell_3 = {\rm odd} &: X_1 X_2 X_3 \in TTB, \, TEB, \, EEB, \, BBB.
    \end{cases}
  \end{aligned}
  \label{eq:CMB_Peven}
\end{align}
The multipoles other than Eqs.~\eqref{eq:CMB_iso} and \eqref{eq:CMB_Peven} can be generated only by breaking isotropy and parity symmetry, respectively (for a detailed discussion see Ref.~\cite{Shiraishi:2016ads} and references therein).

As observed violations of Gaussianity, isotropy and parity invariance are tiny, the statistically isotropic and parity-even two-point correlators ($C_{\ell}^{TT}$, $C_{\ell}^{TE}$, $C_{\ell}^{EE}$ and $C_{\ell}^{BB}$) have been well investigated first. The scalar spectral index $n_s$ representing the tilt of the curvature power spectrum and the tensor-to-scalar ratio $r$ denoting the primordial GW power spectrum divided by the curvature one are main searchable parameters. The latest bounds on $n_s$ and $r$ obtained by the Planck team and BICEP/Keck collaboration give the tight constraint on the inflation models (see Fig.~8 in Ref.~\cite{Planck:2018jri} and also Fig.~5 in Ref.~\cite{BICEP:2021xfz}).
 In particular, simple chaotic inflation models with any monomial power law are now excluded.
Thus, the possible extended inflation models have been extensively discussed in the context of modified gravity,
e.g., Higgs inflation with introducing the non-minimal coupling to gravity (see, e.g., ~\cite{Bezrukov:2007ep}), $R^2$ inflation with adding the higher-order curvature term~\cite{Starobinsky:1980te}, and so on.
Note that in terms of testing the modified theory of gravity responsible for the current accelerated expansion the angular power spectrum on large scales is quite useful.
Depending on the theory of gravity or the property of dark energy, the time-evolution of the gravitational potential at the late time would be changed and it affects the CMB angular power spectrum on large scales, the so-called integrated Sachs-Wolfe (ISW) effect.
In Sec.~\ref{subsec:Boltzmann},
we will briefly discuss this issue by showing the results of calculations for specific models.

On the other hand, the bispectrum information is utilised for not only complementary tests but also for probing non-linear interactions unsearchable by the power spectrum. First, the scalar-mode bispectra have been thoroughly studied. Most of the theoretically-predicted bispectra can be parameterised by three scale-invariant basis templates dubbed as the local, equilateral, and orthogonal bispectrum templates (for review see Sec.~2 in Ref.~\cite{Planck:2019kim} and references therein). The local and equilateral ones are amplified at squeezed ($\ell_1 \ll \ell_2 \sim \ell_3$, $\ell_2 \ll \ell_3 \sim \ell_1$ and $\ell_3 \ll \ell_1 \sim \ell_2$) and equilateral ($\ell_1 \sim \ell_2 \sim \ell_3$) triangles, respectively, and the orthogonal one has the shape orthogonal to the former two. The amplitude parameters of these three templates, $f_{\rm NL}^{\rm local}$, $f_{\rm NL}^{\rm equil}$ and $f_{\rm NL}^{\rm ortho}$, have been measured with the WMAP temperature and Planck temperature and $E$-mode polarisation maps \cite{WMAP:2012fli,Planck:2019kim}. In Ref.~\cite{Planck:2019kim}, the Planck team also analysed some non-standard shapes including unique scale and angular dependencies and oscillatory features in order to probe non-Bunch-Davies vacuum states and exotic particles such as axions and higher spin fields. Since no significant signal has been detected, various non-linear interactions beyond GR are constrained. No detection of any four-point correlator (or trispectrum) signal further reinforces the constraints.

Recently, especially since the first direct detection of GWs by LIGO \cite{LIGOScientific:2016aoc}, there has been growing interest also in the non-Gaussianity of the tensor mode. In analogy with the scalar-mode case, there are already (not so many but) various theoretical predictions of sizable tensor-mode bispectra with characteristic shapes (for quick review see Ref.~\cite{Shiraishi:2019yux} and references therein). So far, a few scale-invariant templates for squeezed and equilateral tensor auto bispectra and a squeezed tensor-scalar-scalar cross bispectrum have been tested with the WMAP temperature and Planck temperature and $E$-mode polarisation maps \cite{Shiraishi:2013wua, Shiraishi:2014ila, Shiraishi:2017yrq, Shiraishi:2019yux, DeLuca:2019jzc, Planck:2019kim}. The resulting upper bounds on the amplitude of the bispectrum placed constraints on axion-gauge field couplings \cite{Planck:2019kim}, Weyl cubic interactions motivated by a higher spin gravity model \cite{DeLuca:2019jzc} and non-linear couplings depending on graviton masses \cite{Shiraishi:2017yrq}.

In addition to Gaussianity, isotropy and parity invariance are also being actively investigated.
Such tests can be done by determining the presence or absence of non-vanishing multipoles other than Eqs.~\eqref{eq:CMB_iso} and \eqref{eq:CMB_Peven}. From the theoretical point of view, anisotropic signatures, i.e., signals outside Eq.~\eqref{eq:CMB_iso}, are consequences of an anisotropic inflationary expansion, and Bianchi type-I cosmological models driven by higher spin fields are working examples \cite{Ackerman:2007nb,Watanabe:2010bu,Shiraishi:2011ph,Bartolo:2014hwa,Bartolo:2017sbu} (for review see, e.g., Refs.~\cite{Soda:2012zm,Maleknejad:2012fw} and references therein). Parity-violating signatures, i.e., signals outside Eq.~\eqref{eq:CMB_Peven}, can arise from chiral GWs that are sourced by, e.g., Chern-Simons gravitational interactions \cite{Lue:1998mq,Soda:2011am,Shiraishi:2011st,Bartolo:2018elp} and axion-gauge field couplings \cite{Sorbo:2011rz,Shiraishi:2013kxa,Bartolo:2014hwa,Shiraishi:2016yun}. So far, the anisotropic signals have been tested with the temperature and $E$-mode polarisation power spectra \cite{Planck:2015igc, Planck:2018jri}. On the other hand, the parity-odd ones have been assessed employing not only the $TB$ and $EB$ power spectra \cite{Saito:2007kt, Gerbino:2016mqb} but also the $TTT$, $TTE$, $TEE$, and $EEE$ bispectra \cite{Shiraishi:2014ila, Planck:2019kim}. The past analyses have reported no evidence of primordial signals, constraining the above models and the modification of the theory of gravity.

Owing to this and next decadal CMB experiments such as the BICEP Array \cite{Hui:2018cvg}, Simons Observatory \cite{SimonsObservatory:2019qwx}, CMB-S4 \cite{CMB-S4:2020lpa} and LiteBIRD \cite{Hazumi:2019lys}, more noiseless or higher resolved $B$-mode data would be available. Moreover, with the progress of computer technology, big data analysis including higher-order ($N \geq 4$) correlators would be feasible. Accordingly, sensitivities to primordial signals would drastically increase \cite{Hui:2018cvg,SimonsObservatory:2019qwx,CMB-S4:2020lpa,Hazumi:2019lys,Shiraishi:2016yun}, yielding better understanding of gravity.


\subsubsection{CMB Lensing}




CMB observations can bring us information not only on primordial fluctuations as initial conditions but also on LSS through its propagation process. One of the observables related to the LSS is called CMB lensing.
CMB photons travelling to us from the CMB last scattering is subject to the weak lensing effect by the gravitational potential of the LSS (see e.g., \cite{Lewis:2006fu, Hanson:2010, Namikawa:2014xga}). This leads to a distortion in the spatial pattern of observed CMB temperature and polarisation maps. CMB lensing is a direct probe of intervening gravitational fields along the line of sight and is one of the most powerful means of testing theories of gravity from CMB observations. 

The lensed CMB anisotropies, $\tilde{X}(\hat{n})$ ($X=T,Q,U$), are given as a remapping of the primary unlensed CMB anisotropies at the last scattering: $\tilde{X}(\hat{n})=X[\hat{n}+\vec{d}(\hat{n})]$ where $\hat{n}$ is the observer line-of-sight direction and $\vec{d}(\hat{n})$ is the so-called deflection angle.\footnote{See e.g., \cite{Namikawa:2021obu} for other secondary effects which are not simply given as a remapping.} We frequently use the lensing convergence defined as 
\begin{align}
	&\kappa = -\frac{1}{2}\vec{\nabla}_n\cdot \vec{d}
	\,, \label{cmbwl:kappa} 
\end{align}
where $\vec{\nabla}_n$ is the covariant derivative on the unit sphere. 
By solving the geodesic equation in the perturbed spacetime, the lensing convergence in the Born approximation is expressed in terms of the gravitational potential as follows \cite{Lewis:2006fu}: 
\begin{align}
	&\kappa(\vec{n}) = \vec{\nabla}^2_n \int_{0}^{\chi_{\rm cmb}} {\rm d}\chi \, 
		\frac{\chi_{\rm cmb}-\chi}{\chi\chi_{\rm cmb}} (\Phi+\Psi)
	\,, \label{cmbwl:phi} 
\end{align}
where $\chi_{\rm cmb}$ is the comoving distance to the CMB last scattering surface. The lensing convergence directly probes the gravitational potential of the LSS along the line-of-sight, $\Phi+\Psi$. Statistics of the lensing convergence including the angular power spectrum and angular bispectrum are predicted differently in theories of gravity which are modified as compared to GR. 

In CMB measurements, we can reconstruct a map of the lensing convergence using the fact that, for a given lensing mass distribution in our Universe, the lensing effect violates the statistical isotropy of the lensed CMB fluctuations \cite{HuOkamoto:2001,OkamotoHu:2003}. 
As introduced above with regard to the test of the statistical anisotropy of the primordial fluctuations,
this statistical anisotropy due to the lensing effect leads to a coupling between different angular scales of CMB anisotropies: 
\begin{align}
    \langle a^{\tilde{X}_1}_{\ell_1 m_1} a^{\tilde{X}_2}_{\ell_2m_2}\rangle_{\rm cmb} \not= 0 \text{ for } (\ell_1,m_1) \neq (\ell_2,-m_2) \,, 
\end{align}
where $\langle\cdots\rangle_{\rm cmb}$ is the ensemble average over the primary CMB realisations under a realisation of the lensing mass distribution. This correlation has been used to reconstruct the lensing map in multiple CMB experiments including ACTPol \cite{Sherwin:2017,Darwish:2020}, BICEP/Keck~\cite{BKVIII}, Planck \cite{Planck:2013:lens,Planck:2015:lens,Planck:2018:lens}, \textsc{Polarbear}~\cite{POLARBEAR:2019:lens,POLARBEAR:2019:cross}, and \textsc{SPTpol} \cite{Wu:2019}. 
The reconstructed lensing convergence, $\hat{\kappa}$, is already quadratic in the CMB anisotropies, and its angular power spectrum, $C_L^{\hat{\kappa}\hat{\kappa}}$, is a four-point correlation of the CMB anisotropies. The four-point correlation is, in general, decomposed into the disconnected and connected parts. The disconnected part exists even in the absence of lensing. On the other hand, the connected piece of the four-point correlation is dominated by the lensing convergence power spectrum, $C_L^{\kappa\kappa}$. Therefore, the disconnected part must be subtracted from the measured four-point correlation, $C_L^{\hat{\kappa}\hat{\kappa}}$, to extract the lensing power spectrum. Similarly, the bispectrum of the reconstructed lensing convergence map is a six-point correlation and the disconnected part of the six-point correlation must be subtracted from observed $\hat{\kappa}$ bispectrum \cite{Namikawa:2017iak}. 


Multiple works have explored the impact of the modified gravity on the lensing power spectrum \cite{Calabrese:2009,Barreira:2014jha,Hojjati:2015qwa}. For example, Ref.~\cite{Planck:2015bue} employed the lensing power spectrum to constrain modified gravity theories. 
The cross angular power spectra between the lensing convergence and other mass tracers, including the ISW \cite{Hu:2013,Munshi:2014tua,Planck:2015bue}, quasars \cite{Zhang:2020vru} and galaxies \cite{Pullen:2015vtb}, are also useful to constrain modified gravity theories.


An alternative new way of testing modified gravity theories from CMB observations is to analyse the angular bispectrum of the CMB lensing mass distribution as recently proposed by \cite{Namikawa:2018erh}. The angular bispectrum of the CMB lensing convergence is detectable in upcoming CMB experiments such as Simons Observatory and will be a novel tool to constrain cosmology in future CMB observations \cite{Namikawa:2016jff}. 
Aiming at detecting the bispectrum in upcoming experiments, Ref.~\cite{Kalaja:2022xhi} recently explores possible biases in reconstructing the CMB lensing bispectrum. 
Observations of the CMB lensing bispectrum offer an interesting way to constrain deviations from GR in a broad class of scalar-tensor theories of gravity called ``beyond Horndeski", including GLPV and DHOST (see Sec.~\ref{sec:Scalar-tensor theories}). Ref.~\cite{Namikawa:2018erh} proposes a novel analytic model of the lensing convergence bispectrum in modified gravity theories by extending the fitting formula of \cite{Gil_Mar_n_2012} which is derived in GR. Then, they forecast the expected constraints on generic parameters describing the deviations from GR for future CMB experiments. The results indicate that an accurate non-linear correction of the matter bispectrum in the modified gravity considered is necessary when the bispectrum is used to probe scales beyond a multipole $L\gtrsim 1500$. However, the results are insensitive to details of the implementation of the screening mechanism at very small scales. 
Following on \cite{Namikawa:2018erh}, \cite{Namikawa:2018bju} tests analytical predictions of the CMB lensing bispectrum against full-sky lensing simulations and discusses their validity and limitation in detail.\footnote{The bispectrum code is available at \cite{2021ascl.soft04021N}.} They show that, while analytical predictions based on fitting formulas for the matter bispectrum agree reasonably well with simulation results, the precision of the agreement depends on the configurations and scales considered. For instance, the agreement is at the $10\%$-level for the equilateral configuration at multipoles up to $L\sim2000$, but the difference in the squeezed limit raises to more than a factor of two at $L\sim2000$. This discrepancy partially comes from limitations in the fitting formula of the matter bispectrum. To circumvent this situation, \cite{Takahashi:2019hth} recently proposed a new fitting formula of the matter bispectrum in the non-linear regime calibrated by the high-resolution cosmological $N$-body simulations of \cite{Nishimichi:2018etk}. The matter bispectrum for modified gravity theories with simulation is further investigated by \cite{Bose:2019wuz}, showing that the model proposed in \cite{Namikawa:2018erh} achieves the best overall performance among models of the matter bispectrum in the modified gravity theories they considered.

%
%
%

\subsection{Large-Scale Structure}
\label{subsec:LSS}
Large scale structure (LSS) of the Universe, the spatial distribution of galaxies in cosmological scales that looks like a web consisting of galaxy clusters and groups, voids, filaments, and walls, is one of the powerful probes of cosmology. LSS is a consequence of two competing effects, i.e., gathering matter through gravity mainly sourced by dark matter and pulling matter apart through cosmic acceleration, which are embedded in the growth of the structure. Thus, measurements of LSS and its evolution enables us to test GR and modified gravity.

Galaxy surveys, both photometric and spectroscopic, provide us with a large amount of data that allows for measuring LSS probes. Such probes include two-point statistics based on weak lensing and galaxy number density fluctuations, redshift space distortions (RSDs), and non-Gaussian statistics, which are described in detail in the following subsections. Here, we, first, review a summary of recent, ongoing, or upcoming photometric and spectroscopic surveys which are categorised as Stage-III and Stage-IV surveys in Dark Energy Task Force (DETF)~\cite{Albrecht:2006um}.
Then, we give a brief review of relevant LSS observables in terms of the constraint on the modified gravity.

\begin{figure*}
\centering
\includegraphics[width=0.85\textwidth]{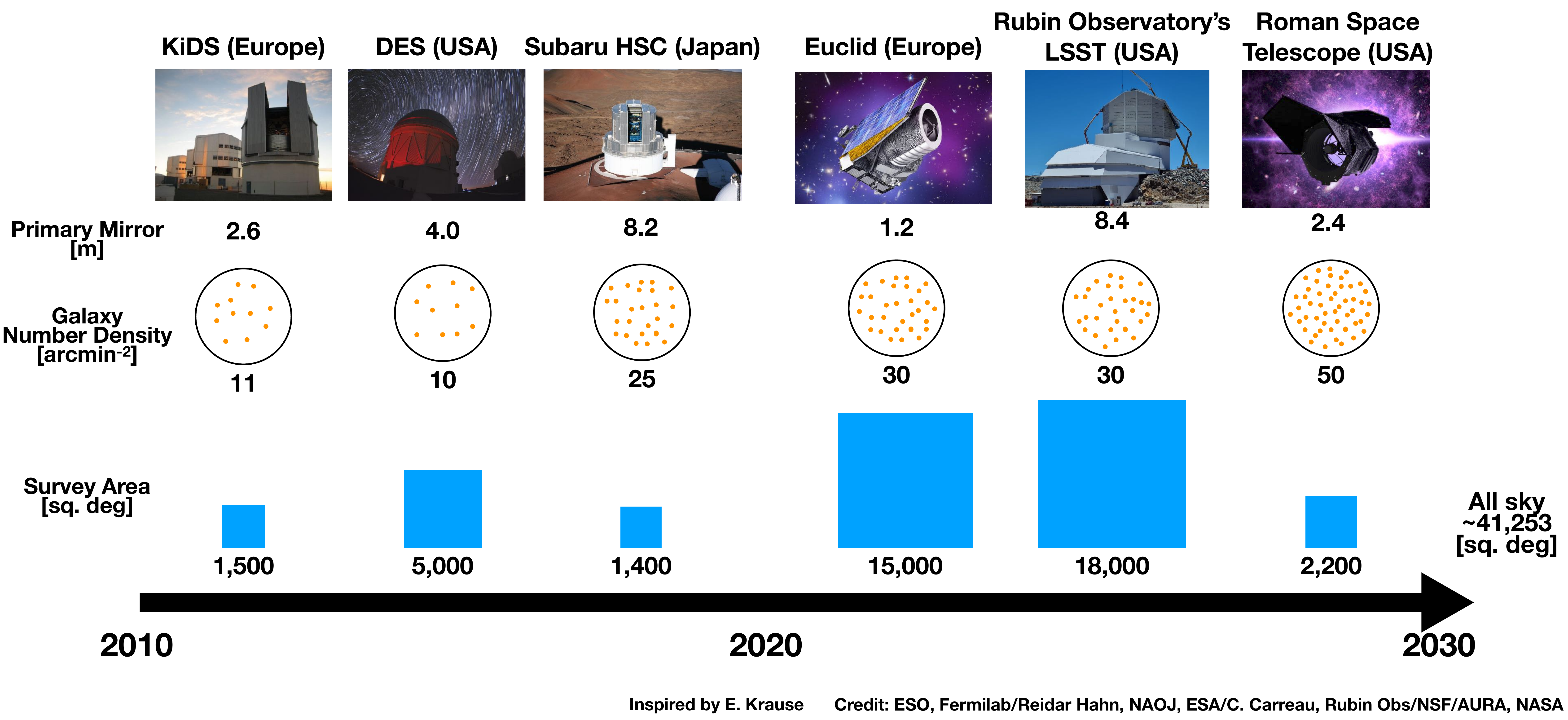}
\label{fig:LSS_imaging}

\vspace{1.0cm}

\includegraphics[width=0.85\textwidth]{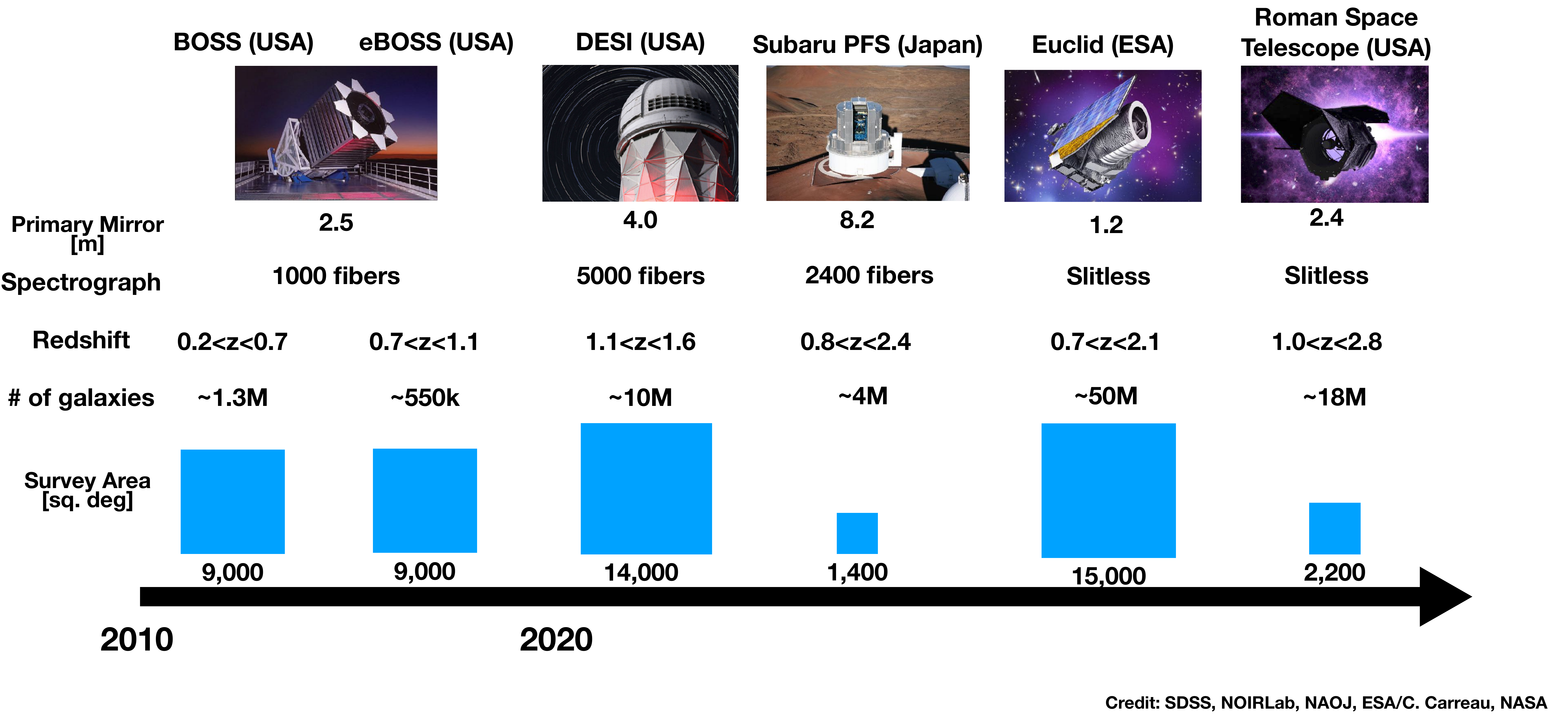}
\caption{Summary of previous, ongoing and upcoming galaxy imaging surveys (top) and spectroscopic surveys (bottom) since 2010.}
\label{fig:LSS_spectroscopic}
\end{figure*}

\subsubsection{Summary of the recent, ongoing, or upcoming photometric and spectroscopic surveys}

\paragraph{Photometric galaxy surveys}

\begin{itemize}
    \item Kilo-Degree Survey (KiDS)\footnote{\url{https://kids.strw.leidenuniv.nl}}: KiDS is one of the Stage-III surveys that started in 2011. Using the newly-developed OmegaCAM, which covers a full sq. degree, on the 2.6-m VLT Survey Telescope (VST), KiDS observed 1,500 sq. degrees of the sky in $u$, $g$, $r$, $i$-band down to the limiting magnitude of $r_{\rm lim}=25.2$ ($5\sigma$, $2^{\prime\prime}$ aperture), which aims to reach the median redshift of $z_{\rm med}\sim0.7$. The same region of the sky was also observed by VISTA Kilo-degree Infrared Galaxy Public Survey (VISTA) in $Z$, $Y$, $J$, $H$, $K$-band. The combination of optical and near-infrared bands allows for more secure photometric redshift estimates than other Stage-III surveys. For details of KiDS, see \cite{deJong:2012zb}.
    \item Dark Energy Survey (DES)\footnote{\url{https://www.darkenergysurvey.org}}: DES is also categorized as a Stage-III survey. DES started in 2013, using the newly-developed Dark Energy Camera (DECam), which covers the field of view to a diameter 2.2 degrees, mounted on the Blanco 4-m Telescope. DES completed observations of 5,000 sq. degrees of the sky in $g$, $r$, $z$, $Y$-bands down to the limiting magnitude of $i_{\rm lim}\sim23.5$ (10$\sigma$, $1.95^{\prime\prime}$ aperture). DES is the widest survey among the Stage-III surveys.
    \item Subaru Hyper Suprime-Cam (HSC) Survey\footnote{\url{https://hsc.mtk.nao.ac.jp/ssp/}}: The HSC survey is one of the Stage-III surveys that started in 2014. The HSC survey completed observations of ${\sim} 1{,}100$~sq. degrees of the sky in $g$, $r$, $i$, $z$, $y$-band, using the newly-developed prime focus camera at the 8.2-m Subaru Telescope (HSC) that covers the field-of-view to a diameter 1.5 degrees. The limiting magnitude is $i_{\rm lim}\sim26$ ($5\sigma$, $2^{\prime\prime}$ aperture), which aims at reaching the median redshift of $z_{\rm med}\sim1.0$. The HSC is the deepest survey and has the most superb image quality among the Stage-III surveys. For details of the HSC survey, see \cite{Aihara:2017paw}.
    \item Vera C. Rubin Observatory's Legacy Survey of Space and Time (LSST)\footnote{\url{https://www.lsst.org}}: LSST is categorised as the Stage-IV survey and is the ground-based photometric survey by the Rubin Observatory which consists of an integrated system of the 8.4-m Simonyi Survey Telescope and the LSST Camera that covers 9.6 sq. degrees of field-of-view with 189 4K~$\times$~4K CCDs. The LSST broadband filters consist of $u$, $g$, $r$, $i$, $z$, $y$-bands. LSST will start the 10-year survey that covers 18,000 sq. degrees of the sky in 2024. The limiting magnitude of the survey is $i_{\rm lim}\sim26.8$ (5$\sigma$, point source). For details, see \cite{LSST:2008ijt}.
\end{itemize}

\paragraph{Spectroscopic galaxy surveys}

\begin{itemize}
    \item SDSS-III Baryon Oscillation Spectroscopic Survey (BOSS)\footnote{\url{https://www.sdss.org/surveys/boss/}}/SDSS-IV eBOSS\footnote{\url{https://www.sdss.org/surveys/eboss/}}: BOSS and eBOSS are Stage-III galaxy spectroscopic surveys. They use the same optical (360~nm--1000~nm) spectroscopic instrument, which covers 7~deg$^2$ by 1000 fibers with the spectral resolution $R\sim2000$, mounted on the 2.5-m SDSS telescope. BOSS observed about $\sim$10,000~deg$^2$ of the sky and obtained spectra of $\sim$1.5 million luminous red galaxies at $0.15<z<0.7$ and Lyman-$\alpha$ forest spectra of $\sim$160,000 quasars at $2.2 < z < 3$ during 2009--2014, which is inherited by eBOSS which obtained $\sim$300,000 luminous red galaxies over $\sim$6,000~deg$^2$ at $0.6 < z < 1.0$, $\sim$175,000 emission line galaxies over 1,000~deg$^2$ at $0.6 < z < 1.1$, and $\sim$500,000 quasars over 6,000~deg$^2$ at $0.8 < z < 3.5$ during 2014--2019.
    \item Dark Energy Spectroscopic Instrument (DESI)\footnote{\url{https://www.desi.lbl.gov}}: DESI is the Stage-IV spectroscopic survey that started in 2021. DESI is mounted on the 4-m Mayall Telescope on top of Kitt Peak in the Sonoran Desert. DESI's field-of-view is 3~degrees in diameter which is filled with 5,000 robotic fiber positioners.  DESI's spectrographs cover 360~nm to 980~nm with the spectral resolution of $R=2{,}000$ to $R=5{,}000$. DESI will observe 14,000~deg$^2$ of the sky and obtain spectra of $\sim$8 million luminous red galaxies at $0.4<z<1.0$, $\sim$40 million emission line galaxies at $0.6<z<1.6$, and quasars at $0.9 < z < 2.1$ and Ly-$\alpha$ forests in quasar spectra at $z>2.1$ with the number density of $\sim$200 quasars per sq. degree.
    \item Subaru Prime Focus Spectrograph (PFS)\footnote{\url{https://pfs.ipmu.jp}}: PFS is the Stage-IV spectroscopic survey which will start in 2023. PFS will be mounted at the prime focus of the Subaru Telescope which provides a wide field-of-view ($\sim$1.25~sq.~degrees) covered with $\sim$2,400 robotic fibers. PFS's spectrographs covers 380~nm--1260~nm with the spectral resolution of $R=2{,}300$ to $R=5{,}000$. PFS will observe the HSC footprint and obtain $\sim$4 million emission line galaxies at $0.8<z<2.4$. For details of PFS, see \cite{Takada:2014}.
\end{itemize}

\paragraph{Imaging \& spectroscopic galaxy surveys}
\begin{itemize}
    \item Euclid\footnote{\url{https://www.euclid-ec.org/?page_id=2581}}: ESA/NASA's Euclid is the first space telescope dedicated to cosmology with the 1.2-m primary mirror. Euclid is planned to launch in 2023. Euclid is categorised as one of the Stage-IV surveys, which covers 15,000 sq. degrees in a single, wide optical band and three near-infrared bands ($Y$, $J$, $H$) with the limiting magnitude in the optical band $\rm{VIS}_{\rm lim}\sim27.4$ ($5\sigma$). Euclid will also conduct a spectroscopic survey of the same area, using the near-infrared (1100~nm--2000~nm) slitless spectroscopy with the spectral resolution $R=250$. For details of Euclid, see \cite{EUCLID:2011zbd}.
    \item Nancy Grace Roman Space Telescope (Roman)\footnote{\url{https://roman.gsfc.nasa.gov}}: Roman Space Telescope is NASA's next flagship mission which is planned to be launched in 2026. Roman has a 2.4-m primary mirror and hosts a Wide Field Instrument (WFI) which covers the 0.28 sq. degrees of field-of-view with 8 imaging filters from 0.43~$\mu$m to 2.3~$\mu$m. WFI also has a medium spectral resolution ($R=461$) grism that covers 1.0~$\mu$m to 1.93~$\mu$m. The imaging survey, the High Latitude Imaging Survey (HLIS), will cover $\sim$2,000~deg$^2$ of the sky down to $J\sim26.7$. The spectroscopic survey, the High Latitude Spectroscopic Survey (HLSS), will observe the same area of the sky and will obtain spectra of $\sim$10 million H$\alpha$ emission line galaxies at $1<z<2$ and $\sim$2 million [OIII] emission line galaxies at $2<z<3$ \citep{Wang:2021oec}. For details of Roman, see \cite{Spergel:2015}.
\end{itemize}



\subsubsection{Two-point correlations among galaxies and weak lensing}
\label{subsubsec:shaer}
Two-point correlations among galaxy and weak lensing, i.e., cosmic shear, galaxy-galaxy lensing, and galaxy-galaxy clustering, are the powerful probes of structure growth. The combination of the three probes is commonly referred to as ``3x2pt,'' and has been de-facto standard in the Stage-III photometric surveys \citep{Eifler:2020vvg}. In this subsection, we focus on the projected clustering for galaxy-galaxy clustering; the three-dimensional clustering is described in Sec.~\ref{subsubsec:RSD}.

Weak lensing is the coherent, subtle distortion of the shapes of distant galaxies due to the gravitational potential produced by the spatial fluctuations of intervening matter between an observer and the distant galaxies. The weak lensing distortion, or the so-called shear, depends also on the geometry of the Universe, i.e., the angular diameter distances between an observer, intervening matter, and source galaxies.

Cosmic shear enables to probe the matter power spectrum, $P_{\rm mm}(k)$,
which is given by
\begin{align}
\langle\delta ({\bf k})\delta ({\bf k}^\prime)\rangle
=(2\pi)^3\delta_{\rm D}^{(3)}({\bf k}+{\bf k}^\prime)P_{\rm mm}(k),
\end{align}
with the matter density fluctuations $\delta$,
and
$\delta_{\rm D}^{(n)}$ being the $n$-dimensional Dirac delta function,
through the auto-correlation of lensing shear. Several estimators have been proposed in the literature, such as aperture mass variance \cite{Schneider:1997ge} and COSEBIs \citep{Schneider:2010pm}, but one of the most popular estimators is the two-point correlation function that can be directly compared against measurements;
\begin{equation}
    \xi_{+,-}^{ab} (\theta) = \frac{1}{2\pi}\int {\rm d}\ell \ell J_{0,4}(\ell\theta)P_{\kappa}(\ell )^{ab},
\end{equation}
where $a$ and $b$ denote tomographic redshift bins, and $J_{0,4}(x)$ is the zeroth-order and fourth-order Bessel function for the first kind for $\xi_+$ and $\xi_-$, respectively. Using the flat-sky approximation and the Limber approximation, the convergence power spectrum,$P_{\kappa}(\ell )$, is written as
\begin{equation}
    P_{\kappa}(\ell )^{ab} = \int_0^{\chi_{\rm H}} {\rm d}\chi \frac{q^a(\chi) q^b(\chi)}{\chi^2}P_{\rm mm}\left(\frac{\ell}{\chi}\right),
\end{equation}
where $\chi$ is the comoving distance and $\chi_{\rm H}$ is the comoving Hubble radius, $q(\chi)$ is the lensing kernel, the aforementioned geometric dependence of lensing shear, which is written as 
\begin{equation}
q^a(\chi)=\frac{3}{2}\Omega_{\rm m}\left(\frac{H_0}{c}\right)^2\int_\chi^{\chi_{\rm H}}{\rm d}\chi^\prime p^a(\chi^\prime)(1+z)\frac{\chi(\chi-\chi^\prime)}{\chi^\prime}
\end{equation}
Here $p^a(\chi)$ is the normalised redshift distribution of source galaxies in the $a$-th tomographic bin.
Since the first detection of cosmic shear in 2000~\cite{Wittman:2000tc,Bacon:2000sy,vanWaerbeke:2000rm} and the subsequent measurements with deeper or wider data \cite{Rhodes:2001rt,Refregier:2002ux,Bacon:2002va,Hamana:2002yd}, a number of measurements have been made using imaging survey data, such as CFHTLens~\cite{Heymans:2013fya}, Deep Lens Survey \cite{Jee:2012hr,Jee:2015jta}, DES~\citep{DES:2017qwj,DES:2021bvc,DES:2021vln} KiDS~\citep{Kohlinger:2017sxk, Hildebrandt:2018yau, KiDS:2020suj}, HSC~\citep{Hamana:2019etx,HSC:2018mrq}, and the combination of the Stave-III surveys~\citep{LSSTDarkEnergyScience:2018bov,LSSTDarkEnergyScience:2022amt}.

The other two-point statistics, galaxy-galaxy clustering and lensing (the so-called 2x2pt), are used as a combination to break the degeneracy between galaxy bias and cosmological parameters. Galaxy-galaxy clustering, which is the auto-correlation of the fluctuations of galaxy number density field, is sensitive to $\Omega_{\rm m}$ and $\sigma_8$ the root-mean-square of the matter density fluctuations smoothed on $8~h^{-1}$Mpc at the present), but the fact that galaxies are a biased tracer of underlying dark matter distribution prevents us from extracting these cosmological parameters solely from the galaxy-galaxy clustering. On the other hand, galaxy-galaxy lensing enables us to measure the cross-correlation between the fluctuations of galaxy number density and underlying dark matter distribution, and thus we can extract the cosmological parameters from the combination of galaxy-galaxy clustering and galaxy-galaxy lensing. 

Using galaxy-galaxy power spectrum $P_{\rm gg}(k)$ and the zero-th order of spherical Bessel function $j_0(x)$, the projected galaxy-galaxy clustering signal is modelled as
\begin{equation}
    w_{\rm p}(R)=2\int_0^{\pi_{\rm max}}{\rm d}\Pi\  \xi_{\rm gg}\left(\sqrt{R^2+\Pi^2}\right),
\end{equation}
where $\xi_{\rm gg}(r)$ is the three-dimensional galaxy clustering;
\begin{equation}
    \xi_{\rm gg}(r)=\int_0^\infty\frac{k^2{\rm d}k}{2\pi^2} P_{\rm gg}(k) j_0(kr).
\end{equation}
Note that the projected correlation function is not sensitive to RSDs due to the peculiar velocity of galaxies (see Sec.~\ref{subsubsec:RSD}), if a sufficiently large maximum projected length $\pi_{\rm max}$ is adopted.
Using galaxy-matter power spectrum $P_{\rm gm}$, galaxy-galaxy lensing signal is modelled as
\begin{equation}
    \Delta\Sigma_{\rm gm}(R)=\rho_{\rm 0}\int \frac{k{\rm d}k}{2\pi^2}P_{\rm gm}(k)J_2(kR),
\end{equation}
where $J_2(x)$ is the second-order Bessel function.

At sufficiently large scales, which corresponds to $r\simgt10~\hiMpc$, the galaxy-galaxy power spectrum and galaxy-matter power spectrum are approximately related to matter power spectrum via $P_{\rm gg}=b^2P_{\rm mm}$ and $P_{\rm gm}=b\,r_{\rm cc}P_{\rm mm}$, respectively, where $b$ is the linear galaxy bias and $r_{\rm cc}$ is the cross-correlation coefficient defined as $r_{\rm cc}:= P_{\rm gm}(r)/[P_{\rm gg}(r)P_{\rm mm}(r)]^{1/2}$. Assuming $r_{\rm cc}\sim1$, which is the case at large scales\footnote{The convergence of $r_{\rm cc}\sim1$ to unity is better in real space. For detailed discussions, see \cite{Baldauf:2009vj} and \cite{Mandelbaum:2012ay}.}, we can break the degeneracy between the amplitude of matter power spectrum $\sigma_8$ and the galaxy bias parameter $b$, which enables the extraction of cosmological information from the combined probes. This technique has been widely used for cosmological inference, including SDSS \citep{Mandelbaum:2012ay}, DES \cite{DES:2017myr,DES:2021bpo, DES:2021zxv}, and HSC \cite{Sugiyama:2021axw}.

The scales smaller than $r\simlt10~\hiMpc$ can be used if the non-linear regimes of $P_{\rm gg}(k)$ and $P_{\rm gm}(k)$ are properly modelled. One approach is using the halo model prescription \cite{Seljak:2000gq,Peacock:2000qk,Scoccimarro:2000gm} with the transition regime between the 1-halo and 2-halo term calibrated against $N$-body simulations \cite{vandenBosch:2012nq,More:2012xa,Cacciato:2012gv, Miyatake:2013bha, More:2015ufa}. Recently, the halo-model based on a cosmic emulator built from $N$-body simulations,  which provides power spectra as a function of cosmological parameters in a non-parametric way, is emerging as a more robust technique to extract cosmological inference from non-linear regime \cite{Nishimichi:2018etk, Miyatake:2020uhg, Miyatake:2021sdd} (for details see Section~\ref{sec:emulator}). 

Finally, the combination of cosmic shear, galaxy-galaxy lensing, and galaxy-galaxy clustering, i.e., 3x2pt, enables us to maximally extract cosmological information from the two-point correlations among galaxies and weak lensing. The 3x2pt analysis has become standard for the Stage-III surveys, as reported by KiDS \citep{vanUitert:2017ieu, Heymans:2020gsg}\footnote{Note that \cite{Heymans:2020gsg} used three-dimensional galaxy clustering rather than the projected galaxy clustering.} and DES \citep{DES:2017myr, DES:2021wwk}. As mentioned in Section~\ref{sec:Signatures of modified gravity against LCDM model}, the 3x2pt measurements have been used for constraining the modified gravity parameters $(\mu, \Sigma)$ \citep{Simpson:2012ra, Joudaki:2016kym, DES:2018ufa, DES:2022ygi}.

\subsubsection{Redshift-space distortions}
\label{subsubsec:RSD}

The three-dimensional distributions of galaxies (or other tracers, such as quasars) over the LSS observed by spectroscopic surveys are apparently distorted due to the peculiar motions of galaxies.
By the Doppler effect of light emitted from a galaxy, the observed redshift $z_\mathrm{obs}$ differs from its true value $z = 1/a-1$ arising from the homogeneous and isotropic expansion of the Universe:
\begin{align}
    z_\mathrm{obs} = z + \frac{v_\mathrm{LoS}}{a},
\end{align}
where $v_\mathrm{LoS}$ is the line-of-sight component of the physical peculiar velocity of the galaxy.
This shift of redshift leads to that of the comoving coordinates, calculated from the observed redshift:
\begin{align}
    {\bf s} = {\bf x} + \frac{v_\mathrm{LoS}}{a H(a)} \hat{{\bf e}}_\mathrm{LoS},
\end{align}
where $H(a)$ is the Hubble parameter at the scale factor $a$, and $\hat{{\bf e}}_\mathrm{LoS}$ is the unit vector along the line-of-sight direction.
This modulation induces an anisotropy along the line-of-sight in the observed galaxy distributions, called the redshift-space distortions (RSDs) \cite{Kaiser:1987qv,Hamilton:1997zq}.

Due to the RSD effect, the galaxy-galaxy power spectrum observed in spectroscopic surveys also becomes anisotropic.
In linear theory, the galaxy power spectrum in redshift space $P_\mathrm{gg}(k,\mu)$ is related to the 
matter power spectrum 
$P_{\rm mm}(k)$ as \cite{Kaiser:1987qv}
\begin{align}
    P_\mathrm{gg}(k,\mu) = \left(b + f \mu^2 \right)^2 
    P_{\rm mm}(k),
\end{align}
where 
$\mu$ is the cosine of angle between ${\bf k}$ and the line-of-sight direction, $b$ is the linear galaxy bias and 
$f$ is the linear growth rate of the structure formation given by Eq.~\eqref{eq:LGR}. 
Note that all of $b,~f,$ and $P_{\rm mm}(k)$ depend on redshift $z$. 
This model describes that the redshift-space clustering is enhanced compared to its real-space counterpart by the coherent motions of galaxies towards overdensity regions, which is called the Kaiser effect.
Since the anisotropy due to the RSD effect does not break the azimuthal symmetry of galaxy clustering around the line-of-sight direction, it is common to expand this anisotropy with the Legendre polynomials, $\mathcal{L}_\ell(\mu)$, as
\begin{align}
    P_\mathrm{gg}(k,\mu) = \sum_{\ell} P_{{\rm gg}, \ell}(k) \mathcal{L}_\ell(\mu), \text{ where } 
    P_{{\rm gg}, \ell}(k) = \frac{2\ell+1}{2} \int_{-1}^1 \mathrm{d}\mu~P_\mathrm{gg}(k,\mu) \mathcal{L}_\ell(\mu),
\end{align}
and these multipole moments are measured from the galaxy samples in spectroscopic surveys \cite{BOSS:2016psr,BOSS:2013uda}. 
While the linear model given above is quite simple, it breaks down even on large scales $k \lesssim 0.1\,h\,\mathrm{Mpc}^{-1}$ \cite{Carlson:2009it,Kobayashi:2019jrn}, due to the non-linearity of the density and velocity correlations. 
Hence, in recent survey analyses, perturbation-theory-based models have been used \cite{Taruya:2010mx,Nishimichi:2011jm}. 

\begin{figure*}
\centering
\includegraphics[width=0.85\textwidth]{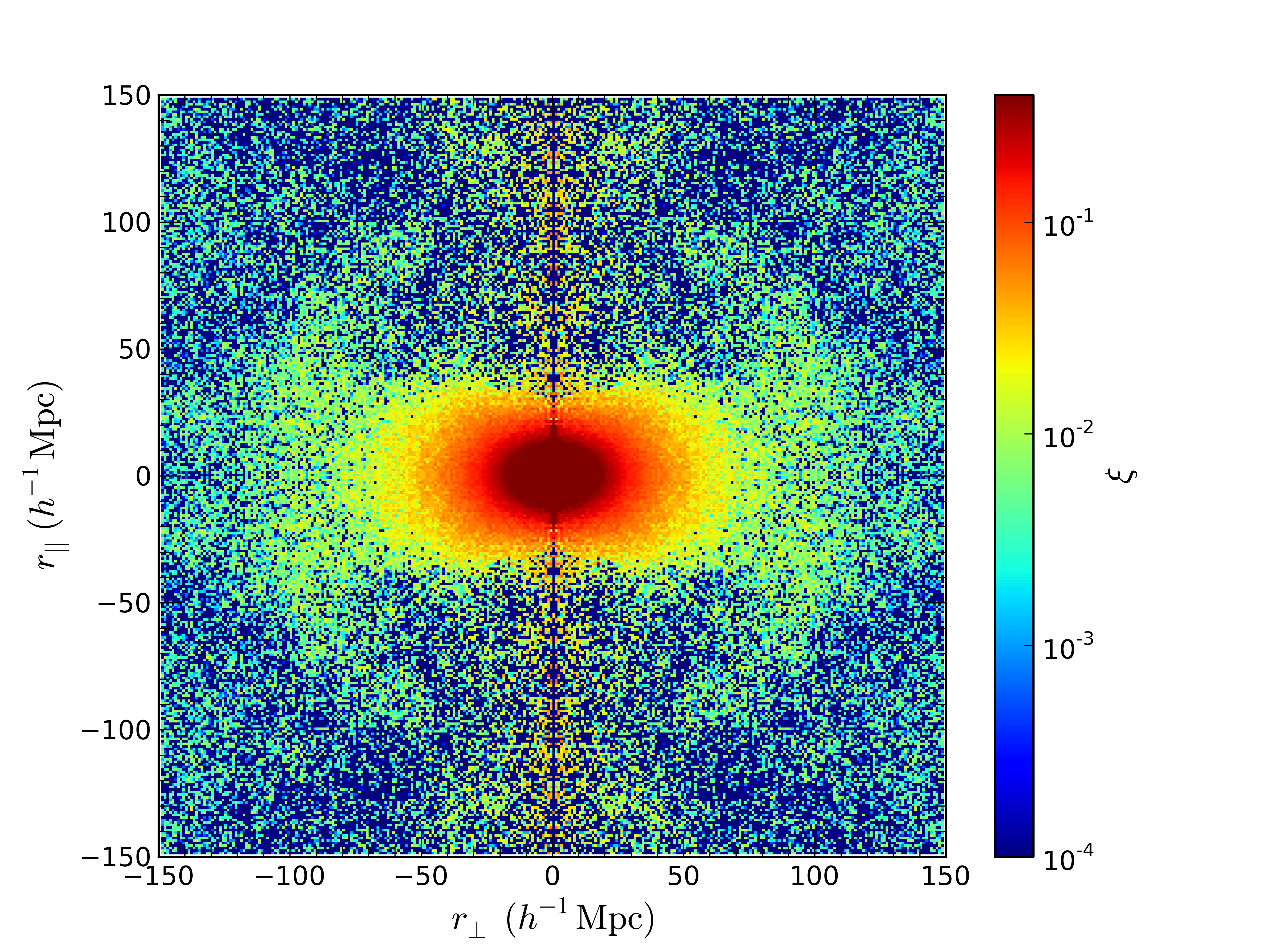}
\caption{
The two-point correlation function $\xi(r_\perp,r_\parallel)$ measured from the SDSS-III BOSS DR11 CMASS galaxy sample (see Figure~2 in Samushia et al., Mon.~Not.~Roy.~Astron.~Soc., 439 (2014)~\cite{Samushia:2013yga}).
}
\label{fig:1312_4899_xi2dmeasured}
\end{figure*}

We also measure the two-point correlation function in configuration space. 
It also has a dependence on the line-of-sight direction, and hence the two-point correlation function in redshift space is expressed as a function of the separation of two galaxies perpendicular to and along the line-of-sight, i.e., $\xi(r_\perp, r_\parallel)$. 
Figure \ref{fig:1312_4899_xi2dmeasured} shows the two-point correlation function of galaxies measured from the SDSS-III BOSS Data Release (DR) 11 \cite{Samushia:2013yga}, as a function of $(r_\perp, r_\parallel)$. 
It clearly shows the anisotropic nature of the redshift-space clustering with respect to the line-of-sight direction.  

\begin{figure*}
\centering
\includegraphics[width=0.85\textwidth]{
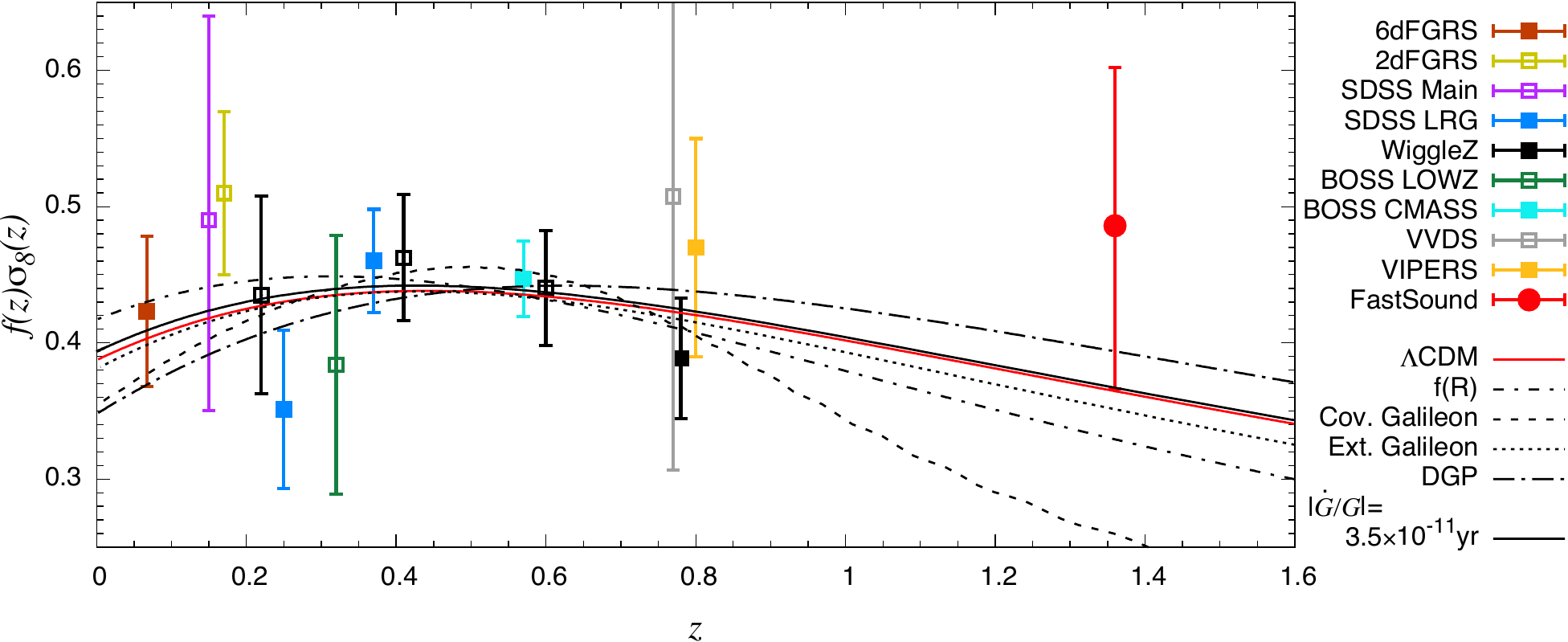}
%
\caption{
The constraints on $f(z) \sigma_8(z)$ as a function of redshift $z$, from galaxy spectroscopic surveys are listed in the legend. 
The theoretical predictions for several modified gravity models are shown as different curves. 
Note that these constraints are typically based on the theoretical templates built in the $\Lambda$CDM+GR, while $f \sigma_8$ is treated as a free parameter at each redshift. 
The figure is taken from Ref.~\cite{Okumura:2015lvp}.
}
\label{fig:1511_08083_fsigma8_table_lss_2}
\end{figure*}

In galaxy spectroscopic surveys, we typically estimate the combination $f(z) \sigma_8(z)$, as well as the geometric parameters determined through the Alcock-Paczynski effect~\cite{Alcock:1979mp,Matsubara:1996nf}, and nuisance parameters that capture the galaxy bias uncertainty.
The measurement of $f(z) \sigma_8(z)$ value is used to detect the deviation from the $\Lambda$CDM+GR picture from the perspective of modified gravity and/or dark energy models since $f(z)$ quantifies how the large-scale structure grows under the gravity
as discussed in Sec.~\ref{sec:Signatures of modified gravity against LCDM model}. 
Figure~\ref{fig:1511_08083_fsigma8_table_lss_2} shows the measurements of $f(z) \sigma_8(z)$ values in different spectroscopic surveys
with theoretical curves for several modified gravity models, taken from 
 Ref.~\cite{Okumura:2015lvp}. 
In current survey analyses, constraints on $f(z)\sigma_8(z)$ are usually obtained using the theoretical template of the galaxy clustering (mainly two-point correlation function or power spectrum) based on $\Lambda$CDM+GR, while $f(z)\sigma_8(z)$ for each redshift bin is varied as a free parameter. 
Therefore, strictly speaking, it is not suitable for comparing different modified gravity models, and it contributes only to searching for an anomaly in the $\Lambda$CDM+GR picture. 

Aside from constraining $f\sigma_8$ using the $\Lambda$CDM+GR template, there are some studies that focus on specific models of modified gravity.  
For instance, Ref.~\cite{Taruya:2013quf} models the power spectrum and correlation function for a class of modified gravity models, which includes the $f(R)$ gravity and the DGP model (see Sec.~\ref{sec:Scalar-tensor theories}), using the resummed perturbation theory, and shows a comparison with $N$-body simulations.
This model is applied to an analysis of the BOSS DR11 galaxy correlation function to estimate parameters of the $f(R)$ gravity and the scale-dependent structure growth \cite{Song:2015oza}. 
On the other hand, Ref.~\cite{Barreira:2016ovx} studies the impact of $f\sigma_8$ estimation from the galaxy clustering wedges, under the stable normal branch of the DGP model. 
To do this, they run the $N$-body simulation under the DGP cosmology and create mock catalogues of BOSS galaxies following a halo occupation distribution (HOD) model. 
It shows that the theoretical template of clustering wedges assuming $\Lambda$CDM+GR raises no biased estimation of $f \sigma_8$ even in the case of the DGP model. 
It is contrasted to the case of $f(R)$ gravity, which could be coming from the difference between scale-dependent growth in the $f(R)$ gravity and the scale-independent growth in the DGP gravity. 

While the modified gravity theories vary the growth of structure with respect to GR, there is no unified description of the variation. 
In current galaxy survey analyses, constraints on the modified gravity models tend to be limited to models with phenomenological parametrisation.
In the BOSS DR12 analysis \cite{Mueller:2016kpu}, they show different levels of parametrisations: the parametrisation of $f(z)$ by growth index $\gamma$ (given by Eq.~\eqref{eq:f parametrisation}), the modification of the Newtonian gravitational and lensing potentials by $\mu$ and $\Sigma$ according to Eqs.~(\ref{eq:Poisson eq}) and (\ref{eq:lensing eq}) ($G_M$ and $G_L$ in their notation), and the Bertschinger-Zukin parametrisation \cite{Bertschinger:2008zb} of scalar-tensor theories. 
For all of these cases, the current constraints show no necessity to include the gravity models beyond $\Lambda$CDM+GR.

\subsubsection{\texorpdfstring{$E_G$}{EG} statistics}
$E_G$ statistics, which is originally proposed by \cite{Zhang2007}, is a cosmological probe combining galaxy clustering and lensing signals. 
Combining clustering and lensing together can measure the difference in the response of galaxy and light to gravity: the motion of a galaxy is a response to gravitational potentials, whereas the path of light is bent by gravity, as showed in Sec.~\ref{sec:Signatures of modified gravity against LCDM model}. 
GR and other theories of gravity make specific predictions for how these two gravitational effects are related, which allows us to test the theory of gravity on cosmological scales.
The original definition of the $E_G$ is given by 
\begin{align}
    E_G(\ell)  = & \frac{C_{\kappa{\rm g}}(\ell)}{3H_0^2 a^{-1} 
        C_{{\rm g} v}(\ell)}\notag\\
            = & \frac{\nabla^2(\Psi + \Phi)}{3H_0^2 a^{-1} f \delta}, 
\end{align}
where $C_{\kappa{\rm  g}}$ is the cross-spectrum between convergence and galaxy positions (see Sec.~\ref{subsubsec:shaer}), $C_{{\rm g} v}$ is the cross-spectrum between galaxy positions and velocities. 
The second line of the equation is the expectation value of $E_G$ in terms of the metric perturbations and the matter density fluctuations.
In GR, the $E_G$ on large scale is predicted to be $\Omega_{\rm m}/f(z)$ (see Sec.~\ref{sec:Signatures of modified gravity against LCDM model}).

The first observational measurement of $E_G$ statistics is done by \cite{Reyes2010}. 
This observational $E_G$ statistics used in \cite{Reyes2010} has a slightly different definition than the original form proposed by \cite{Zhang2007}.
The observationally motivated $E_G$ is defined in real space rather than in Fourier space such as
\begin{eqnarray}
    E_G(R) & = & \frac{\Upsilon_{\rm g m}(R)}{\beta \Upsilon_{\rm g g}(R)}, 
\end{eqnarray}
where $R$ is the transverse
distance from the lens galaxy and $\beta = f(z)/b$ with $b$ being the galaxy bias on linear scales, which is measured by using the RSD effect as discussed in Sec.~\ref{subsubsec:RSD}. 
$\Upsilon_{\rm g m}(R)$ and $\Upsilon_{\rm g g}(R)$ are galaxy-matter and galaxy-galaxy annular differential surface densities (see the definitions in \cite{Reyes2010}).

While the motivation of taking the ratio between clustering and lensing in the original form is to cancel out the scale-dependence of galaxy bias and cosmic variance, the observationally motivated form lost these purposes because it uses different scales to measure $\beta$ and $\Upsilon_{\rm g m}(R)/\Upsilon_{\rm g g}(R)$. Things that are crucial to consider to measure $E_G$ properly from observational data are the scale-dependence of galaxy bias, the difference in clustering and lensing window, different redshift weighting of lensing and clustering (details are in \cite{Alam2016}), and magnification bias of the galaxy sample \cite{Pullen2018}.
The observationally motivated definition has been used and measured by numerous measurements using available spectroscopic samples at multiple redshifts \cite{Reyes2010, Blake2020, Alam2016, Amon2018}. Some works use quasars \cite{Zhang:2020vru} and CMB lensing \cite{Singh2018,Pullen:2015vtb}. All these measurements so far show consistency with the prediction from GR.
Note that because the $E_G$ statistics assume the value of $\Omega_{\rm m}$ to test the modified theories of gravity, in order to simultaneously constrain $\Omega_{\rm m}$ and parameters related to the modification of theories of gravity, this probe has to be measured at multiple redshifts.


\subsubsection{Non-Gaussian statistics} 

As has been discussed so far, LSS arises from the gravitational growth of tiny initial density perturbations.
Non-linear processes in gravitational growth inevitably make 
LSS observables non-Gaussian (see Sec. \ref{sec:Signatures of modified gravity against LCDM model}).
Hence, non-Gaussian statistics of LSS would allow us to 
study the non-linear gravitational growth of cosmic density fields in detail.
We here introduce two representative non-Gaussian statistics which have been studied in the literature.

\paragraph{Three-point correlations}

Three-point correlation functions (3PCFs) are among the most popular non-Gaussian statistics in studies of LSS.
For the three-dimensional cosmic mass density field in real space, it is defined by
\begin{align}
&\zeta({\bf r}_1, {\bf r}_2) = \langle \delta({\bf x}_1)\delta({\bf x}_2)\delta({\bf x}_3) \rangle
\,. \label{eq:def_3PCF}
\end{align}
In this equation, we assume that the Universe is statistically homogeneous,
allowing to write the 3PCF as a function of two relative coordinates among three vectors of 
${\bf x}_{i}\, (i=1,2,3)$: ${\bf r}_1 = {\bf x}_1 - {\bf x}_3$ and ${\bf r}_2 = {\bf x}_2 - {\bf x}_3$.
Using the Fourier transform of $\delta({\bf x})$, we also define the bispectrum $B$ as
\begin{align}
&\langle {\delta}({\bf k}_1){\delta}({\bf k}_2){\delta}({\bf k}_3) \rangle = (2\pi)^3\, 
\delta^{(3)}_\mathrm{D}({\bf k}_1+{\bf k}_2+{\bf k}_3)\, B({\bf k}_1, {\bf k}_2, {\bf k}_3)
\,. \label{eq:def_bispectrum}
\end{align}

For an initial Gaussian density field, 
the standard perturbation theory predicts that the bispectrum can be expressed as (see \cite{Bernardeau:2001qr} for a review of the standard perturbation theory)
\begin{align}
&B({\bf k}_1, {\bf k}_2, {\bf k}_3) = 2 F_2 ({\bf k}_1, {\bf k}_2) P_\mathrm{L}(k_1) P_\mathrm{L}(k_2) + (2\, \mathrm{perms}) + O(\delta^6_\mathrm{L})
\,, \label{eq:tree_level_bispectrum}
\end{align}
where $F_2$ is a kernel function for the second-order solution in the standard perturbation theory (see Eq.~\eqref{eq:F2}),
$\delta_\mathrm{L}$ represents the linear perturbation of the density contrast,
and $P_\mathrm{L}$ is the power spectrum of $\delta_\mathrm{L}$.
The kernel function $F_2$ controls the non-linear mode coupling of $\delta_\mathrm{L}$ between 
two different Fourier modes and depends on the theory of gravity~\cite{Tatekawa:2008bw, Koyama:2009me, Takushima:2013foa, Takushima:2015iha, Hirano:2018uar, Burrage:2019afs, Crisostomi:2019vhj} (see Sec.~\ref{sec:Signatures of modified gravity against LCDM model}).

Hence, the 3PCFs are a complementary probe of the theory of gravity to the standard two-point correlations, but there are several limitations in actual observations of the 3PCFs.
As in cosmological analyses with two-point correlations, 
the galaxy bias prevents measuring the kernel functions (e.g., $F_2$ in Eq.~\eqref{eq:tree_level_bispectrum})
from direct observables in galaxy spectroscopic surveys.
In the spectroscopic surveys, we also observe galaxy density fields in redshift space
and the distortion along a line of sight in the observed galaxy's density, that is, RSD makes the analysis of the 3PCFs 
more complicated \cite{Sugiyama:2018yzo, Sugiyama:2020uil}.
On the other hand, in the imaging surveys, cosmic shear, which is  the gravitational lensing effects of galaxy shapes by their foreground LSS, 
provides unbiased information on underlying cosmic mass density fields.
Nevertheless, the cosmic shear bispectrum suffers from projection effects 
of LSS and encodes 
the information of density perturbations at a highly non-linear regime 
where perturbation-theory approaches can be invalid \cite{Takada:2003ef, Valageas:2011kx}.
Also, the correlation of galaxy shapes before being lensed, 
known as the intrinsic alignment of galaxies (see Sec.~\ref{sec:Intrinsic alignments of galaxies}), 
can introduce a serious systematic effect for the cosmic shear bispectrum if ignored \cite{Semboloni:2008xa}.

From a practical point of view, the number of data elements for the 3PCF
tends to become larger than that in the two-point correlation analysis.
This indicates that a huge set of mock observations is required to set the covariance matrix (or statistical errors) of the observed 3PCF, but the mock observation of galaxy surveys usually relies on time-consuming cosmological $N$-body simulations \cite{Colavincenzo:2018cgf}.
Analytic and semi-analytic approaches may be complementary to the simulation-based method for the covariance estimation \cite{Taruya:2018jtk, Sugiyama:2019ike}.

\paragraph{Minkowski Functionals}

Minkowski Functionals (MFs) are a set of morphological descriptors.
According to Hadwiger's theorem, 
the morphological properties of $n$-dimensional pattern are fully described by $n+1$ MFs.
Hence, MFs have drawn much attention as a comprehensive probe of all orders of $N$-point statistics of cosmic density fields at once.
For a three-dimensional matter density contrast $\delta({\bf x})$ 
(or galaxy density contrasts), the MFs $V_{i}\, (i=0,1,2,3)$ are commonly computed from 
the excursion set of $\delta$ with a certain threshold $\nu$.
All MFs can be interpreted as known geometric quantities, 
the volume fraction $V_0(\nu)$, the total surface area $V_1(\nu)$, the integral mean curvature $V_2(\nu)$, and the Euler characteristic $V_3(\nu)$.

For a weakly non-Gaussian random field with the statistical isotropy, 
a multivariate Edge-worth expansion around a Gaussian distribution can provide an analytic expression of MFs by a power series of the root-mean-square of the field of interest \cite{Matsubara:1994wn, Matsubara:2003yt}, 
while the Gram-Charlier expansion has been examined to compute full-order non-Gaussian corrections in MFs \cite{Gay:2011wz}.
In actual galaxy spectroscopic surveys, observed galaxy density fields exhibit an anisotropy due to the RSD.
An analytic model of MFs for an anisotropic random field with weak non-Gaussinities
has been developed in \cite{Codis:2013exa}.

Since MFs are commonly defined in some smoothed density field with a smoothing scale $R_\mathrm{sm}$, they are less efficient to study the scale-dependence of the non-Gaussianity compared to higher-order correlation functions such as the 3PCF.
Nevertheless, there are some advantages to studying the MFs as a probe of the non-Gaussianity.
First, MFs are additive and motion-invariant, making the analysis less prone to observational effects (e.g., the survey shape \cite{SDSS:2003xnk}).
Moreover, MFs are found to be less sensitive to systematic effects in galaxy surveys \cite{Benson:2001ew, Park:2009ja, Wang:2010ug, Blake:2013noa} (At least, MFs should be insensitive to the linear galaxy bias).
MFs can be more effective to explore under-density regions, i.e., voids,
in the Universe, allowing to study the screening mechanisms in modified gravity (see Sec.~\ref{sec:screening}).
It would be worth noting that the measurement of MFs is much easier than that of
higher-order correlation functions in general.

\subsubsection{Intrinsic alignments of galaxies} 
\label{sec:Intrinsic alignments of galaxies}


In photometric galaxy surveys, we can measure the shapes of galaxies in addition to their positions as explained in Sec.~\ref{subsubsec:shaer}. Combining the redshift information obtained from spectroscopic measurements, the three-dimensional distribution of the galaxy shape field can be constructed, and this would provide additional cosmological information complementary to the conventional galaxy clustering data. The observed correlation of galaxy shapes involves the intrinsic correlation induced by LSS, before being affected by the lensing effect, which is called ``intrinsic alignments'' (IA) \cite{Catelan:2000vm,Heavens:2000ad,Hirata:2004gc}.
The simplest model to explain this correlation assumes that galaxy shapes are determined by large-scale tidal fields; in particular, in linear order galaxy shapes are related to large-scale tidal fields as 
\begin{align}
    \gamma_{ij}({\bf x};z) = b_K(z)K_{ij}({\bf x};z),
    \label{eq:IA_LA}
\end{align}
where $\gamma_{ij}$ characterises the galaxy shape as ellipsoid, $K_{ij}:= (\partial_i\partial_j/\partial^2 -  \delta^{\rm K}_{ij}/3)\delta$ is the tidal field, and $b_K$
is the linear shape bias.
This relation is analogue to $\delta_{\rm g} = b \delta$ in the density case and is called the linear alignment (LA) model.
It is noteworthy that $b_K$ is usually negative, meaning that the galaxy's major axis is perpendicular to the largest eigenvector of the tidal field because density perturbation tends to collapse first in the direction of the largest potential curvature.

Just as galaxy number density fluctuations are the counterpart of temperature fluctuations in CMB,
one can think of the galaxy shape as the counterpart of the polarisation of CMB.
As in the CMB polarisation, we can only observe the projected shape field, not the three-dimensional shape field introduced in Eq.~\eqref{eq:IA_LA}, and we can apply the $E/B$-decomposition to the projected shape field in Fourier space as 
\begin{align}
    E({\bf k},\hat{n}) \pm i B({\bf k},\hat{n}) := {_{\pm2}\gamma}({\bf k},\hat{n})e^{\mp 2i\phi_k},
\end{align}
where
\begin{align}
    {_{\pm2}\gamma}({\bf k},\hat{n}) := m_\mp^i(\hat{n}) m_\mp^j (\hat{n}) \gamma_{ij}({\bf k}),
\end{align}
with ${\bf m}_{\pm}:= (1, \mp i, 0)/\sqrt{2}$ and the line-of-sight direction $\hat{n}$ being $z$-direction.
Assuming the LA model, we get
\begin{align}
    E(k,\mu) & = b_K(1-\mu^2)\delta({\bf k}),
    \label{eq:IA_LA_E}
    \\
    B(k,\mu) & = 0,
\end{align}
where $\mu$ is the cosine between the line-of-sight and the wavevector.
Note that here we have projected shape field itself ($\gamma_{ij}$) but we do not have projected their positions ({\bf x}); 
it is possible to obtain the three-dimensional distribution of the shape field by combining photometric and spectroscopic surveys.
Notice also that for the LA model (i.e., the linear perturbation theory) $B$-mode is zero while $E$-mode is nonzero since scalar perturbations can only contribute to $E$-mode like in the CMB polarization.
In fact, beyond linear order, there appears $B$-mode from scalar perturbations, and vector and tensor perturbations can generate $B$-mode even in linear order.


The factor $(1-\mu^2)$ in Eq.~\eqref{eq:IA_LA_E} implied that the mode parallel to the line-of-sight does not have any impact on the observables.
This makes sense because by the projection we lose the information about the shape in the line-of-sight direction.
This is in contrast to the RSD case, where only the transverse mode contributes to the observables.
In this sense, IA carries complementary information to RSD.

Given the expression of $E$-mode, power spectra of IA in redshift space in the LA model are given by \cite{Taruya:2020}
\begin{align}
P_\mathrm{gE}(k,\mu) & = b_K (1-\mu^2) (b+f\mu^2) P_\mathrm{mm}(k), \label{eq:PgE} \\
P_\mathrm{EE}(k,\mu) & = b_K^2 (1-\mu^2)^2 P_\mathrm{mm}(k),\label{eq:PEE}
\end{align}
where $P_\mathrm{EE}$ is the auto-power spectrum of $E$-mode and $P_\mathrm{gE}$ is the cross-power spectrum of $E$-mode and the galaxy density field. Their configuration-space counterparts, correlation functions, have somewhat more complicated forms, as derived in Ref.~\cite{Okumura:2020}. 
The factor of $(b+f\mu^2)$ in the first equation is induced by RSD (see Sec.~\ref{subsubsec:RSD}). 
The factor does not appear in $P_\mathrm{EE}$ because the ellipticity field is not affected by RSD in linear theory. 
The $f$ parameter is further parameterised as
Eq.~\eqref{eq:f parametrisation}.
In this way, one can expect the intrinsic alignment of galaxies to be a sensible probe of gravity via the measurement of $P_\mathrm{gE}$ and $P_\mathrm{EE}$ is used to constrain the amplitude of IA, $b_K$. 

\begin{figure*}
\centering
\includegraphics[width=0.4\textwidth]{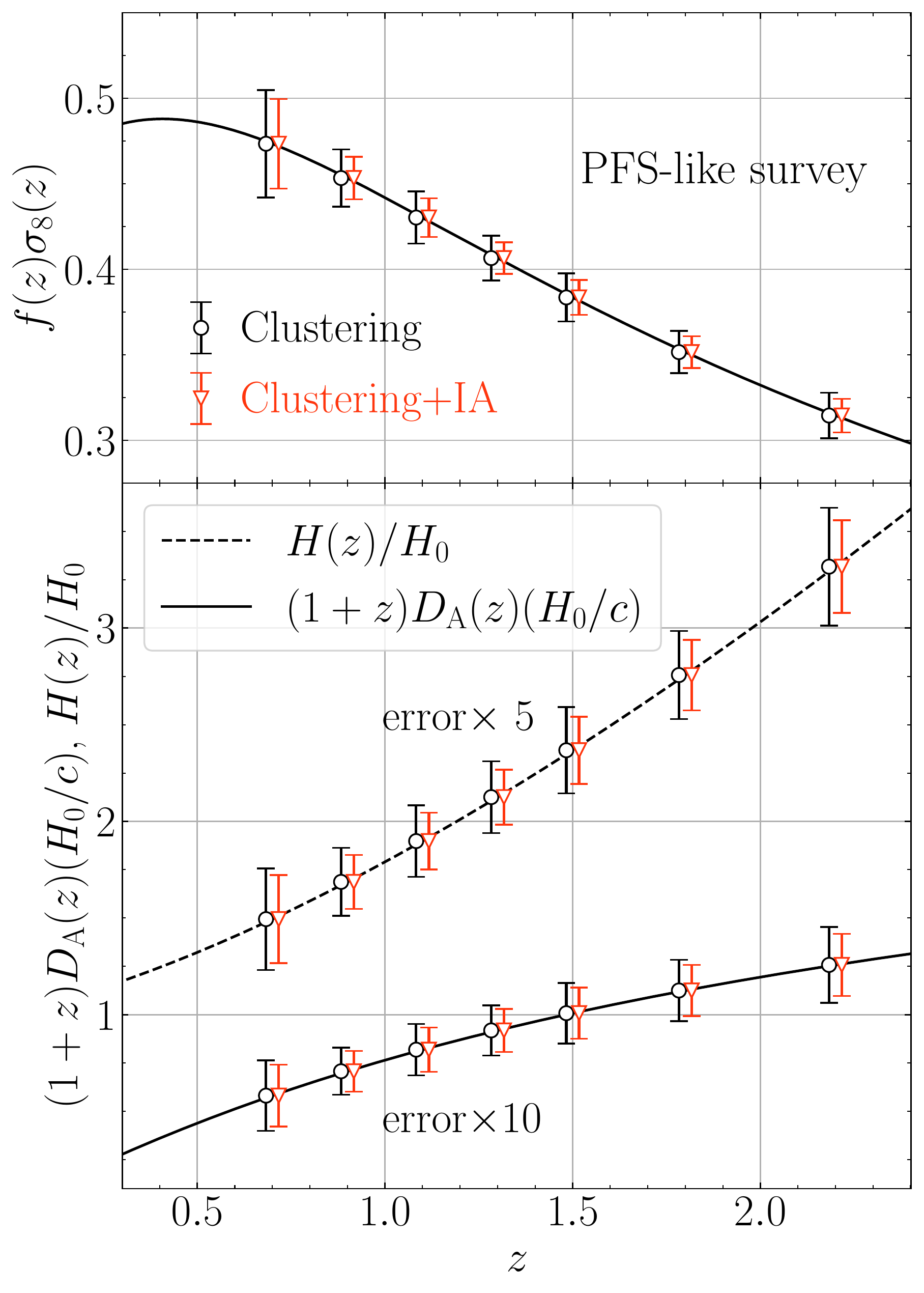}
\ \ \ 
\includegraphics[width=0.55\textwidth]{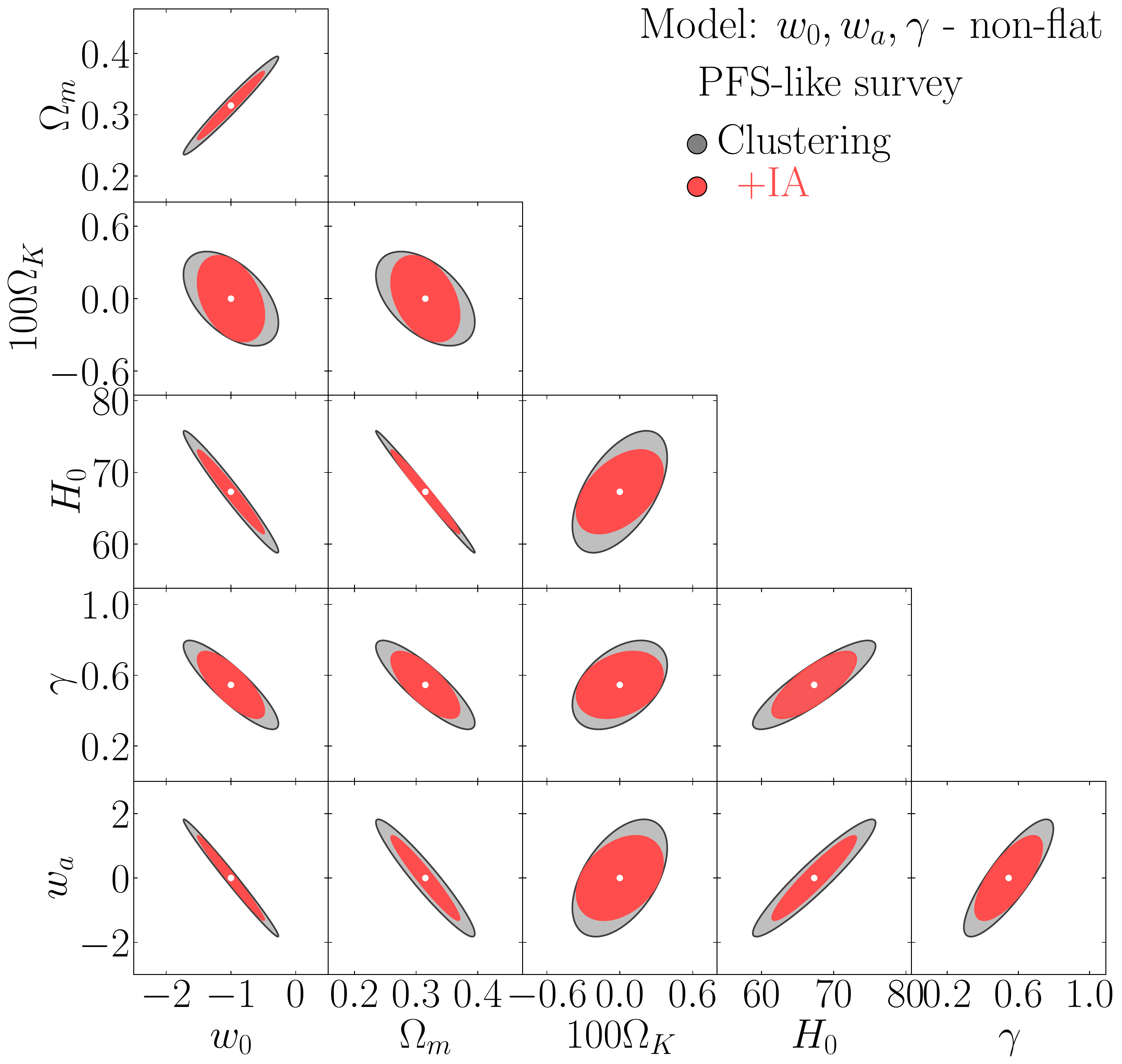}
\caption{
{\it Left set:} one-dimensional constraints on the growth rate $f(z)\sigma_8(z)$ (upper panel) and geometric distances (lower panel), $D_A(z)$ and $H(z)$, as a function of redshift expected from the upcoming PFS-like survey. {\it Right set:} two-dimensional constraints on a non-flat dynamical dark energy model which also allows deviation of gravitational law from general relativity, converted from the dynamical and geometric constraints obtained on the left set. 
Image reproduced and modified from Ref.~\cite{Okumura:2022} with permission by the authors.
}
\label{fig:2110_11127}
\end{figure*}

To quantify the constraining power for the gravity models in future surveys, 
Refs.~\cite{Taruya:2020,Okumura:2022} used the Fisher matrix formalism. 
The analysis considered a set of five parameters, namely 
the dynamical and geometric parameters $(f\sigma_8,H,D_A)$ and nuisance amplitude parameters $(b, b_K)$, for the three power spectra, $P_\mathrm{gg}$, $P_\mathrm{gE}$ and $P_\mathrm{EE}$. 
For the forecast, the combination of the Subaru HSC and PFS surveys is considered as an example \cite{Takada:2014}. 

The left panel of Fig.~\ref{fig:2110_11127} shows the one-dimensional constraints on each of the dynamical and geometric parameters after marginalising over all the other parameters.
It shows that combining the intrinsic alignment with the conventional galaxy clustering analysis improves the constraints on both the growth of structure and geometric distances. 
Although the galaxy clustering still outperforms the intrinsic alignment, systematic effects in each probe come to play differently. Thus considering intrinsic alignment effects would help addressing systematic-related issues such as the $\sigma_8$ and Hubble tensions (cf.~\cite{Verde:2019ivm,DiValentino:2021izs}). 

The obtained constraints on the geometric and dynamical quantities can be translated into specific cosmological model constraints by projecting the Fisher matrix into a new parameter space of interest. 
We further add the prior information on $\Omega_\mathrm{cb}$ and $D_\mathrm{M}(1090)/r_\mathrm{d}$ from the CMB experiment, to constrain cosmological parameters.
Adopting the non-flat $w$CDM model which allows both time variation of the dark energy equation-of-state and deviation of the gravity law from GR, the right panel of Fig.~\ref{fig:2110_11127} shows the expected cosmological constraints from the PFS-like deep galaxy survey. 
In this specific cosmological model, compared to the clustering-only analysis, the simultaneous analysis with the clustering and intrinsic alignment improves the constraints by 25\% for the modified gravity parameter $\gamma$ and ${\sim} 30\%$ for others except for $\Omega_K$. 

In the above analysis, constraining power for the gravity models with the intrinsic alignment of galaxies has been explored based on the $\Lambda$CDM model. There are a few works that studied the effect of modified gravity models on galaxy shapes and their alignment with surrounding LSS. 
Based on the Fisher matrix formalism, Ref. \cite{Reischke:2022} studied the effect of the Horndeski-type modified gravity (Sec.~\ref{sec:Scalar-tensor theories}) on the intrinsic alignment of galaxy shapes, although they considered only the angular (projected) statistics and did not utilise the full three-dimensional information. 
More realistic studies of the effects of modified gravity encoded on the galaxy or halo shapes can be performed utilising numerical simulations. 
Using $f(R)$ simulation, Ref.~\cite{LHuillier:2017} investigated how different gravity models affect basic internal properties of dark matter halos, such as spins and shapes. They found that at higher redshifts, $z=1$, halo shapes in $f(R)$ gravity are more elongated than those in $\Lambda$CDM, which could be because more masses undergo infalling along the filaments into halo centers in $f(R)$ gravity. Towards $z = 0$ where the structure growth becomes more non-linear, the difference of the axial ratio distributions between $f(R)$ gravity and $\Lambda$CDM models becomes less prominent. These trends have been confirmed by Ref.~\cite{Chuang:2022}. 
Ref.~\cite{Chuang:2022} presented the first measurements of intrinsic alignment of halo shapes in $f(R)$ gravity simulations using galaxy position-intrinsic ellipticity (GI) correlation and intrinsic ellipticity-ellipticity (II) correlation functions, the Fourier counterparts of Eqs. (\ref{eq:PgE}) and (\ref{eq:PEE}), respectively. 
They found that the GI and II correlations are useful in distinguishing between $\Lambda$CDM and $f(R)$ gravity models, and intrinsic alignment statistics would enhance the detectability of the imprint of $f(R)$ gravity on LSS by ${\sim} 40 \%$ when combined with the conventional halo clustering.

\subsubsection{Galaxy clusters}

Galaxy clusters (GCs) are the most massive gravitationally self-bound objects in the Universe. These clusters form at the rare high peaks of the primordial density fluctuations, and they subsequently trace the growth of structure in the Universe as they grow in mass and abundance. As such, clusters constitute a natural cosmological probe for constraining the properties of the primordial fluctuations as well as cosmological parameters including the nature of dark energy. 

GCs can be identified by a wide variety of observations, such as optical, X-ray, and CMB surveys.
However, in general, the mass of GCs is not a direct observable, 
and an alternative observable (mass proxy) suitable for each observation should be used as a measure of mass.
Optical clusters are identified through the overdense regions of red-sequence galaxies. The mass proxy for such optical clusters is richness, which is the number counts of red-sequence galaxies weighted by their membership probabilities.
Although a sample of the optical clusters has the highest completeness and contains relatively lower mass clusters (i.e., better $S/N$), the systematic issues such as projection effects (i.e., misidentification of interlopers as members) and miscentering effects (i.e., misidentifying the center of the clusters) are quite serious for the identification of optical clusters. 
GCs identified from X-ray data, X-ray clusters, are less sensitive to systematics compared with the optical clusters, whereas the completeness of X-ray clusters drops at higher redshift (i.e., mass limits increase with redshift). The mass proxy for X-ray clusters is the flux or temperature of the intracluster medium.
From CMB observations, GCs can be identified through the Sunyaev-Zeldovich (SZ) effect (see e.g., Ref.~\cite{Planck:2015lwi}).
Such SZ clusters are typically massive and 
their completeness is independent of the redshift.
The mass proxy for SZ clusters is the so-called Compton-$y$ parameter.

As mentioned above, it is difficult to directly measure the mass of GCs in the above surveys. Thus, to compare the theoretically estimated halo mass function with observations, it should be necessary to derive the relationship between mass and its proxy for each observation and the biggest systematics for GCs cosmology
is coming from the uncertainties in such mass-observable relations.
In the face of such difficulties, GCs cosmology using the weak lensing measurements which directly probe the matter distribution in LSS (see Sec.~\ref{subsubsec:shaer}) has recently been attracting attention. Thanks to this, the study on the test of the modified gravity theory by GCs cosmology has been developing.
%
%
In fact, in many models in modified gravity, a screening mechanism (see Sec.~\ref{sec:screening}) is employed and it makes GCs particularly interesting. This is because the cluster outskirts lie at the transition scale from screened, where GR is recovered, to unscreened regions, where gravity is universally enhanced. Hence, the clusters would be able to detect the signature of modified gravity.

To probe modified gravity, the following observables are often used. The first observable is the shape of the halo mass function. The halo mass function predicted from modified gravity theories differs from the one from $\Lambda$CDM. For the case of $f(R)$ gravity, the increase in the number of halos happens at a particular mass scale depending on the model parameters.

The second observable is galaxy infall kinematics/phase spacediagrams/mass-temperature relation. Looking at galaxy infall kinematics, phase space diagrams and mass-temperature relation of GC combined with weak lensing measurement offer a powerful cosmological test of modified gravity. This is because estimates of dynamical mass systematically deviate from the GR due to the change of the effective gravitational coupling to matter $G_{\rm matter} = \mu G_{\rm N}$ in the unscreened region. On the other hand, weak lensing measurements trace the gravitational coupling of photon $G_{\rm light}  = \Sigma G_{\rm N}$. As we shall show in Sec.~\ref{sec:Density perturbations in scalar-tensor theories}, scalar-tensor theories generically predict $G_{\rm matter} \neq G_{\rm light}$ (equivalently $\mu \neq \Sigma$), namely the breaking of the weak equivalence principle. This implies that a difference between the two estimators of a cluster mass from the measurements of kinematics and weak lensing can quantify the validity of GR at which the two estimators must be equivalent.\footnote{There is a certain subclass that satisfies $G_{\rm matter} = G_{\rm light}$ called no-slip gravity~\cite{Linder:2018jil}. A gravity theory in the no-slip type cannot be distinguished by the mass estimation of a cluster.} Hence an equivalence of the masses estimated from the kinematics and weak lensing gives a stringent confirmation of the weak equivalence principle. For the case of $f(R)$ and Galileon, the infall velocity of galaxies around clusters exhibits 100--200\,[km/s] enhancement at $r=5\,[{\rm Mpc/h}]$ and the radial and tangential velocity dispersions are 50--100\,[km/s] broadened at $z=0.25$~\cite{Zu:2013joa}.\footnote{According to this paper, the author said photons does {\it not} feel the effects from the extra scalar field. However, this is not true in generic scalar-tensor theory such as Horndeski theory, because the gravitational coupling for photon $G_{\rm light}$ is not equivalent to that of the Newton constant in GR.} Such deviations are detectable when combined with cluster weak lensing mass measurements. 

The third observable is splashback radius. The splashback radius is the location of the first turn-around of the particles and is considered a physical boundary of halos (see, e.g., Refs.~\cite{Diemer:2014xya, More:2015ufa}). Due to the transition from screened regions to unscreened ones occurring at the scale of cluster outskirts, the splashback radius offers an excellent probe for modified gravity. The enhanced gravitational forces increase galaxy infall velocities and therefore the size of the splashback radius by at most 10$\%$ compared with GR~\cite{Adhikari:2018izo}. Additionally, the effect of dynamical frictions, which makes the splashback radius for massive subhalos (or luminous galaxies) smaller than the splashback radius for DM particles, is reduced for the same reason.

One of the major obstacles in utilising clusters for testing cosmology and modified gravity is that we need to properly understand potential sources of systematics such as selection bias and projection effects.
Selection bias happens as the unknown impact of modified gravity on galaxy formation compared to GR. It is unclear how the interplay between environments and modified gravity alters galaxy properties. For example, there is a possibility that galaxies with different colors have different infall velocities. The projection effects are caused by the misidentification of interloper galaxies along the line-of-sight as genuine members of the cluster. Such misidentification of member galaxies can lead to false measurements of infall velocity or velocity dispersions. Furthermore, the projection effects are known to cause a smaller splashback radius for optical clusters~\cite{Sunayama:2019ssn, Murata:2020enz}. To test the theory of modified gravity, it is crucial to properly model the projection effects for optical clusters or use X-ray or SZ clusters which are less susceptible to the projection effects.

\section{Linear perturbations in modified gravity}
\label{sec:Perturbative predictions in modified gravity}

In this section, we provide a review of the basic formulation of cosmological perturbations in modified gravity theories. Given a theory of modified gravity, this will allow us to
give concrete predictions on cosmological observables, possibly with the help of numerical methods overviewed in the next section.
On the one hand, the phenomenological parameters introduced in Sec.~\ref{sec:Signatures of modified gravity against LCDM model} serve as a model-independent test of gravity at the cosmological scales. On the other hand, this approach is unsatisfactory due to at least the following two reasons. Firstly, we need to compute the observables for a given theory to perform a consistent analysis and make a concrete constraint on the given theory. Secondly, the constraints can be biased by the parameterisations of the modification of gravity. The phenomenological parameters may have an unphysical parameter space that cannot be achieved by a concrete theory, and this parameter space should be removed as a theoretical prior. As we will show shortly, the parameters $\mu$ and $\Sigma$ have complicated dependence on theory parameters, and the simple parameterisations cannot correctly capture the effects of the modification. We need to develop a more theory-based approach for precision tests of gravity. 

We briefly summarise the current results from cosmological perturbations in scalar-tensor theories, massive gravity theories, vector-tensor theories, metric-affine gravity, and cuscuton/minimally modified gravity. Particularly in the case of the Horndeski theory with $G_5=0$, we explicitly illustrate the field equations for the background and perturbed field equations in the limit of the quasi-static approximation with the explicit expressions of the phenomenological parameters.

\subsection{Perturbations in scalar-tensor theories}
\label{sec:Density perturbations in scalar-tensor theories}


Given a Lagrangian for a scalar-tensor theory, 
one can study the evolution of the homogeneous and isotropic background
and the dynamics of cosmological perturbations
on the basis of the modified gravitational field equations
derived from the Lagrangian.
While the manipulation is more involved
in the case of DHOST theories, it is rather straightforward to derive
the background and linear perturbation equations in the case of the Horndeski theory
in which all the field equations are of second order from the beginning.
Let us now demonstrate how, starting from a Lagrangian of a scalar-tensor theory
within the Horndeski family, one can obtain the basic equations
governing the evolution of the homogeneous background and
matter density perturbations on subhorizon scales.
To simplify the expressions without any loss of essential information,
for the moment we focus on the case with $G_5=0$.

The background metric is given by the flat FLRW: ${\rm d}s^2=-{\rm d}t^2+a^2(t)\delta_{ij}{\rm d}x^i{\rm d}x^j$,
with the homogeneous scalar field $\phi=\phi_0(t)$.
We consider
non-relativistic matter minimally coupled to gravity,
whose energy density is given by $\rho=\rho_0(t)$.
The background evolution is governed by
the gravitational field equations,
\begin{align}
    &2XG_{2X}-G_2-2XG_{3\phi}+6\dot\phi_0
    \left(X G_{3X}-2G_{4\phi X}-  G_{4\phi}\right)H
    \notag \\ &
    -6H^2\left(G_4-4XG_{4X}-4X^2G_{4XX}\right)=-\rho_0,\label{Horn-00-b}
    \\
    &G_2-2X\left(G_{3\phi}+\ddot\phi_0 G_{3X}\right)+
    2\left(3H^2+2\dot H\right)G_{4}-4\left(3H^2X+H\dot X+2\dot HX\right)G_{4X}
    \notag \\ &
    -8HX\dot XG_{4XX}+2\left(\ddot\phi_0+2H\dot\phi_0\right)G_{4\phi}+4XG_{4\phi\phi} 
    +4X\left(\ddot\phi_0-2H\dot\phi_0\right)G_{4\phi X}=0,
    \label{Horn-ij-b}
\end{align}
and the equation of motion for the scalar field,
\begin{align}
    \dot J+3HJ=G_{2\phi}-2X\left(G_{3\phi\phi}+\ddot\phi_0 G_{3\phi X}\right)
    +6\left(2H^2+\dot H\right)G_{4\phi}
    +6H\left(\dot X+2HX\right)G_{4\phi X},\label{s-eom-b}
\end{align}
where $X=\dot\phi_0^2/2$ and 
\begin{align}
    J=\dot\phi_0G_{2X}+6HXG_{3X}-2\dot\phi_0 G_{3\phi}+6H^2\dot\phi_0
    \left(G_{4X}+2XG_{4XX}\right)-12HXG_{4\phi X}.
\end{align}
We also have the standard energy conservation equation,
$\dot\rho_0+3H\rho_0=0$.

\begin{table}[t]
\caption{Various gravitational couplings in the Horndeski theory.}
\label{table:coupling}
  \centering
  \begin{tabular}{cl}
  \hline
  Gravitational coupling & Description \\
  \hline \hline 
  $G_{\textrm{cos}}$ &  Gravitational coupling in the Friedmann equation. \\
  $G_{\textrm{local}}$ & Coupling in the Newtonian limit (in screened regions). \\
  $G_{\rm N}=G_{\textrm{local}}|_{\textrm{present}}$ & The Newtonian constant measured by solar-system experiments. \\
  $G_{\textrm{matter}}=\mu G_{\rm N}$ & Coupling to non-relativistic matter field in unscreened regions. \\
    $G_{\textrm{light}}=\Sigma G_{\rm N}$ & Coupling to photons in unscreened regions. \\
        $G_{\text{GW}}$ & Coupling between gravitational waves and matter. \\
    \hline
  \end{tabular}
\end{table}

Some comments are now in order. 
As summarised in Table~\ref{table:coupling}, there are various gravitational coupling ``constants'' in the Horndeski theory.
In general, there is no unique way to define
the ``effective dark energy density'' $\rho_{\textrm{DE}}$ 
and the ``effective gravitational constant'' $G_{\textrm{cos}}$ in
the modified Friedmann equation~\eqref{Horn-00-b}, because
there is no clear distinction between the scalar field contribution and
the gravitational field contribution in the above equations.
In fact, neither the ``effective dark energy density'' $\rho_{\textrm{DE}}$ nor the ``effective gravitational constant'' $G_{\textrm{cos}}$ is directly observed and only
the measurable quantity at the background level would be the Hubble expansion rate. Nonetheless, it is sometimes convenient to introduce the notion of the ``effective dark energy density'' to discuss the property of dark energy. The general strategy for this is that we use a gravitational coupling which is physically meaningful in a certain context and then define the associated effective dark energy density (and the pressure) through the modified Friedmann equation. For instance, it would be natural to use the gravitational coupling appearing in the Newtonian limit, i.e.,~the gravitational coupling measured in local experiments, as the effective coupling, $G_{\textrm{cos}}=G_{\textrm{local}}=1/8\pi M^2_{\textrm{local}}$. In the Horndeski theory $(G_5=0)$ with the Vainshtein screening mechanism, such a gravitational coupling is given by~\cite{Kimura:2011dc}
\begin{align}
    &M^2_{\textrm{local}}:=2\left(G_4-4XG_{4X}-4X^2G_{4XX}\right).
    \label{Mcos}
\end{align}
One then writes Eq.~\eqref{Horn-00-b} as
\begin{align}
    &3M^2_{\textrm{local}}H^2=\rho_0+\rho_{\textrm{DE}},
    \\
    &\rho_{\textrm{DE}}:=
    2XG_{2X}-G_2-2XG_{3\phi}+6\dot\phi_0
    \left(X G_{3X}-2G_{4\phi X}-  G_{4\phi}\right)H.\label{rhoDE}
\end{align}
where $\rho_{\textrm{DE}}$ is the effective dark energy density associated with $M^2_{\textrm{local}}$. The Newtonian ``constant'' \eqref{Mcos} is generically a function of time. The measured Newtonian constant should be understood as its present value, $G_N=G_{\textrm{local}}|_{\rm present}$. Hence, as an alternative, one could define the effective dark energy density,
by using the constant $M_{\textrm{Pl}}=1/\sqrt{8\pi G_{N}}$, as
\begin{align}
    &3M_{\textrm{Pl}}^2H^2=\rho_0+\rho'_{\textrm{DE}},
    \\
    &\rho'_{\textrm{DE}}:=
    2XG_{2X}-G_2-2XG_{3\phi}+6\dot\phi_0
    \left(X G_{3X}-2G_{4\phi X}-  G_{4\phi}\right)H
    \notag \\ &\quad \quad\quad 
    -6H^2\left(G_4-4XG_{4X}-4X^2G_{4XX}-M_{\textrm{Pl}}^2/2\right).
\end{align}
In this case, all the effects of the modification of the Friedmann equation, even the time evolution of $M^2_{\textrm{local}}$, are interpreted as the effective dark energy density $\rho'_{\textrm{DE}}$.
One of the other possible ways to define the effective dark energy density is the use of the effective gravitational coupling between
gravitational waves and matter, which corresponds to $M^2=1/8\pi G_{\textrm{GW}}={\cal G}_T$, where ${\cal G}_T$
is defined below in Eq.~\eqref{action:densitypert} [see also \eqref{relationM2}].
We again emphasise that the choice of the ``effective gravitational coupling'' in the Friedmann equation is just a matter of definition although each gravitational coupling has certain physical meaning.  
In any case, one can solve the set of the background equations
to determine the evolution of the homogeneous background.
However, one should be aware of which definition is used when discussing the property of the ``dark energy density''.

Let us then consider linear perturbations around the homogeneous background.
For the present purpose, it is convenient to work in the Newtonian gauge
Eq.~\eqref{eq:conformal Newton metric},
and the scalar field is now given by $\phi=\phi_0(t)+\pi$.
The matter energy density is given by $\rho=\rho_0(t)+\delta\rho$.
Since we are interested in the quasi-static evolution of the perturbations
on subhorizon scales, we make the approximation,
\begin{align}
    \partial^2\Phi/a^2\gg H^2\Phi \sim H\dot \Phi\sim \ddot \Phi ,
\end{align}
and similarly for $\Psi$ and $\pi$. 
Substituting the Horndeski action to the perturbed metric and scalar field,
expanding it to quadratic order in perturbations, and making the quasi-static approximation,
we obtain
\begin{align}
    S^{(2)}&=\int{\rm d}t {\rm d}^3x \bigl\{a\bigl[{\cal F}_T(\partial\Psi)^2-2{\cal G}_T\partial\Psi\partial\Phi 
    +b_0(\partial \pi)^2-a^2 m^2\pi^2
    \notag \\ & \qquad \qquad \qquad 
    -2b_1\partial\pi \partial \Psi-2b_2\partial\pi\partial\Phi 
    \bigr]-a^3\Phi\delta\rho\bigr\},
\end{align}
where 
\begin{align}
    {\cal F}_T:=2G_4,\quad {\cal G}_{T}:=2\left(G_4-2XG_{4X}\right),
    \quad 
    b_1:=\frac{1}{\dot\phi}\left[
    \dot{\cal G}_T+H\left({\cal G}_T-{\cal F}_T\right)
    \right],\label{action:densitypert}
\end{align}
while $b_0$ and $b_1$ depend on $G_3$ and $G_4$ in a more complicated way.
The explicit expression for the effective mass term, $m^2$, is also lengthy:
\begin{align}
    m^2:=-G_{2\phi\phi}+\left(\ddot\phi+3H\dot\phi\right)G_{2\phi X}+\cdots
    -6\left(2H^2+\dot H\right)G_{4\phi\phi}+\cdots.
\end{align}
Note that if the Lagrangian is invariant under a constant shift of
the scalar field, $\phi\to\phi+c$, then $m^2=0$. See, e.g., Ref.~\cite{DeFelice:2011hq}
for the explicit form of the coefficients.

From Eq.~\eqref{action:densitypert} one obtains the field equations
(in the Fourier space) as 
\begin{align}
    k^2\left({\cal F}_T\Psi
    -{\cal G}_T\Phi -b_1\pi 
    \right)&=0,
    \\
    k^2\left(
    {\cal G}_T\Psi +b_2\pi 
    \right)&=-\frac{a^2}{2}\delta\rho,
    \\
    \left(b_0k^2-a^2m^2\right)\pi -b_1k^2\Psi-b_2k^2\Phi&=0.
\end{align}
These equations can be solved algebraically to give
\begin{align}
    \frac{k^2}{a^2}\Phi &= -
    \left[\frac{b_0{\cal F}_T-b_1^2-{\cal F}_T(a^2m^2/k^2)}%
    {b_0{\cal G}_T^2+2b_1b_2{\cal G}_T+b_2^2{\cal F}_T-{\cal G}_T^2(a^2m^2/k^2)}\right]
    \frac{\delta\rho}{2}=:-4\pi G_{\textrm{matter}}(t,k)\delta\rho ,\label{eq:P-Phi}
    \\
    \frac{k^2}{a^2}\Psi &= -
    \left[\frac{b_0{\cal G}_T+b_1b_2-{\cal G}_T(a^2m^2/k^2)}%
    {b_0{\cal G}_T^2+2b_1b_2{\cal G}_T+b_2^2{\cal F}_T-{\cal G}_T^2(a^2m^2/k^2)}\right]
    \frac{\delta\rho}{2},\label{eq:P-Psi}
    \\
    \frac{k^2}{a^2}\pi& = -
    \left[\frac{b_1{\cal G}_T+b_2{\cal F}_T}%
    {b_0{\cal G}_T^2+2b_1b_2{\cal G}_T+b_2^2{\cal F}_T-{\cal G}_T^2(a^2m^2/k^2)}\right]
    \frac{\delta\rho}{2}.
\end{align}
Eq.~\eqref{eq:P-Phi} is the generalisation of the Poisson equation.
Using Eq.~\eqref{eq:P-Psi}, one can derive
the ratio of the two metric potentials, $\eta(t,k)=\Psi/\Phi$.
Since we assume that matter is minimally coupled to gravity,
$\delta\rho$ obeys the usual fluid equations. 
Combining 
Eq.~\eqref{eq:delta eq} with
the generalised Poisson equation~\eqref{eq:P-Phi},
we obtain the evolution equation for the density perturbation $\delta$.

Let us present an example. In the case of the k-essence theory
with non-minimal coupling to the Ricci scalar, we have
$G_2=P(\phi,X)$, $G_3=0$, and $G_4=f(\phi)$, and thus
${\cal F}_T={\cal G}_T=2f$, $b_1=2f_{\phi}=-2b_2$, $b_0=-P_X/2$, leading to
\begin{align}
    8\pi G_{\textrm{matter}}=\left[\frac{fP_X+4f_\phi^2+2f(a^2m^2/k^2)}{fP_X+3f_\phi^2+2f(a^2m^2/k^2)}\right]
    \frac{1}{2f},
    \quad 
    \eta=\frac{fP_X+2f_\phi^2+2f(a^2m^2/k^2)}{fP_X+4f_\phi^2+2f(a^2m^2/k^2)}.
\end{align}
Here, the effective mass is given by
\begin{align}
    m^2=-P_{\phi\phi}+\left(\ddot\phi+3H\dot\phi\right)P_{\phi X}+2XP_{\phi\phi X}+2X\ddot\phi P_{\phi XX}
    -6\left(2H^2+\dot H\right)f_{\phi\phi}.
\end{align}

So far we have focused on the evolution of the density perturbations
in the quasi-static regime.
Relaxing the quasi-static approximation and deriving the complete form of
the perturbation equations are straightforward~\cite{Kobayashi:2011nu}.
Moreover, to provide new insights into the modification of the gravity theory that would not be imprinted in
the linear perturbation theory, we need to extend the analysis to the quasi-nonlinear regime
because the modifications to the theory of gravity typically alter the clustering properties
of LSS.
Exploring such a quasi-nonlinearity due to the non-linear growth of structure 
by observing the higher-order spectra of galaxies and CMB 
are discussed in the literature~\cite{Takushima:2013foa,Takushima:2015iha,Yamauchi:2017ibz,Hirano:2018uar,Hirano:2020dom,Yamauchi:2021nxw,Namikawa:2018erh,Yamauchi:2022fss}.
\\


In the formulation presented above, one can proceed to give theoretical predictions
straightforwardly,
given a concrete form of the Horndeski functions $G_a=G_a(\phi, X)$.
Alternatively,
one can consider a rather simpler characterisation stimulating physical intuition.
Let us reformulate perturbations of the metric and matter around a spatially flat FLRW Universe
on the basis of the effective field theory (EFT) of dark energy. 
The metric is now written in the Arnowitt-Deser-Misner (ADM) form as
\begin{align}
	{\rm d}s^2=-N^2{\rm d}t^2+h_{ij}\left({\rm d}x^i+N^i{\rm d}t\right)\left({\rm d}x^j+N^j{\rm d}t\right)
	\,.
\end{align}
Choosing the time coordinate so that the constant time hypersurfaces coincide with the uniform scalar-field hypersurfaces (i.e., the unitary gauge), 
we introduce the perturbations of the three-dimensional geometric quantities:
the perturbation of the lapse function, $\delta N=N-1$,
the perturbation of the extrinsic curvature, $\delta K^i{}_j=K^i{}_j-H\delta^i{}_j$,
and the three-dimensional spatial curvature, ${}^{(3)}R$. 
The minimum set of parameters (functions)
for determining completely the dynamics of cosmological perturbations to
linear order in the Horndeski class of gravity theories is a set of three independent functions of time
that are conventionally labelled by $\{\alpha_{\rm T},\alpha_{\rm K},\alpha_{\rm B}\}$,
in addition to the Hubble parameter $H$ and the effective Planck mass $M$~\cite{Bellini:2014fua}.
If gravity is described by the more general DHOST class of gravity theories, 
we need to introduce additional four time-dependent functions, $\alpha_{\rm H}$, $\beta_1$, $\beta_2$, and 
$\beta_3$, three of which should be related through certain conditions to ensure the propagation
of a single scalar degree of freedom (rather than two or more)~\cite{Langlois:2017mxy}.
In the case of the Horndeski class, the effective Lagrangian is expressed as
\begin{align}
	{\cal L}_{\rm EFT}
		=\sqrt{h}\frac{M^2}{2}\biggl[\delta K^i{}_j\delta K^j{}_i-\delta K^2+\left( 1+\alpha_{\rm T}\right){}^{(3)}R+H^2\alpha_{\rm K}\delta N^2+4H\alpha_{\rm B}\delta K\delta N+{}^{(3)}R\delta N\biggr]
	\,.
\end{align}
We have thus introduced the three time-dependent
parameters characterising the effective quadratic Lagrangian.
In addition to them, it is convenient to introduce a parameter characterising the time-variation of the effective Planck mass as $\alpha_{\rm M}:={\rm d}\ln M^2/{\rm d}\ln a$.
Note that $M^2$ is defined as the coefficient of $\delta K^i{}_j \delta K^j{}_i$ which leads to the kinetic term of the gravitational waves (the tensor perturbations). Therefore, $M^2$ should be understood as the effective Planck mass for the gravitational waves and is related to $G_{\textrm{GW}}$ in Table~\ref{table:coupling} via $G_{\textrm{GW}}=1/8\pi M^2$.
One can compare the previous results
in the Newtonian gauge
where the scalar perturbation is exposed through 
the time-coordinate transformation $t\to t -\pi (t,{\bf x})/\dot\phi (t)$.
As a result of this transformation, the homogeneous scalar field in the unitary gauge acquires spatial dependence.
In the framework of the EFT and under the quasi-static approximation,
the phenomenological functions defined in Eqs.~\eqref{eq:Poisson eq} and \eqref{eq:lensing eq}
in the limit of $k/a\gg m$ 
can be written in terms of
the EFT parameters as~\cite{Pogosian:2016pwr,Gleyzes:2015rua}
\begin{align}
	&\mu =\frac{M_{\rm Pl}^2}{M^2}\left( 1+\alpha_{\rm T}+\beta_\xi^2\right)
	\,,\\
	&\Sigma =\frac{M_{\rm Pl}^2}{M^2}\biggl[1+\frac{1}{2}\left(\alpha_{\rm T}+\beta_\xi^2+\beta_{\rm B}\beta_\xi\right)\biggr]
	\,.
\end{align}
Here we have introduced
\begin{align}
	&\beta_\xi^2 =\frac{2}{c_{\rm s}^2\alpha}\Bigl[\alpha_{\rm B}(1+\alpha_{\rm T})+\alpha_{\rm T}-\alpha_{\rm M}\Bigr]^2
	\,,\ \ \ 
	\beta_{\rm B}^2=\frac{2\alpha_{\rm B}^2}{c_{\rm s}^2\alpha}
	\,,
\end{align}
where the denominator of $\beta_{\rm B}^2$ is defined as
\begin{align}
	c_{\rm s}^2\alpha =-2\biggl\{(1+\alpha_{\rm B})\biggl[\frac{\dot H}{H^2}-\alpha_{\rm M}+\alpha_{\rm T}+\alpha_{\rm B}(1+\alpha_{\rm T})\biggr]+\frac{\dot\alpha_{\rm B}}{H}+\frac{\rho_0}{2H^2M^2}\biggr\}
	\,,
\end{align}
with $c_{\rm s}^2$ being the propagation speed of the scalar mode.

The functions appearing in the effective action are related explicitly to
the Horndeski functions through
\begin{align}
	&M^2=2\left(G_4-2XG_{4X}
            \right)
	\,,\label{relationM2}\\
	&M^2\alpha_{\rm T}=4XG_{4X}
	\,,\\
	&HM^2\alpha_{\rm B}=\dot\phi\left( XG_{3X}-G_{4\phi}-2XG_{4\phi X}\right)
				+4HX\left(G_{4X}+2XG_{4XX}
				\right)
	\,.
\end{align}
We do not show the explicit form of $\alpha_{\rm K}$ because it is so lengthy that
it is practically impossible to constrain $\alpha_{\rm K}$ with observational data on quasi-static scales.
So far we have discussed linear perturbations;
in order to take into account the second-order perturbations, 
we need to introduce additional two time-dependent functions
$\alpha_{\rm V1}$ and $\alpha_{\rm V2}$
to characterise their nature~\cite{Yamauchi:2017ibz,Bellini:2015wfa}.



\subsection{Perturbations in massive gravity theories}
\label{sec:Perturbations in massive gravity theories}
\paragraph{dRGT massive gravity}
The dRGT theory described by the action~\eqref{action:dRGT}
admits an open FLRW solution 
with the following St\"{u}ckelberg configuration \cite{Gumrukcuoglu:2011ew}:
\begin{equation}
	\phi^0 = f(t)\,\sqrt{1-K(x^2+y^2+z^2)}\,,
	\qquad
	\phi^i = f(t)\,\sqrt{-K}\,x^i\,,
	\label{eq:Stuckelberg}
\end{equation}
where $K$ is the spatial curvature of the three-dimensional manifold. (Note that flat and closed FLRW solutions are not allowed.) After solving the constraint equation for $f(t)$, it turns out that the Friedmann equation obeys the same equation as the $\Lambda$CDM model, i.e., the graviton mass exactly behaves as the cosmological constant $\Lambda$ which leads to de Sitter expansion. Unfortunately, the scalar and vector perturbations around this cosmological background suffer from serious instabilities due to the absence of the kinetic terms \cite{Gumrukcuoglu:2011zh,DeFelice:2012mx}. The quasi-dilaton theory \eqref{action:quasi-dilaton} has the same problem \cite{DAmico:2013saf}.

\paragraph{Generalised dRGT massive gravity} The generalised theory of dRGT theory described by
the action~\eqref{action:GMG} also admits only an open FLRW solution with the same St\"uckelberg
configuration~\eqref{eq:Stuckelberg}~\cite{Gumrukcuoglu:2020utx}. In this case,
the action depends explicitly on $X = \phi^a\phi_a$, which introduces a deviation from the cosmological constant, and hence the mass term plays the role of dynamical dark energy with the time-varying equation of state parameter $w$. This deviation from the cosmological constant ensures the presence of the kinetic terms in the scalar and vector sectors and the instabilities appearing in the original dRGT theory can thus be avoided~\cite{Kenna-Allison:2019tbu}.
According to the linear perturbation analysis under the quasi-static approximation, the new scalar degree of freedom in massive graviton invokes anisotropic stress, leading to an amplification of the effective gravitational constant in the matter perturbation equation and the growth rate of matter density perturbations~\cite{Kenna-Allison:2020egn}. The authors in~\cite{Gumrukcuoglu:2021gua} have confirmed that the Vainshtein screening mechanism works in this theory.

\paragraph{Projected massive gravity} Similarly to the generalised dRGT massive gravity, the projected massive gravity~\eqref{action:PMG} again admits only an open FLRW universe with the same St\"uckelberg configuration~\eqref{eq:Stuckelberg}~\cite{Gumrukcuoglu:2020utx}. In the case of the minimal coupling, $G=1$, and the mass term up to quadratic order, the equation of state parameter of the effective dark energy component arising from the graviton mass varies in time and approaches $-1$ in the future from $-1/3$ in the past, whose intermediate behaviour can be controlled by the model parameters. Any perturbative instabilities can be avoided due to the same reason as in the case of generalised dRGT massive gravity. Interestingly, the perturbation equations for the matter field on subhorizon scales are identical to those in GR
(i.e., ${\ddot \delta} + 2H {\dot \delta} -4\pi G \rho \delta = 0$)
due to vanishing scalar perturbations,
and the Poisson equation also takes the same form as the Newtonian one~\cite{Manita:2021qun}. 
Thus, the modification due to the scalar graviton appears only through the different expansion rate due to the background scalar mode, and, as a result, the evolution of density fluctuations slightly changes at present time. In contrast, the growth rate on superhorizon scales tends to be large. The constraints on the concrete model of this theory from the Pantheon and RSD data can be found in~\cite{Manita:2021qun}. 

\paragraph{Massive bigravity theory} In Hassan-Rosen
massive bigravity~\eqref{action:HR}, there are two branches of
cosmological solutions with the ansatz: 
$g_{\mu\nu} \mathrm{d}x^\mu \mathrm{d}x^\nu = -\mathrm{d}t^2 + a^2 \delta_{ij} \mathrm{d}x^i \mathrm{d}x^j$
and
$f_{\mu\nu} \mathrm{d}x^\mu \mathrm{d}x^\nu = -n^2 \mathrm{d}t^2 + \alpha^2 \delta_{ij} \mathrm{d}x^i \mathrm{d}x^j$,
where $n$ is a lapse function and $a$ and $\alpha$ are scale factors in the $g$ and $f$ metrics, respectively~\cite{Comelli:2012db,DeFelice:2014nja}. One of the branches corresponds to the one in dRGT theory, which contains unstable perturbations in the scalar and vector sectors. In the other branch, the late-time self-accelerating solution can be realised and all perturbation modes are stable. When $m \sim H_0$, the perturbations around cosmological background suffer from either ghost or gradient instabilities at early times~\cite{Comelli:2012db}. It has been argued, however, that this problem can be resolved by the non-linear effects of the scalar graviton~\cite{Aoki:2015xqa}. When $m \gg H_0$, all modes are stable, though a fine-tuning of coupling constants is needed~\cite{DeFelice:2014nja}. 
The equation of state parameter of the effective dark energy component arsing from the graviton mass can be less or more than $-1$ depending on model parameters~\cite{Koennig:2013fdo}. 
In the specific model given in~\cite{Koennig:2014ods},
the gravitational slip parameter $\eta$
and effective gravitational constant deviate from unity, leading to the modification of the evolution of the density perturbation. The constraints on the model parameters from the
measured growth data and type Ia supernovae has been investigated in~\cite{Koennig:2014ods}.
Some recent studies can be found in \cite{Hogas:2021fmr, Hogas:2021lns, Caravano:2021aum}.

\paragraph{Lorentz-violating massive gravity} As mentioned in Sec.~\ref{Sec.th.LVMG},
one of the Lorentz-violating massive gravity theories called the minimal theory of massive gravity
(MTMG)~\cite{DeFelice:2015moy} yields the same background equations for a FLRW universe as
those in GR. 
The linear perturbation analysis \cite{DeFelice:2015moy} shows that the effective gravitational constant $G_{\rm matter}$ and the gravitational slip parameter $\eta$ can deviate from unity despite the absence of a new degree of freedom in the theory \cite{DeFelice:2015hla,DeFelice:2015moy}. 
This fact allows us to put constraints on the graviton mass of the MTMG theory by using type Ia Supernovae, CMB, and RSD data  \cite{Bolis:2018vzs,DeFelice:2016ufg,deAraujo:2021cnd,DeFelice:2021trp}.

\subsection{Perturbations in vector-tensor theories}
\label{sec:pert_VT}
For simplicity, we focus on the generalised Proca (GP) theory, though the arguments may be applied to (at least most of) the extended classes as well. Linear perturbations in the GP theory can be analysed in a similar way to the Horndeski theory and they share common features~\cite{Tasinato:2014eka,Tasinato:2014mia,DeFelice:2016yws,Heisenberg:2016wtr}. This is easily anticipated because the GP theory is reduced to the shift-symmetric Horndeski theory when $A_{\mu}=\partial_{\mu}\phi$ is substituted. However, as we have mentioned in Sec.~\ref{sec:Vector-tensor theories}, there are several differences due to the vectorial nature~\cite{DeFelice:2016uil,deFelice:2017paw,Nakamura:2018oyy,Domenech:2018vqj,Nakamura:2019phn,DeFelice:2020icf}. For cosmological tests of gravitational theories, it is important to understand how these differences can be seen at the level of cosmological perturbations, which is achieved by using the EFT approach~\cite{Aoki:2021wew}.

In the case of vector-tensor theories, gravity is modified by a vector field $A_{\mu}$ which has four independent components. Since we are interested in the dynamics of the matter density perturbations, we focus on the scalar-type perturbations
\begin{align}
A_{\mu}=(\bar{A}_0(t)+\delta A_0(t,{\bf x}) + \partial_0 \pi(t,{\bf x}), \partial_i \pi(t,{\bf x}) ) = (\bar{A}_0(t)+\delta A_0(t,{\bf x}), {\bf 0}) + \partial_{\mu} \pi(t,{\bf x}) \,,
\end{align}
where $\bar{A}_0(t)$ is the background while $\delta A_0$ and $\pi$ are perturbations. If one were to impose $\delta A_0=0$ by hand, the vector would be given by the form $A_{\mu}=\partial_{\mu}\phi$ with $\phi=\int {\rm d}t \bar{A}_0(t) +\pi(t,{\bf x})$ and then the action would be reduced to the Horndeski action. Adopting the Newtonian gauge, for instance, the gravitational part of the quadratic-order action is given by a functional of the four variables $\{\Phi, \Psi, \delta A_0, \pi \}$, i.e., one has an additional variable in comparison to scalar-tensor theories. However, the theories are constructed to be ghost-free with one scalar dynamical degree of freedom. This implies that the additional variable $\delta A_0$ is an auxiliary variable that can be eliminated by using a constraint equation. Once the constraint is solved, the quadratic-order action formally takes the same form as that of the Horndeski theory, but the coefficients must be corrected since the solution to the constraint gives $\delta A_0\neq 0$ in general.
 
The above argument implies that the cosmological perturbations of scalar-tensor theories and vector-tensor theories can be unified into a single framework.
 Note that the corrections of the coefficients must be scale-dependent because the constraint is generically given by an elliptic differential equation. Hence, the corrections cannot be absorbed by redefinitions of the coefficients and we need an additional parameter(s) that determines how the coefficients are corrected in a scale-dependent way. As we have mentioned, the minimum set of parameters to specify the linear perturbations within the Horndeski theory is $\{\alpha_{\rm T}, \alpha_{\rm K}, \alpha_{\rm B} \}$ on top of $H$ and $M$. The inclusion of the GP theory requires an additional time-dependent function $\alpha_{\rm V}(t)$ that fully parameterise differences between the GP theory and the Horndeski theory at linear order in perturbations. For instance, the uncorrected $\alpha_{\rm B}(t)$ (the coefficient of the EFT action before integrating out $\delta A_0$) and the corrected one $\tilde{\alpha}_{\rm B}(t,k)$ (the coefficient after integrating out $\delta A_0$) are related by
 \begin{align}
     \tilde{\alpha}_{\rm B}(t,k)=\frac{k^2/(a^2H^2)}{\alpha_{\rm V}(t)\alpha_{\rm K}(t)+k^2/(a^2H^2)} \alpha_{\rm B}(t)
     \,.
 \end{align}
 All the coefficients have to be consistently corrected and the complete map between the uncorrected ones and the corrected ones is found in~\cite{Aoki:2021wew}. 
 The function $\alpha_{\rm V}$ has to be positive to avoid a ghost and the Horndeski theory is recovered in the limit $\alpha_{\rm V} \to 0$ (the decoupling limit) in which $\tilde{\alpha}_{\rm B}=\alpha_{\rm B}$. Note that $\{\alpha_{\rm T}, \alpha_{\rm K}, \alpha_{\rm B}\}$ in the GP theory and the shift-symmetric Horndeski theory cannot have independent time-dependency to be consistent with the requirement of shift symmetry, called consistency relations~\cite{Aoki:2021wew} (see also~\cite{Finelli:2018upr}).

By the use of the $\alpha$-parameters, the effective gravitational coupling and the gravitational slip parameter under the quasi-static approximation are written as
\begin{align}
    \mu(t) &=\frac{M_{\rm Pl}^2}{M^2(t)}\left[ 1+ \alpha_{\rm T} -2\alpha_{\rm T}^2 \alpha_{\rm V} + \frac{2}{V_S}( \alpha_{\rm M} + \mathcal{A}-2\mathcal{A}\alpha_{\rm T} \alpha_{\rm V})^2 \right]
    \,, \label{mu:H=0} \\
    \eta(t) &=\frac{1}{\mu(t)} \frac{M_{\rm Pl}^2}{M^2(t)}\left[ 1+ \frac{2(\mathcal{A}+\alpha_{\rm T})}{V_S (1+\alpha_{\rm T})}
    ( \alpha_{\rm M} + \mathcal{A}-2\mathcal{A}\alpha_{\rm T} \alpha_{\rm V}) \right]
    \,, 
\end{align}
where 
\begin{align}
\mathcal{A}=-\alpha_{\rm T}-\alpha_{\rm B}(1+\alpha_{\rm T})
\,, \quad
V_S=4\alpha_{\rm V}\mathcal{A}^2+(\text{terms independent of } \alpha_V)\,,
\end{align}
are computed from the $\alpha$-parameters (See \cite{Aoki:2021wew} for the precise definition of $V_S$). The $\alpha$-parameters are not completely arbitrary as they are related to stability conditions. The conditions relevant here are
\begin{align}
M^2>0\,, \quad \alpha_{\rm V}>0\,, \quad V_S>0 \,, \quad 1+\alpha_{\rm T}>0\,.
\end{align}
The only negative contribution to $\mu$ is provided by the third term of \eqref{mu:H=0} which is proportional to $\alpha_{\rm V}$. The GP theory with $(\alpha_{\rm V}>0)$ weakens gravity in comparison with its Horndeski counterpart $(\alpha_{\rm V}=0)$. The same conclusion is reached even when $\alpha_{\rm T}=0$ because $\alpha_{\rm V}$ increases $V_S$ and thereby decreases the last term of~\eqref{mu:H=0}.

In the Horndeski theory, the gravitational coupling on scales associated with 
the LSS tends to get larger than that in GR at low redshifts (cf. Sec.~\ref{subsubsec:demonstration}).
On the other hand, as we have seen, the existence of intrinsic vector modes of the GP theory can prevent such 
an enhancement of gravitational coupling~\cite{DeFelice:2016uil,deFelice:2017paw,Aoki:2021wew}. 
This property realises a possibility to evade the 
ISW-galaxy anti-correlation 
incompatible with the observational data~\cite{Nakamura:2018oyy} even in the case of 
the cubic-order GP theories $(\alpha_{\rm T}=0)$ in which the tensor propagation speed is strictly the same 
as that of light, and the observational constraint from the gravitational-wave event 
GW170817 is automatically satisfied (see also 
Refs.~\cite{Domenech:2018vqj,Nakamura:2019phn,DeFelice:2020icf} 
for the related works).

\subsection{Perturbations in metric-affine gravity}
\label{sec:pert_MA}
In metric-affine gravity (MAG), the connection is promoted to be an independent variable. In limited classes of theories, e.g.,~theories with the non-dynamical connection or theories having the Einstein frame, the theories can be reformulated in the Riemannian geometry by solving the equations of motion for the connection. The problem is then reduced to studying perturbations in the corresponding classes of gravity such as scalar-tensor theories in the Riemannian geometry. In contrast, if the connection is directly perturbed, analysing cosmological perturbations in generic MAG is quite involved since the general connection $\Gamma^{\mu}{}_{\nu\rho}$ has 64 independent components in four dimensions (see e.g.~\cite{Kubota:2020ehu}). For this reason, the cosmological perturbation theory in MAG has not been well formulated especially in the context of dark energy.

\subsection{Perturbations in cuscuton and minimally modified gravity}
\label{sec:pert_cus}
The cosmological perturbations in cuscuton and minimally modified gravity are recently investigated in~\cite{Aoki:2020oqc,DeFelice:2020cpt,DeFelice:2020prd,Iyonaga:2021yfv,Hiramatsu:2022ahs}. As the cuscuton theory and the minimally modified gravity theories can be regarded as a special class of scalar-tensor theories, the theory of cosmological perturbations can be similarly formulated. It is, however, important to stress that the scalar field is non-dynamical in minimally modified gravity and its predictions may be quantitatively different from the conventional scalar-tensor theories. For instance, the paper~\cite{Iyonaga:2021yfv} constructed a theory that gives the same prediction as GR in small scales and recovers the standard $\Lambda$CDM cosmology at the background level. Nonetheless, the linear perturbations are modified at cosmological distances which can be tested by CMB observations~\cite{Hiramatsu:2022ahs}. We will discuss this model in Sec.~\ref{subsubsec:demonstration}.

\section{Numerical tools for theoretical predictions }
\label{sec:numerical}
In this section, we introduce numerical tools in order to calculate realistic predictions of the CMB and LSS observables. In the former half, we explain the Bolzmann code that is utilised for computing the CMB anisotropies. As a demonstration, we show the angular power spectrum of the CMB temprerature fluctuations for certain subclasses of DHOST theory. In the latter, we explain a state-of-the-art technique called {\it emuration} that predicts gravitational interactions and galaxy biases in non-linear regimes in computationally less expensive way. We briefly summarise the current studies on non-linear structure formations in the framework of modified gravity.

\subsection{Boltzmann code}
\label{subsec:Boltzmann}
\subsubsection{Overview}
\label{subsubsec:overview}

The cosmological history after the Big Bang is embedded in the angular power spectrum of the CMB anisotropy (see Sec.~\ref{sec:CMB_angular}).
In these three or four decades, much effort has been devoted to extracting fruitful information from it using the large-scale observational data obtained by the space-based observatories such as COBE~\cite{Bennett:1996ce}, WMAP~\cite{WMAP:2012nax} and Planck~\cite{Planck:2018vyg}, and the ground-based experiments such as BICEP/Keck~\cite{BICEPKeck:2020hhe} and ACTPol~\cite{Mallaby-Kay:2021tuk}.\footnote{A full list of the active and completed CMB observatories and experiments is available at {\tt\url{https://lambda.gsfc.nasa.gov/product/expt/}}}
As a result, they have ruled out the structure formation scenario starting from a small seed produced by topological defects such as the cosmic strings (e.g., \cite{Vilenkin:2000jqa}). Instead of the defects, the inflationary scenario has been strongly supported.

In the radiation-dominant epoch, the photons tightly couple to baryons and charged particles (electrons and protons). The Thomson scattering between the photons and electrons is the most important physical process for the CMB photons. Hence, the dominant contribution to the collision term in the Boltzmann equations for the CMB photons comes from the Thomson scattering. The final expression of the Boltzmann equation will be shown in the later subsection.

To compute the angular power spectrum of the CMB anisotropies, we need to follow the co-evolution of the CMB photons, massless neutrinos, baryons (electrons)\footnote{The term `baryon’ represents ordinary matter in contrast with dark matter. In the community of CMB, we misuse this term. The original meaning in particle physics is, of course, the composite particle made of three quarks.}, CDM, and gravitational perturbations. The Boltzmann equation is represented as a set of recursive differential equations labelled by the angular index $\ell$, which is called the Boltzmann hierarchy. As it is highly tedious to solve their governing equations simultaneously in an analytic way, we resort to numerical computation using the so-called Boltzmann solver.

The pioneering numerical work~\cite{Sugiyama:1994ed} required huge computational resources at that time to solve $\mathcal{O}(1{,}000)$ of equations. Seljak and Zaldarriaga developed a sophisticated technique for the fast integration of the Boltzmann hierarchy, called the {\it line-of-sight formula} \cite{Seljak:1996is,Zaldarriaga:1999ep}. They released a Boltzmann solver implementing the formula, {\tt CMBFAST}\footnote{{\tt\url{ https://lambda.gsfc.nasa.gov/toolbox/tb\_cmbfast\_ov.cfm}}}, in 1996. We then became able to compute the angular power spectrum up to $\ell\sim\mathcal{O}(1{,}000)$ even with insufficient computational resources. After that, Lewis and Challinor released {\tt CAMB}\footnote{{\tt\url{https://camb.info/}}} in 1999~\cite{Lewis:1999bs}, which was written in Fortran 90 and employs the covariant formalism. {\tt CAMB} is the de facto standard code to estimate the cosmological parameters from the observational data, and the derivative numerical codes of this software have been developed by many people, such as {\tt MGCAMB}\footnote{{\tt\url{ https://github.com/sfu-cosmo/MGCAMB}}} \cite{Zhao:2008bn,Hojjati:2011ix,Zucca:2019xhg} and {\tt EFTCAMB}\footnote{{\tt\url{http://eftcamb.org/}}} \cite{Hu:2013twa} incorporating gravity theories beyond GR. Recently, Lesgourgues and his colleagues developed another Boltzmann solver named {\tt CLASS}\footnote{{\tt\url{ http://class-code.net/}}} \cite{Lesgourgues:2011re,Blas:2011rf}, which is written in C language. This numerical code is also widely used and there are some derivative works such as {\tt hi\_class}\footnote{{\tt\url{ http://miguelzuma.github.io/hi\_class\_public/}}} \cite{Zumalacarregui:2016pph,Bellini:2019syt} implementing the Horndesky theory \cite{Horndeski:1974wa,Deffayet:2011gz,Kobayashi:2011nu} (for a review, see \cite{Kobayashi:2019hrl} and the references therein). Similarly, {\tt COOP}\footnote{{\tt\url{ https://zenodo.org/record/61166\#.YTxIFjpUthE}}} implements the effective theory including the beyond Horndeski theory~(GLPV theory)~\cite{Gleyzes:2014dya,Gleyzes:2014qga}. The open-source Boltzmann solvers have been comprehensively tested in Ref.~\cite{Bellini:2017avd}. In addition, in Ref.~\cite{Hiramatsu:2020fcd}, one of the authors developed a Boltzmann solver incorporating the Type-I DHOST theory \cite{Langlois:2015cwa,Langlois:2015skt,BenAchour:2016cay,Crisostomi:2016czh}, a theory with only two tensorial degrees of freedom \cite{Hiramatsu:2022ahs}, and the vector-tensor EFT~\cite{Aoki:2021wew}.

The angular power spectrum of the CMB anisotropies depends on the cosmological parameters such as the amplitude of the dimensionless curvature power spectrum, $A_s$\footnote{See Ref.~\cite{Planck:2013pxb} for the parametrisation of the primordial power spectrum.}, the spectral index, $n_s$, the optical depth, $\tau$, the density parameter of baryons, $\Omega_{\rm b}$, and that of CDM, $\Omega_{\rm c}$, and the reduced Hubble parameter, $h$. In the case of extended theories of gravity beyond GR, the model parameters controlling the theories get involved. As the dimension of the parameter space is large, we usually employ the Markov-Chain Monte-Carlo (MCMC) analysis to estimate the confidence ranges of the parameters from the angular power spectrum. There are some open-source packages for the MCMC analysis, such as {\tt CosmoMC}\footnote{\tt\url{ https://cosmologist.info/cosmomc/}} \cite{Lewis:2002ah}, {\tt MontePython}\footnote{\tt\url{ https://https://baudren.github.io/montepython.html}} \cite{Audren:2012wb} and {\tt emcee}\footnote{{\tt\url{ https://emcee.readthedocs.io/en/stable/}}} \cite{Foreman-Mackey:2012any}.

A Boltzmann solver also computes the linear power spectrum of the non-relativistic matter fluctuations. The amplitude of matter fluctuations is usually characterised by $\sigma_8$ or $S_8:= \sigma_8(\Omega_{\rm m}/0.3)^{0.5}$, the amplitude on $8\, h^{-1}~$Mpc scale. The combination $f\sigma_8$, where $f$ is the growth rate, can be observed from the measurement of the RSD \cite{Kaiser:1987qv} from the galaxy survey like eBOSS \cite{Bautista:2020ahg}, 6dFGRS \cite{Jones:2009yz}, WiggleZ \cite{Parkinson:2012vd} (see Sec.~\ref{subsubsec:RSD}). In the context of extended theories of gravity, $f\sigma_8$ is a good indicator of the deviation from GR (e.g., \cite{Kazantzidis:2018rnb}). Furthermore, the linear power spectrum is the main building block to compute the non-linear power spectra taking into account the loop corrections (see \cite{Bernardeau:2001qr} for review).

There have been enormous numbers of works and experiments on the CMB in the past three or four decades, and many people have developed various schemes to analyse the CMB data. In this decade, many people have studied extended theories of gravity beyond GR. A motivation to study them is to explain the present acceleration of the cosmic expansion without dark energy at $z\lesssim 1$. In this case, the effects beyond GR on the angular power spectrum appear only at the large scales, $\ell<\mathcal{O}(10)$, through the ISW effect as we will see later. Making good use of the excellent legacies, we are fascinated to excavate slight hints for such theories from the CMB observations. 

Furthermore, the recent observational breakthrough of the gravitational waves since GW150914 \cite{LIGOScientific:2016vbw,LIGOScientific:2016lio} and the rapid development of LSS surveys such as BOSS and Subaru HSC can promote the studies on extended theories of gravity. These new surveys can put constraints on the theoretical parameters in different ways from the CMB observations. Hence, the joint analyses will make a definitive result on these theories and the unresolved issues in cosmology.

\subsubsection{Formalism}
\label{subsubsec:formalism}

In this section, we briefly review the governing equations for the photons, massless neutrinos,
CDM, baryons, and gravity to compute the angular power spectrum of the CMB anisotropy.
As in Sec.~\ref{sec:Signatures of modified gravity against LCDM model}, the metric perturbations are defined by
%
\begin{equation}
 \mathrm{d}s^2 = a^2\left[ -(1+2\Phi)\mathrm{d}\eta^2 + (1-2\Psi)
 \delta_{ij}\mathrm{d}x^i\mathrm{d}x^j\right],
\end{equation}
%
where $\eta$ is the conformal time.\footnote{Note that $\eta$ has been used as a gravitational slip parameter in the previous sections, but in this section it is used as a conformal time.}
The basic formulation for the propagation of photons from the last-scattering surface to us 
has been developed in Refs.~\cite{Bond:1993fb,Hu:1994jd,Hu:1994uz,Seljak:1994yz,Hu:1995en}.
In Ref.~\cite{Hu:1997hp}, the author developed an alternative representation named
the total angular momentum (TAM) method.
Hereafter we follow this method.

We begin with the Stokes parameters of electromagnetic radiation, $(I,Q,U,V)$,
which describe the intensity ($I$), the linear polarisations ($Q,U$), and the circular polarisation ($V$).
The temperature anisotropy of photons is equivalent to the anisotropy of their intensity, 
and the Fourier transform of the combination, $Q\pm iU$, corresponds to the $E$- and $B$-mode polarisations 
\cite{Hu:1997hp}. The photons follow the Bose-Einstein distribution, and 
we expand the Boltzmann equation for the Stokes parameters in terms of their small fluctuations.
In particular, for the scalar modes, we obtain the coupled recursive differential equations
(for details, see \cite{Hu:1997hp}),
%
\begin{align}
 \dot{\Theta}_\ell
   &= k\left[\frac{{}_0\kappa_\ell}{2\ell-1}\Theta_{\ell-1}
           - \frac{{}_0\kappa_{\ell+1}}{2\ell+3}\Theta_{\ell+1}
        \right]
        + \dot{\tau}\Theta_\ell + S_\ell, \label{eq:TH:Theta} \\
 \dot{E}_\ell
   &= k\left[\frac{{}_2\kappa_\ell}{2\ell-1}E_{\ell-1}
           - \frac{{}_2\kappa_{\ell+1}}{2\ell+3}E_{\ell+1}
        \right]
        + \dot{\tau}\left[E_\ell+\sqrt{6}P\delta_{\ell2}\right],\label{eq:TH:E} 
\end{align}
%
with
%
\begin{alignat}{5}
 S_0&=-\dot{\tau}\Theta_0+\dot{\Psi},
 &\quad
 S_1&=-\dot{\tau}v_b+k\Phi,
 &\quad
 S_2&=-\dot{\tau}P, 
 &\quad
 {}_s\kappa_\ell &= \sqrt{\ell^2-s^2}, 
 &\quad
 P &= \frac{1}{10}\left[\Theta_2-\sqrt{6}E_2\right].
\end{alignat}
%
In this section, we use the dot as the derivative with respect to the conformal time.
Note that the $B$-mode polarisation, $B_\ell$, is not induced by the scalar perturbations at the linear level.
We obtain similar equations for the massless neutrinos, $\mathcal{N}_\ell$, by 
setting $\dot{\tau}=0$ in the Boltzmann hierarchy for the photons;
$\tau(\eta)$ is the optical depth, $\dot{\tau} = -n_{\rm e}\sigma_{\rm T}a$, induced by the Thomson scattering.
The infinite number of equations should be truncated at some finite $\ell$ when we solve them numerically.
The simplest treatment is to set $\Theta_{\ell}=E_{\ell}=B_{\ell}=0$ for $\ell \geq \ell_{\rm max}+1$.
However, this treatment requires a large $\ell_{\rm max}$ to suppress the propagation of the truncation errors
to the low-$\ell$ modes. Instead, many numerical codes use an improved boundary condition 
proposed by Ma and Bertschinger~\cite{Ma:1995ey}. 

The baryons and CDM are described as fluids, which satisfy the continuity equation
and the Euler equation. Introducing the matter density perturbation, $\rho_{i}=\rho_{i0}(1+\delta_{i})$,
and the velocity perturbation, $v_{i}$, for $i=b,c$ (baryons and CDM, respectively),
their governing equations are given by
%
\begin{alignat}{2}
\dot{\delta}_b &= -kv_b + 3\dot{\Psi}, 
&\quad
\dot{v}_b &= -\mathcal{H} v_b + k\Phi + \frac{\dot{\tau}}{R}(v_b-\Theta_1),\\
\dot{\delta}_c &= -kv_c + 3\dot{\Psi}, 
&\quad
\dot{v}_c &= -\mathcal{H} v_c + k\Phi,
\end{alignat}
%
where $\mathcal{H} := aH$.
These equations should be modified if the matter content non-minimally couples to gravity (e.g., \cite{Bettoni:2011fs, Bettoni:2015wla}).
In addition, we need to know the time evolution of the electron's number density, $n_{\rm e}$, involved in the recombination history of hydrogen. The basic ideas and formulations are provided in Weinberg's textbook \cite{Weinberg:2008zzc}, and a well-known open source code, {\tt recfast}\footnote{{\tt\url{ https://www.astro.ubc.ca/people/scott/recfast.html}}} \cite{Seager:1999bc,Seager:1999km}, is also available.

The metric perturbations obey the perturbed Einstein equations sourced by the
matter density perturbations and the monopole moment of the photons and massless neutrinos.
In GR, they satisfy
%
\begin{align}
 \mathcal{H}\dot{\Psi} +\frac{k^2}{3}\Psi + \mathcal{H}^2\Phi 
&= -\frac{4\pi Ga^2}{3}\left(\rho_b\delta_b + \rho_c\delta_c
  + 4\rho_\gamma\Theta_0+ 4\rho_\nu \mathcal{N}_0\right)
, \label{eq:TH:Ein1}\\
 k^2(\Phi-\Psi) &=  -32\pi G a^2\left(\rho_\gamma\Theta_2+\rho_\nu \mathcal{N}_2\right).\label{eq:TH:Ein2}
\end{align}
%
In extended theories of gravity, only the left-hand sides are modified as long as the matter contents minimally couple to gravity.

Going back to the Boltzmann hierarchy in Eqs.~(\ref{eq:TH:Theta}) and~(\ref{eq:TH:E}), 
we have to solve a large number of ordinary differential equations.
We employ the line-of-sight formula developed by Seljak and Zaldarriaga~\cite{Seljak:1996is,Zaldarriaga:1996xe} (and see \cite{Lin:2004xy} for review) instead of directly solving the hierarchy. The Boltzmann equation has a formal solution,
%
\begin{align}
 \Theta_\ell(k,\eta_0) 
  = \int_0^{\eta_0}\!d\eta\,&
    \left\{
     g(\eta)\left[\Theta_0(k,\eta)+\Phi(k,\eta)+\frac{1}{2}P(k,\eta)\right]j_\ell(x) \right. \notag \\  &
     -g(\eta)v_b(k,\eta)j'_\ell(x)
     +\frac{3}{2}g(\eta)P(k,\eta)j''_\ell(x)  \notag \\ &\left. 
     +e^{-\tau}\left[\dot{\Phi}(k,\eta)+\dot{\Psi}(k,\eta)\right]j_\ell(x)
\right\}, 
\end{align}
%
where $x=k(\eta_0-\eta)$, $j_\ell(x)$ is the spherical Bessel function of the first kind, and
$g(\eta):=-\dot{\tau}e^{-\tau}$ is the visibility function determining the location of the last-scattering surface. Similarly, one can obtain the same integral form for the polarisations.
Note that the last line of the above equation depends on the time derivative of $\Phi$ and $\Psi$. 
Hence, it makes a non-negligible contribution when these metric perturbations are time-dependent, which could be caused in the Universe where the dark energy component is dominant.
This contribution is called the integrated Sachs-Wolfe (ISW) effect.

To evaluate $\Theta_{\ell}$ for large $\ell$, we need only the modes with $\ell\leq 2$ on the right-hand side.
Hence, it is required to compute the time evolution of the modes with $\ell\leq 2$ with high accuracy.
That can be done if we truncate the Boltzmann hierarchy at $\ell \sim 10$ with the 
boundary condition proposed in Ref.~\cite{Ma:1995ey}. 
Then, to obtain the angular power spectrum up to $\ell\leq 2000$, we can reduce the 
total number of equations from ${\sim} 6000$ (including the polarisations) to ${\sim} 40$.
The line-of-sight formula has been extended to include the 2nd-order effects such as the weak lensing \cite{Saito:2014bxa,Fidler:2014zwa}.\footnote{The authors of Ref.~\cite{Fidler:2014zwa} developed a second-order Boltzmann solver, {\tt SONG}, available at {\tt{\url{ https://github.com/coccoinomane/song}}}}

At the initial time, we can decompose the initial perturbations into the {\it adiabatic mode} and 
the {\it isocurvature mode}.
The isocurvature mode is further classified into {\it the baryon isocurvature mode},
{\it the CDM isocurvature mode}, {\it the neutrino density isocurvature mode}, and {\it the neutrino velocity isocurvature mode}.
The general initial condition for the scalar perturbations is a superposition of these five modes \cite{Bucher:1999re}.
The fraction of the isocurvature modes has been constrained to be less than ${\sim} 10\%$ of the total power of the 
perturbations on large scales ($k=0.002~{\rm Mpc}^{-1}$) \cite{Planck:2018jri}.
The adiabatic initial condition on the superhorizon scales is given as
%
\begin{align}
 \Psi &= -\frac{10+4f_\nu}{15+4f_\nu}\zeta, 
 \quad 
 \Phi = -\frac{10}{15+4f_\nu}\zeta, \label{eq:Psi0} \\
 \delta_c &= \delta_b = 3\Theta_0 = 3\mathcal{N}_0 = -\frac{3}{2}\Phi,
 \quad
 v_c = v_b = -3\Theta_1 = -3\mathcal{N}_1 = -\frac{k}{2\mathcal{H}}\Phi,
 \quad
 \mathcal{N}_{2} = -\frac{\Phi-\Psi}{12f_\nu}\frac{k^2}{\mathcal{H}^2},
\end{align}
%
where $f_\nu:=\rho_\nu/(\rho_\gamma+\rho_\nu)$ and
$\zeta(\boldsymbol{k})$ is the curvature perturbation generated in the inflationary epoch. 
For the detailed derivation of this result, see the standard textbook~\cite{Dodelson:2003ft}.

The temperature and $E$/$B$-mode fluctuations at the present time can be decomposed as 
$X(\eta_0,k) = \widehat{X}(\eta_0,k)\zeta(k)$ for $X=\Theta, E, B$. The transfer function, $\widehat{X}(\eta_0,k)$, is computed by a Boltzmann solver. 
Then, the angular power spectrum can be computed as
%
\begin{align}
 C_\ell^{XY} &= \frac{1}{(2\ell+1)^2}\frac{2}{\pi}
  \int\!dk\,k^2 \widehat{X}_\ell^{*}(\eta_0,k)\widehat{Y}_\ell(\eta_0,k)P_\zeta(k).
\end{align}
%
The primordial power spectrum, $P_{\zeta}(k)$, is usually parameterised as
\begin{align}
P_{\zeta}(k) = \frac{2\pi^2}{k^3}A_s\left(\frac{k}{k_{\rm pivot}}\right)^{n_s-1}\,,
\end{align}
where
$A_s=2.101^{+0.031}_{-0.034}\times 10^{-9}$ and $n_{s}=0.965\pm 0.004$ at $k_{\rm pivot}=0.05\,{\rm Mpc}^{-1}$
from the Planck 2018 results \cite{Planck:2018vyg}.
We can define the angular power spectra for the vector and tensor perturbations in a similar way \cite{Hu:1997hp}.

\subsubsection{Demonstrations}
\label{subsubsec:demonstration}


The DHOST theory is an extension of the Horndeski theory and
is so far the most general systematically constructed theory having
a single scalar and two tensorial degrees of freedom.
The action of the DHOST theory contains free functions of the scalar field, $\phi$, and its kinetic term, $X:=-(1/2)\partial^\mu\phi\partial_\mu\phi$ (see Sec.~\ref{sec:DHOST}). Fixing the functions specifies a particular theory in the DHOST family.
The Boltzmann solvers such as {\tt hi\_class}, {\tt COOP}, {\tt MGCAMB}, and {\tt CMB2nd} implement
the EFT description of the Horndeski and DHOST
theories~\cite{Langlois:2017mxy,Gleyzes:2015pma,DAmico:2016ntq,Bloomfield:2012ff,Gubitosi:2012hu,Gleyzes:2013ooa,Bloomfield:2013efa,Piazza:2013coa,Gleyzes:2014rba}.
The dictionary translating between the DHOST Lagrangian and the EFT description
is given in the previous section.

The quadratic action of the scalar perturbations in the gravity sector is given on the basis of the ADM formalism, and we add possible operators to the action according to our assumed symmetries. As discussed in the previous section, the coefficients of the operators such as $K_{ij}K^{ij}, {}^{(3)}R$ are parameterised by the time-dependent functions. A phenomenologically useful parameterisation is the $\alpha$-basis parametrisation~\cite{Bellini:2014fua} (for a review, see \cite{Frusciante:2019xia}). As shown in the previous section, the Horndeski theory can be represented by four $\alpha_i$ parameters ($i={\rm K},{\rm B},{\rm T},{\rm M}$), and $\alpha_{\rm H}$ needs to be added in the GLPV
theory~\cite{Gleyzes:2014rba,Gleyzes:2014qga}.
Going beyond the GLPV theory, the DHOST theory\footnote{In this section, we focus only on the quadratic DHOST theory that contains up to $(\partial\partial\phi)^2$. The cubic DHOST theory containing $(\partial\partial\phi)^3$ has also been formulated \cite{BenAchour:2016fzp}. However, the cubic DHOST theory has no viable cosmological solutions \cite{Ezquiaga:2017ekz,Baker:2017hug,Creminelli:2017sry,Minamitsuji:2019shy}.}
 needs further parameters $\alpha_{\rm L}, \beta_1, \beta_2$ and $\beta_{3}$ \cite{Langlois:2017mxy}, and 
the theories including higher spatial derivatives accounting for Lorentz violation
require extra ones, $\alpha_{K^2}, \alpha^{\rm GLPV}_{\rm B}$ \cite{Frusciante:2016xoj}.
The time dependence of the $\alpha$-parameters is in principle arbitrary. However, it is frequently assumed that
$\alpha_i(t) = \alpha_{i,0}\Omega_{\rm DE}(t)$ (or $\alpha_i(t) = \alpha_{i,0}\Omega_{\rm DE}(t)/\Omega_{{\rm DE},0}$) in the literature, where $\Omega_{\rm DE}(t)=1-\Omega_{\rm m}(t)$ is the density parameter of (effective) dark energy and the subscript '0' means the value at the present time.

\begin{figure}[t]
\centering{
 \includegraphics[width=7cm]{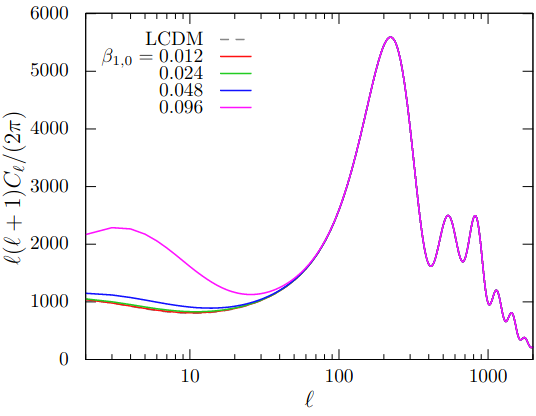}
 \includegraphics[width=7cm]{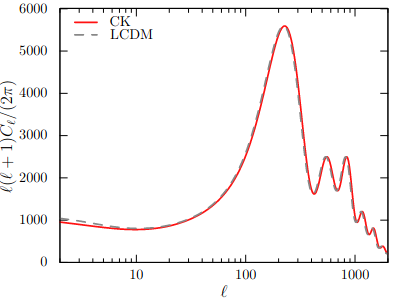}
}
 \caption{Angular power spectrum of the temperature anisotropies, $C^{TT}_\ell$, in EFT with varying $\beta_{1,0}$ ({\it left}) \cite{Hiramatsu:2020fcd} and that in the Crisostomi-Koyama model with with $(c_2,c_3,c_4,\beta)=(3.0,5.0,1.0,-5.3)$ ({\it right})  \cite{Hiramatsu:2020fcd}.}
 \label{fig:TH:beta1}
\end{figure}

As a demonstration of the Boltzmann solver for the DHOST theory~\cite{Hiramatsu:2020fcd}, 
we show $C^{TT}_\ell$ for different values of $\beta_{1,0}$ in the left panel of Fig.~\ref{fig:TH:beta1}, where we set $\alpha_{{\rm K},0}=1$ and other $\alpha$-parameters to be zero. We assume the background evolution to be identical to that of the $\Lambda$CDM model. 
The change in the angular power spectrum from the $\Lambda$CDM case is manifest only on large scales, where the scalar degree of freedom mediating gravity modifies
the ISW effect.
One can find a similar feature in the GLPV theory where $\alpha_{{\rm H},0}\ne 0$ and $\beta_{1,0}=0$ \cite{DAmico:2016ntq} and in the kinetic braiding \cite{Deffayet:2010qz} where $\alpha_{{\rm B},0}\ne 0$ and other $\alpha$-parameters are set to be zero \cite{Zumalacarregui:2016pph}.

Next, we consider another approach in which we begin with the original action of the DHOST theory described by the free functions of the scalar field $\phi$ and its kinetic term $X$.
The action contains eight arbitrary functions, $P,Q,f_2,a_1,\cdots,a_5$. Here we adopt 
a simple model proposed by Ref.~\cite{Crisostomi:2017pjs},
\begin{align}
P=-2c_2X,
\quad Q=-2c_3\frac{X}{\Lambda^3}, \quad f_2=\frac{m_{\rm pl}^2}{2}+4c_4\frac{X^2}{\Lambda^6}, \quad a_1=0, \quad a_3 = -\frac{\beta+8c_4}{\Lambda^6}.
\label{eq:TH:CK}
\end{align}
The remaining parameters, $a_2, a_4$ and $a_5$, are determined from the degeneracy conditions~\cite{Langlois:2015cwa,Langlois:2015skt}. The variation of the action yields the equations for the background and the perturbations.

In the right panel of Fig.~\ref{fig:TH:beta1}, we show the angular power spectra with $(c_2,c_3,c_4,\beta) = (3.0, 5.0, 1.0, -5.3)$.
From the view of the $\alpha$-basis EFT,
this parameter set yields $\beta_{1,0}\sim 0.06$ and $\alpha_{{\rm H},0} \sim -0.2$ in the EFT language.
As the effects of these parameters degenerate, the effects arising from the two parameters on large scales are almost cancelled and thus the spectrum becomes close to that in the $\Lambda$CDM model (dashed line).

In conjunction with the Markov-Chain Monte-Carlo (MCMC) method, we can estimate
the viable parameter ranges for the model parameters characterising the extended theories of gravity as well as the standard cosmological parameters, $(A_s, n_s, h, h^2\Omega_{\rm c}, h^2\Omega_{\rm b}, \tau)$ \cite{Hiramatsu:2022fgn}. 
For simplicity, we define a subset of the DHOST theories where $\alpha_{{\rm H},0}=\alpha_{{\rm M},0}=\alpha_{{\rm T},0}=\alpha_{{\rm B},0}=0$ dubbed as ``EFT'', and another subset, the Crisostomi-Koyama model, given in Eq.~(\ref{eq:TH:CK}) where we fix $c_2=1$ and $c_4=-8\beta$ leading to $a_{3}=0$, dubbed as ``cCK''. We vary one parameter, $\beta_{1,0}$, in EFT, and two parameters, $c_3$ and $c_4$, in cCK. As a result of the MCMC analysis, we obtain the best-fit values with their uncertainties at a 68\% confidence level in EFT and cCK, as shown in Table~\ref{tab:TH:bestfit}. We put constraints on the model parameters characterising the gravity theories,
%
\begin{align}
\beta_{1,0} = 0.032_{-0.016}^{+0.013} \quad ({\rm EFT}); \quad c_3=1.59_{-0.28}^{+0.26}, \quad (0<)~c_4 < 0.0088 \quad ({\rm cCK})~~~(68\%\,{\rm c.l.}). 
\end{align}
%

In Ref.~\cite{Hiramatsu:2022ahs}, we employ a cuscuton-like theory, i.e., a ``scalarless''
scalar-tensor theory. The theory is equipped with a non-dynamical scalar field and the
usual two tensorial degrees of freedom corresponding to gravitational waves.
The action in terms of the ADM variables is given by
%
\begin{align}
S=\frac{M_{\rm Pl}^2}{2}\int{\rm d} t{\rm d}^3x\, \sqrt{\gamma}N\left[
K_{ij}K^{ij}-\frac{1}{3}
\left(\frac{2N}{\beta+N}+1\right)K^2
+R+\alpha_1+\frac{\alpha_3}{N}
\right]+S_{\mathrm{m}},
\label{2DoF1}
\end{align}
%
with
%
\begin{align}
    \alpha_1(t)&=\frac{6\beta(2+\beta)}{(1+\beta)^2}H_{\mathrm{GR}}^2-2\Lambda,
    \\
    \alpha_3(t)&=4\frac{\rm d}{{\rm d} t}\left(\frac{\beta H_{\mathrm{GR}}}{1+\beta}\right)
    -\frac{6\beta H_{\mathrm{GR}}^2}{(1+\beta)^2}.
    \label{2DoF3}
\end{align}
%
where $N, \gamma_{ij}$, and $K_{ij}$ are the lapse function, the spatial metric, and the extrinsic curvature, respectively. 
We choose $\alpha_1$ and $\alpha_3$ so that the background evolution is the same as that in the $\Lambda$CDM model.
Note that the constant parameter, $\beta$, defined here differs from those in the EFT and cCK model introduced above.
The theory has a particular feature which recovers GR both in the small-scale and the large-scale limits.
Hence, the only possible way to test the theory is to focus on the phenomena involved in the intermediate-scale physics like the CMB and LSS. As a result of the MCMC analysis, we found that the best-ﬁt range for $\beta$ is
\begin{align}
    \beta = -0.0388^{+0.011}_{-0.0083}~~~(68\%\,{\rm c.l.}),
\end{align}
which indicates a ${\sim} 4\sigma$ deviation from GR $(\beta=0)$, and the other parameters can be found in Table~\ref{tab:TH:bestfit} (see ``2DoF'').

\begin{table}[!t]
\centering
\begin{tabular}{c|cl|cl|cl}
\hline
\hline
                      & \multicolumn{2}{c}{2DoF}         & \multicolumn{2}{|c}{EFT}                          & \multicolumn{2}{|c}{cCK} \\
\hline
                      &~~      &~~~                             &~~             &~~~                               &~~$c_3$ &$1.59_{-0.28}^{+0.26}$          \\[0.4em]     
                      &~~$\beta$&$-0.0388_{-0.0083}^{+0.011}$ &~~$\beta_{1,0}$&$0.032_{-0.016}^{+0.013}$      &~~$c_4$ &$<0.0088$                       \\[0.4em] \hline
$10^{9}A_se^{-2\tau}$           & &$1.8902_{-0.0062}^{+0.0090}$    &               &$1.8986_{-0.0048}^{+0.010}$    &        &$1.8816_{-0.0050}^{+0.0055}$    \\[0.4em]
$n_s$                 & &$0.9748_{-0.0025}^{+0.0035}$    &               &$0.9735_{-0.0051}^{+0.0025}$   &        &$0.9909_{-0.0030}^{+0.0028}$    \\[0.4em]
$h$                   & &$0.6853_{-0.0051}^{+0.0027}$    &               &$0.6760_{-0.0073}^{+0.0048}$   &        &$0.8043_{-0.0058}^{+0.0040}$    \\[0.4em]
$h^2\Omega_{\rm c}$      & &$0.11843_{-0.00069}^{+0.0010}$  &               &$0.1194_{-0.0011}^{+0.0016}$   &        &$0.11523_{-0.00072}^{+0.0011}$  \\[0.4em]
$h^2\Omega_{\rm b}$      & &$0.02172_{-0.00012}^{+0.00011}$ &               &$0.02175_{-0.00017}^{+0.00012}$&        &$0.02218_{-0.00012}^{+0.00013}$ \\[0.4em]
$\tau$                & &$0.0497_{-0.0052}^{+0.0052}$    &               &$0.0494_{-0.0045}^{+0.0061}$   &        &$0.0423_{-0.0071}^{+0.0044}$    \\[0.4em]
\hline
$\ln\mathcal{L}$      & &$-1432$                         &               &~~~$-1428$                        &        &~~~$-1463$                         \\[0.4em]
\hline
\hline
\end{tabular}
\caption{The best-fit values with their uncertainties at a 68\% confidence level in 2DoF model, EFT and cCK model from left to right. The inequality indicates that we obtain only the upper limit \cite{Hiramatsu:2022fgn,Hiramatsu:2022ahs}. }
\label{tab:TH:bestfit}
\end{table}

\begin{figure}[t]
\centering{
 \includegraphics[width=7cm]{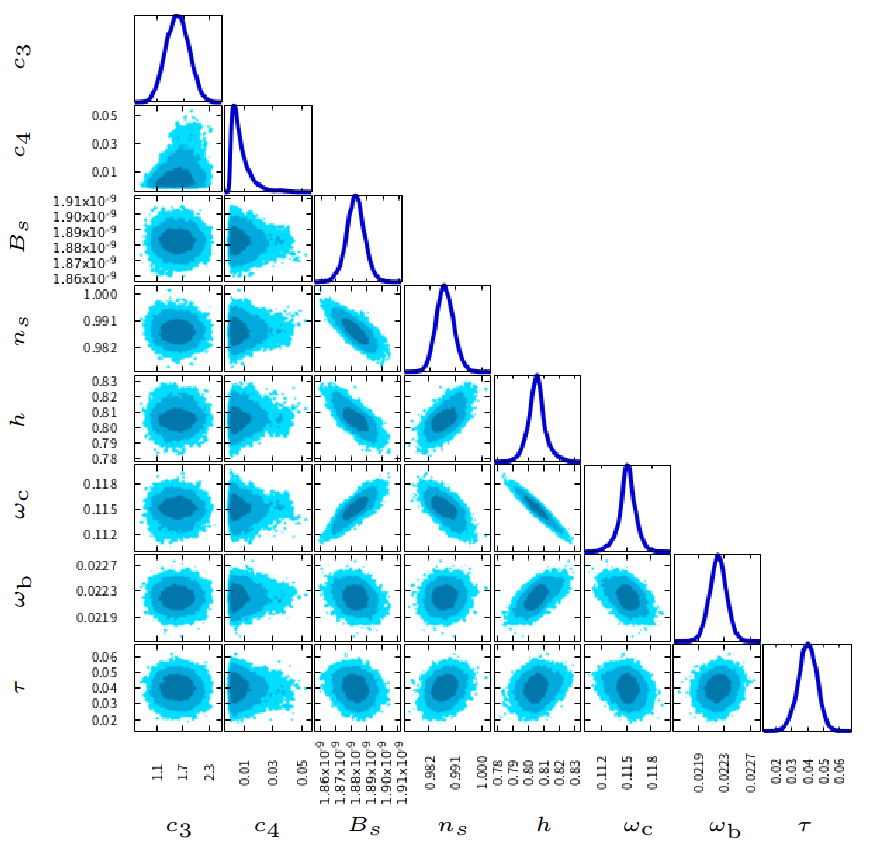}
 \includegraphics[width=7cm]{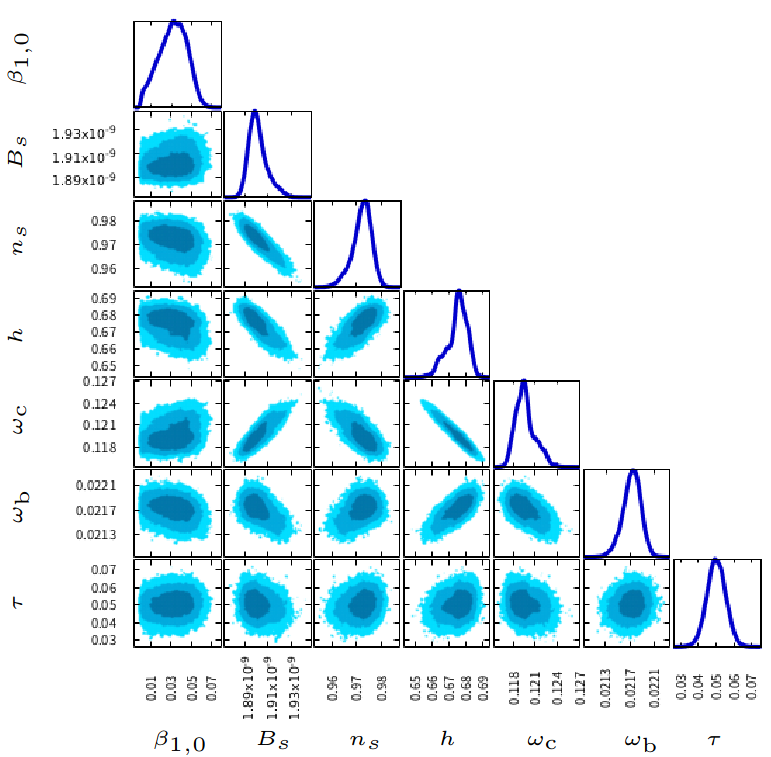}
}
 \caption{$68\%$, $95\%$ and $99\%$ confidence contours in cCK model ({\it left}) and EFT ({\it right}) \cite{Hiramatsu:2022fgn}.}
 \label{fig:TH:contour}
\end{figure}

\subsection{Predicting non-linear structure formation}
\label{sec:emulator}
\subsubsection{Overview}
As explained in Sec.~\ref{subsec:LSS}, various non-linear effects are involved in cosmological LSS, including the non-linearity in the gravitational structure growth, the relation between the underlying density field and the observed galaxies (i.e., galaxy bias), and the RSD (see Sec.~\ref{subsubsec:RSD}), which states the mapping between the actual positions of galaxies and those determined through the measurement of redshifts (see \cite{Bernardeau:2001qr,Desjacques18} for review). These make analytical approaches difficult, and one has to resort to numerical simulations instead. However, for cosmological analyses, i.e., in inferring cosmological models and their model parameters from observations, numerical simulations are often too costly. 

Simulation-based inference is a key ingredient in different domains of scientific and engineering research these days. Interested readers may find how different approaches have been considered and developed in a review article~\cite{Cranmer20}. Here, we focus on emulation as a working example of simulation-based inference in the context of cosmological LSS.

\subsubsection{Emulation}
Emulation is a useful technique that replaces expensive numerical simulations with a much cheaper statistical model. The idea is that one first performs many, but an affordable number of numerical simulations at different input parameters and then uses them as the training data for the statistical model. The statistical model, called an \textit{emulator}, then generalises the prediction to unexplored sets of input parameters. Here, one can consider summary statistics such as the matter power spectrum, or even the full density field, as the target quantity to be emulated as a function of cosmological parameters. The advantage of this approach is that one does not need knowledge of the functional form of the target quantity (e.g., how the power spectrum should depend on $\Omega_\mathrm{m}$), unlike the traditional fitting formula approach. This \textit{non-parametric} property makes emulators attractive for modelling various different quantities. The dependence of the target quantity on the input parameters is instead \textit{learnt} completely in a data-driven way.

This technique was first introduced to the problem of cosmological LSS formation in Refs.~\cite{Heitmann06,Habib07}. The same group has developed an emulator for the matter power spectrum (Coyote Universe: \cite{Heitmann10,Heitmann10,Lawrence10}). They initially considered a five-parameter $w$CDM cosmology, where the value of $h$ is automatically determined to match the observed acoustic scale from CMB, and now is updated to include not only $h$ \citep{Heitmann14}, but also the time-varying equation of state parameter as well as neutrino masses to cover an eight-parameter input space (Mira-Titan Universe: \cite{Heitmann16,Lawrence17,Moran22}). There are other attempts to model the matter power spectrum using emulation \cite{Agarwal12,Agarwal14,EuclidEmu1,EuclidEmu2}. Coyote Universe/Mira-Titan Universe project also develop emulators for halos (the mass function \cite{Bocquet20}, and the concentration mass relation \cite{Kwan13}) or galaxies \cite{Kwan15}. There are also several attempts to model various weak lensing statistics \citep{HarnoisDeraps19,Liu18,Liu19,ZackLi19,Coulton19,Marques19,HarnoisDeraps21,Martinet21,Davies22}. More recently, there appear different groups to model the clustering and lensing signals of biased tracers (Aemulus project: \cite{DeRose19,McClintock19,Zhai19,McClintock19b,Zhai22}, AbacusCosmos~\cite{Garrison18}, AbacusSummit~\cite{Maksimova21} and Dark Quest~\cite{Nishimichi:2018etk,Kobayashi20,Cuesta-Lazaro22}, see Sec.~\ref{subsubsec:DQ} for more detail). Some emulators constructed from these projects are actually used to analyse observational data and derive cosmological parameters (e.g., \cite{Huff14,Liu15,Petri15,HarnoisDeraps21,Zurcher22,Zhai22,Yuan22,Miyatake:2021sdd,Kobayashi22}). 

\subsubsection{\texttt{DarkEmulator}: halo model meets emulation}
\label{subsubsec:DQ}

We introduce an emulator code, \texttt{DarkEmulator}, developed as a part of the Dark Quest project \cite{Nishimichi:2018etk}. This code is aimed at providing theoretical templates in light of recent multi-probe cosmological analyses from imaging and spectroscopic galaxy surveys. \texttt{DarkEmulator} follows the halo model approach (see \cite{Cooray02} for a review). Galaxies are assumed to form only in dark matter halos, which are gravitationally bound mass concentrations, hierarchically formed by gravitational instability from primordial tiny fluctuations. While various complicated baryonic processes are essential for the formation of galaxies, the formation of halos is rather straightforward to model by only considering gravity. The clustering and abundance of halos can therefore be safely modelled by standard gravity-only $N$-body simulations. 

The halo model then provides a natural way to describe the correlation functions (or their Fourier counterpart, polyspectra) of mass or galaxies. The simplest example is the two-point correlation function, which is the excess pair counts as a function of the pair separation compared to a random uniform distribution. One separates the pairs counts into two components: the one-halo term, i.e., pairs in the same halo, and the two-halo term, those in two distinct halos. In this approach, one needs the knowledge of (1) the abundance of halos, (2) the two-point correlation function of halos, and (3) how the object of interest (mass or galaxies) is distributed within a halo.

It is natural to consider that this last ingredient depends on the properties of halos. The major property of a halo is its mass, and others are usually not taken into account in the model.\footnote{In practice, the clustering property of halos is known to be dependent on other halo properties as well, such as the formation time, concentration, ellipticity and the environment (i.e., whether a halo is in an overdense or underdense region). These extra dependencies are often called halo assembly bias (e.g., \cite{Gao05}). This matters when the galaxies of interest preferentially form in halos with certain properties other than the mass.} In the simplest halo model, one, therefore, needs all three ingredients as a function of mass. The first two are the halo mass function, $(\mathrm{d}n_\mathrm{h}/\mathrm{d}M)(M)$, and the halo-halo cross-correlation function for halos with two different masses, $\xi_\mathrm{hh}(M_1,M_2)$. These can be measured and calibrated by $N$-body simulations. If one is interested in the distribution of mass, the third ingredient can also be studied using $N$-body simulation. The Navarro-Frenk-White (NFW) profile is a well-known result from $N$-body simulations for the radial profile of mass in a halo, which is approximately universal over a wide range of halo masses and also for different cosmological models \cite{NFW96}. However, predicting the position and abundance of galaxies in a halo is much more involved. A practical way forward is to employ a simple parametric model, called the halo occupation number (HOD, e.g., \cite{Zheng05}), and vary the HOD parameters together with the cosmological parameters when the model is fitted to observational data. This is the approach on which \texttt{DarkEmulator} is based. It models the ingredients (1) and (2) as functions of the halo mass, redshift, and cosmological parameters. In contrast, the remaining ingredient, (3), is modelled using a simple but flexible parametric model, which is to be fixed when confronted with an actual galaxy catalogue from observation.

Dark Quest simulations are arranged in a six-parameter flat-geometry $w$CDM cosmology centred on the Planck result \cite{Planck:2015cosmo}. Once one defines the cosmological model and the parameter ranges over which to make predictions, the first task is to find an efficient way to generate samples at which simulations will be performed. This is often a difficult problem, especially when the dimensionality of the input parameter space is high (i.e., the curse of dimensionality). An often adopted strategy is based on Latin Hypercube designs (LHD). That is, the hyperrectangle-shaped input space is divided into uniform grid cells, and one asks to put one and only one sample point in any segment in the one-dimensional projection. This ensures a good projection property by construction, while the required sampling point is equal to the number of one-dimensional segments, regardless of the dimensionality. One usually imposes additional conditions to further ensure a good space-filling property. Dark Quest simulations follow a maximin-distance sliced LHD, in which the whole design points are subdivided into ``slices'', each of which separately follows the conditions for an LHD~\cite{Ba15}. The space-filling property is ensured by minimizing a cost function that is designed to maximise the minimum distance between the pair of samples, both within the slices and in the whole sample. $100$ samples are chosen in five slices for Dark Quest. In the subsequent analyses, one slice of 20 cosmological models is left unused in the training process and is instead used as a validation sample.

The next task is to model the dependence of the target quantities on the cosmological parameters, that is, \textit{regression}. \texttt{DarkQuest} employs Gaussian Process (GP)~\cite{Rasmussen06}. It assumes that the target function is a random process controlled by our prior knowledge and the available data points. Assuming that this follows a multivariate normal distribution, any calculation can be done analytically via matrix products. A normal distribution is governed by the mean function and the covariance function. One can design these functions on the basis of prior knowledge. Usually, the mean function is set to zero, and one encodes the known properties, such as the expected smoothness, amplitude, or periodicity, in the covariance function. Usually, a parameterised covariance function is assumed and then faced with the simulation database. The parameters characterising the covariance function can be optimised by asking that the simulation data in hands are the most likely to be generated from the distribution function. In this way, a GP works as a machine learning model, which is trainable in a data-driven manner.

A possible caveat of GP is that it is not straightforward to generalise for multi-output regression problems. In reality, given cosmological parameters, one wishes to predict various statistical quantities of halos at different redshifts, halo masses, pair separations, and so on. This makes the output data vector pretty large. \texttt{DarkEmulator} tries to reduce the dimensionality of the output data vector by applying a weighted Principal Component Analysis (PCA)~\cite{Bailey12}. The PCA weights are individually modelled by GPs. In the follow-up papers of \cite{Nishimichi:2018etk}, the halo statistics are modelled by the Feedforward Neural Network (FFNN) \cite{Kobayashi20,Cuesta-Lazaro22}. This is a natural framework for dealing with multi-input/multi-output regression problems. These studies have found that although the non-linear structure formation is a heavily non-linear process, the dependence of standard statistical quantities, such as the correlation function or the power spectrum, are only weakly non-linear functions of cosmological parameters. Therefore, relatively shallow networks with a few to several hidden layers are sufficient to successfully model the functions.

\subsubsection{Extensions to modified gravity models}

Here, we mention the theoretical modelling of non-linear structure formations in modified gravity.
Different from GR, there are in general extra degrees of freedom in modified gravity, mostly a light scalar degree of freedom, other than the standard massless spin-2 graviton. Then Einstein field equations and the standard fluid equations for baryon and dark matter are augmented as including interactions of the scalar field. 
Hence, it is significantly different from GR when considering gravitational interactions for structure formations in modified gravity, even in non-linear ways.

The non-linearity of the scalar-field equation predicts phenomena that never occur in GR. Virtually, certain screening mechanisms, i.e., Vainshtein screening, Chameleon mechanism, symmetron model, or $k$-mouflage in high-density regions (see in Sec.~\ref{sec:screening}) suppress the effects of the fifth force mediated by the scalar field, recovering Newton law at small scales. The screening mechanisms thus can make a theory evading the stringent constraints from local tests of gravity, e.g., solar-system test and Hulse-Taylor pulsar. At large scales, on the other hand, the fifth force cannot be suppressed as the matter density becomes smaller, printing a signature of the fifth force on the dynamics of structure formations. Typically, the scale where the screening does not work is above the galactic scales.  Hence, it is reasonable to test modified gravity via non-linear structure formations of LSS.

To model non-linear dynamics in modified gravity, it is straightforward to apply $N$-body simulation for the theories, as same as GR. Since the scalar field becomes dynamical at considered scales, one needs to solve the equation of motion of the scalar field in conjunction with the Vlasov-Poisson system. A difficulty lies in solving the non-linear equation of motion of the scalar field, which in general becomes more challenging as containing non-linear terms composed of derivatives.
To this end, $N$-body simulations in modified gravity can be expensive in computation even more than those in GR.

It also matters in modified gravity to take a quasi-static limit to formulate the Vlasov-Poisson equations.  In generic theories, the sound speed of the propagating scalar mode is dependent on a choice of models, and thus there is no {\it a priori} information on whether or not the quasi-static limit is applicable. At present, we assume the quasi-static limit to model non-linear structure formations in modified gravity. Once provided the quasi-static limit, there exist useful parametrisations that characterise the effects of gravitational interactions. For instance, as shown in  the previous section, Horndeski theory or vector-tensor theories precisely predict the values of the gravitational coupling for matter $G_{\rm matter}$ (or $\mu$) and photon $G_{\rm light}$ (or $\Sigma$).

Finally, it is necessary to clarify the baryon feedback or the effect of neutrinos should be consistently implemented accommodating with modified gravity models.
It is uncertain whether such existing effects degenerate with the effects of modified gravity (cf. \cite{Wright:2019qhf} reports the degeneracy between massive neutrinos and the effects of modified gravity in RSDs). Hence it is important to study non-linear modelling with baryons and neutrinos in modified gravity.

As long as currently known, $N$-body simulations in the specific subclass of modified gravity, i.e., $f(R)$ gravity \cite{Oyaizu:2008sr,Li:2011vk} and DGP model \cite{Schmidt:2009sg,Khoury:2009tk} are the most popular. The major algorithm is {\tt ECOSMOG}~\cite{Li:2011vk}, {\tt MGGADGET}~\cite{Puchwein:2013lza}, and {\tt ISIS}~\cite{Llinares:2013jza}.\footnote{{\tt ISIS} code does not take into account baryon physics as same as the other two algorithms, although the original {\tt RAMSES} code \cite{Teyssier:2001cp} that is the basis of {\tt ISIS} can compute the hydrodynamics of baryons by using a second-order Godunov method.} The paper \cite{Winther:2015wla} compares all the existing algorithms for $f(R)$ gravity, DGP, and Symmetron models by using the velocity divergence power spectrum, halo abundances, and halo profiles. Other than the popular $N$-body codes, there is a $N$-body simulation for the minimal theory of massive gravity (MTMG) \cite{Hagala:2020eax}.
Although there are some upgrades to reduce the computational costs of simulations, 
e.g., Refs.~\cite{Ruan:2021wup, Hernandez-Aguayo:2021kuh,Fiorini:2021dzs},
so far, there is no simulation that covers the whole Horndeski theory\footnote{Very recently, {\tt hi-COLA}~\cite{Wright:2022krq} which is based on COLA\cite{Tassev:2013pn}, which is the first model for non-linear structure formation in the framework of Horndeski theory have been developed.} and DHOST theory.


\section{Outlook}\label{sec:outlook}

This section summarises the current status of theories of modified gravity in light of the following three perspectives: (i) {\it Physical motivations}, (ii) {\it Validity and Appealing features}, and (iii) {\it Maturity and Calculability}. {\it Physical motivations} clarify the underlying motivation of a given theory of modified gravity and whether the theory is self-consistent. The theory should evade pathological instabilities and a strong coupling. Otherwise, it loses predictability. In addition, some theories of modified gravity may require a violation of fundamental assumptions such as Lorentz invariance or locality. Even if they are seemingly satisfied at low energies, a violation may be required for a theory to have a self-consistent UV completion. Testing these theories may lead to a test of the fundamental assumptions. {\it Validity and Appealing features} summarise whether the theory is consistent with the existing observational constraints. We already have strong observational constraints on modifications of gravity from solar-system experiments, gravitational waves, and cosmological observations. 
Furthermore, it would be interesting if the theory has a unique observational feature that can be tested by future observations. The third point is {\it Maturity and Calculability}: the theory is already well-developed to compute observables for future observations or not. The three criteria are not independent; for instance, if a theory has a low maturity and calculability, it is not clear whether the theory has an appealing feature for observational tests. The purpose of this classification is to understand which part is missing at the current stage and to provide the outlook for further theoretical and observational developments.

~

\textbf{Horndeski theory}
\begin{itemize}
\item {\it Physical motivations.} 
\indent The Horndeski theory is the most general scalar-tensor theory having the second-order field equations in four dimensions. Well-known theories are included in the Horndeski family: Quintessence, k-essence, $f(R)$ gravity, Jordan-Brans-Dicke theory, Generalised Brans-Dicke theory, Kinetic gravity braiding, Galileon, Einstein-scalar-Gauss-Bonnet gravity, and so on. Various self-accelerating cosmological solutions are known. 
While modifying the gravitational law at large distances $(\mu\neq 1, \Sigma \neq 1)$, GR can be recovered on small scales thanks to one of the screening mechanisms, though whether the mechanism works successfully or not depends on some details of the theory. It should be noted that recent studies~\cite{Bellazzini:2020cot,Tolley:2020gtv,Caron-Huot:2020cmc,Arkani-Hamed:2020blm,Joyce:2014kja,Aoki:2021ffc} suggest that the screenings via non-linear derivative interactions (the Vainshtein screening and the kinetic screening) may contradict the requirements from the positivity bounds, though the validity of the positivity bounds on the cosmological background is an open question~\cite{Baumann:2015nta, Grall:2021xxm, Aoki:2021ffc,Melville:2022ykg,Creminelli:2022onn}.

\item {\it Validity and Appealing features.} 
\indent As the Horndeski theory includes many models of modified gravity and even the conventional dark energy models \cite{Ratra:1987rm,Caldwell:1997ii}, many observational constraints have been discussed. Constraints from the solar-system experiments and the observation of the Hulse-Taylor pulsar can be passed when the screening mechanism is efficient. 
Since the Horndeski theory contains arbitrary functions, cosmological constraints would be biased by the choice of the functions. As a phenomenological approach, cosmological constraints are often discussed by using the EFT parameters with assumed time dependence. The Planck collaboration reported the constraint $\alpha_{{\rm M},0} = -0.015^{+0.019}_{-0.017}$ and $\beta=0.66^{+0.44}_{-0.21}$ for the parametrisation $\alpha_{\rm M} = \alpha_{{\rm M},0}a^\beta$ combined with Planck, baryon acoustic oscillation, redshift-space distortion, and galaxy weak lensing measurements~\cite{Planck:2018vyg}, while the LSS gives $\alpha_{{\rm B},0}=0.20^{+0.20}_{-0.33}$ and $\alpha_{{\rm M},0}=0.25^{+0.19}_{-0.29}$ for $\alpha_i(t)=\alpha_{i,0} \Omega_{\rm DE}(t)~(i = {\rm M},{\rm B})$~\cite{SpurioMancini:2019rxy}. The present value of $\alpha_{\rm T}$ is tightly constrained by the arrival-time difference between the signals of GW170817 and GRB 170817A: $|\alpha_{\rm T}| \lesssim 10^{-15}$~\cite{Sakstein:2017xjx,Creminelli:2017sry,Ezquiaga:2017ekz,Baker:2017hug,Arai:2017hxj}.
It was, however, pointed out
that the energy scales of LIGO observations are close to the
cutoff scale of dark energy/modified gravity models in the Horndeski family~\cite{deRham:2018red}, and hence
more careful investigations are necessary on the constraints from the speed of gravitational waves.
To conclude, there are still viable parameter spaces that may be tested by future observations.

\item {\it Maturity and Calculability.} 
\indent Linear cosmological perturbations around the FLRW metric are well formulated as shown in Sec.~\ref{sec:Perturbative predictions in modified gravity}, and public Boltzmann codes, {\tt hi\_class} and {\tt EFTCAMB}, are available. 
The matter bispectrum has been derived to trace the non-linearity of LSS~\cite{Terukina:2013eqa,Bellini:2015wfa}. As for full non-linear simulations, there have been a few $N$-body simulations except for the major subclasses: Quintessence, DGP and $f(R)$, and Generalised Brans-Dicke theory. Recently, {\tt hi-COLA}~\cite{Wright:2022krq} was released: a fast simulation to compute non-linear structure formations based on COLA algorithms \cite{Tassev:2013pn} in the Horndeski theory with the luminal gravitational wave speed. This is the very first attempt to compute non-linear structure formation in the Horndeski theory, and further developments could reveal interesting observational features that can be tested by forthcoming observations.
\end{itemize}

\textbf{DHOST theory}

\begin{itemize}
\item {\it Physical motivations.} 
\indent The DHOST theory is an interesting extension of the Horndeski theory as it is a counterexample to the common belief that equations of motion have to be second-order differential equations.
A healthy class of the DHOST theory is related to the Horndeski theory via a disformal transformation, while there are other classes that are disconnected to the Horndeski theory. Theories in the latter classes are, however, phenomenologically unacceptable due to gradient instabilities in cosmological perturbations of scalar and tensor modes. The degeneracy conditions would demand fine-tuning of the theory and would not be protected by symmetry in general (see, however,~\cite{Aoki:2019rvi}). Nonetheless, a small violation of the degeneracy conditions is accepted and indeed required in the so-called stealth solutions \cite{Motohashi:2019ymr}, making the DHOST theory a self-consistent EFT.

\item {\it Validity and Appealing features.} 
\indent Although the DHOST theory of our interest is disformally related to the Horndeski theory, it exhibits features distinct from the Horndeski theory in the presence of matter, as a consequence of the derivative coupling between matter fields and gravity in the Horndeski frame. 
In the linear cosmological perturbations, ``beyond Horndeski''
effects of the DHOST theory are described by the parameters (functions) $\{\alpha_{\rm H},\beta_1,\beta_2,\beta_3\}$ (note that these four are not independent). 
One can give constraints on the $\beta$-parameters from CMB, e.g.,
$\beta_{1,0} = 0.032_{-0.016}^{+0.013}$ for the parametrisation $\beta_1(t)=\beta_{1,0}\Omega_{\rm DE}(t)$~\cite{Hiramatsu:2022fgn}. In addition to the luminal propagation of gravitational waves,
the particular relation
$\alpha_{\rm H} + 2\beta_1=0$ is often assumed to prohibit a decay of gravitational waves into dark energy~\cite{Creminelli:2018xsv}.\footnote{On top of the perturbative decay~\cite{Creminelli:2018xsv}, a resonant decay is relevant in a certain range of $\alpha_H + 2\beta_1$~\cite{Creminelli:2019nok}: a change of the gravitational wave signals is observable for $10^{-20} \lesssim \alpha_H + 2\beta_1 \lesssim 10^{-17}$ in the LIGO/VIRGO band and $10^{-16} \lesssim \alpha_H + 2\beta_1 \lesssim 10^{-10}$ in the LISA band.}
Due to the derivative coupling seen in the Horndeski frame,
partial breaking of the Vainshtein screening mechanism occurs inside a matter source, which
helps us to put constraints on the ``beyond Horndeski'' parameters from small-scale
astrophysical observations as well. For example, one obtains the
constraint from stellar physics on one of the ``beyond Horndeski'' parameters as $-7.2\times10^{-3}\leq\Xi_1\leq 4.8\times10^{-3}$~\cite{Saltas:2019ius}, where $\Xi_1$ parameterised
the deviation from Newton's law inside a stellar body (see Eq.~\eqref{eq: mod Phi}). 
In the above-mentioned particular subset of the DHOST theory evading graviton decay
and the constraint on the speed of gravitational waves,
the Vainshtein mechanism and its partial breaking operate in a very different way
as compared with the case of the generic DHOST theory, and in this case
the Hulse-Taylor pulsar gives some constraints on the theory~\cite{Hirano:2019scf,Crisostomi:2019yfo}.
As in the Horndeski theory, a large parameter space still survives and further studies are awaited.

\item {\it Maturity and Calculability.} 
\indent Linear cosmological perturbations have been formulated~\cite{Langlois:2017mxy}. Currently, {\tt CMB2nd} is the only available Boltzmann code for the DHOST theory. The matter bispectrum has been computed~\cite{Hirano:2020dom}, while the one-loop correction of the matter bispectrum suffers from the divergent behaviour in the UV limit. No full non-linear simulations like $N$-body simulations have been developed so far. 
\end{itemize}

\textbf{Massive gravity/bigravity}
\begin{itemize}
\item {\it Physical motivations.} 
\indent It would be interesting to ask whether the graviton is massless or massive, or whether
the graviton \textit{can be} massive.
In the context of dark energy, massive gravity is also attractive because it is technically natural to set the graviton mass to a small value, that is, a small mass remains small under quantum corrections. One of the cosmological constant problems (``why it is so small'') may be answered if the cosmological constant is related to the graviton mass. In addition, GR can be recovered on small scales via the Vainshtein mechanism. However, the cosmological solutions in the simplest ghost-free massive gravity theory, i.e. the dRGT theory, suffer from pathological instabilities, and hence extensions of the theory are required.

\item {\it Validity and Appealing features.}
\indent  Two main modifications in massive gravity are the appearance of the Yukawa force and the modified dispersion relation of gravitational waves due to the mass term (see~\cite{Murata:2014nra,deRham:2016nuf} for
a review about the constraints on the graviton mass). Assuming the linear theory with the single massive graviton, the solar-system experiments give an upper bound $m \lesssim 10^{-23}$~eV~\cite{Bernus:2020szc}. The gravitational-wave observations also provide the comparable bound $m \lesssim 10^{-23}$~eV \cite{LIGOScientific:2020tif}. Note that they are completely independent constraints: the former is the constraint on the Yukawa force, while the latter is the constraint on the modified dispersion relation. Yet, these constraints are far below the cosmologically-motivated graviton mass ($m\sim H_0 \simeq 10^{-33}$~eV), and
cosmological observations are required to probe such a small mass range.
However, cosmological scenarios depend strongly on the models of massive gravity, and the constraints depend on the models accordingly. 

\item {\it Maturity and Calculability.} 
\indent 
The linear perturbation analysis has been studied for each model. 
In the case of the Lorentz-invariant massive gravity such as the generalised dRGT theory, projected massive gravity, and bigravity, the linear perturbation theory breaks down in the early stage of the universe, making computations difficult (see \cite{Hogas:2021fmr, Hogas:2021lns, Caravano:2021aum} for recent developments in bigravity.)
On the other hand, the minimal theory of massive gravity can avoid the theoretical difficulties thanks to the absence of the scalar mode of the graviton and linear calculations~\cite{Bolis:2018vzs,DeFelice:2016ufg,deAraujo:2021cnd,DeFelice:2021trp} as well as a non-linear calculation~\cite{Hagala:2020eax} have been developed.
\end{itemize}

\textbf{Vector-tensor theories}
\begin{itemize}
\item {\it Physical motivations.}
\indent Having developed theories of modified gravity by adding/modifying a spin-0 field and a spin-2 field, it would be natural to explore the possibility of modifying gravity in terms of a spin-1 field. Vector-tensor theories have been developed in close analogy to scalar-tensor theories and from the bottom-up direction. Although their UV origin is still missing, vector-tensor theories might be related to spontaneous Lorentz-symmetry breaking because the vector field is often supposed to have a timelike expectation value. 
Stable self-accelerating cosmological solutions exist and the Vainshtein screening mechanism
can work~\cite{DeFelice:2016cri}, analogously to the scalar-tensor theories.


\item {\it Validity and Appealing features.}
\indent There is a surviving parameter space in the vector-tensor theories even if the speed of the gravitational waves is assumed to coincide with that of light. 
Notably, the vectorial nature can prevent the enhancement of the gravitational coupling $G_{\rm matter}$ in contrast to
the scalar-tensor theories~\cite{DeFelice:2016uil,deFelice:2017paw,Aoki:2021wew}. Thanks to this effect, the observational constraints from the ISW-galaxy correlation data can be evaded~\cite{Nakamura:2018oyy}. 
It would be interesting if there exist unique observational signatures to distinguish the vector-tensor theories from GR and the scalar-tensor theories.

\item {\it Maturity and Calculability.}
\indent Linear perturbations have been studied only in some specific models of vector-tensor theories~\cite{Nakamura:2018oyy,DeFelice:2020icf,DeFelice:2020sdq},
and a unified approach for perturbations of generic vector-tensor theories has not been well developed so far.
The EFT approach~\cite{Aoki:2021wew} is one of the promising ways in
this direction. A Boltzmann solver based on {\tt CMB2nd} is currently
under development. Further studies are needed to go beyond the linear-order perturbations.
\end{itemize}

\textbf{Metric-affine gravity}
\begin{itemize}
\item {\it Physical motivations.}
\indent Metric-affine gravity and its subclasses are mainly motivated by theoretical considerations and the underlying idea would be interesting in light of the successes of the geometric interpretation of gravity and the gauge principle in particle physics. As long as the higher curvature terms are ignored, metric-affine gravity can be recast in other classes of gravity in the Riemannian geometry, though matter couplings may be modified. 
A dynamical connection might give rise to interesting phenomena and some late-time cosmological models have been proposed~\cite{Nikiforova:2016ngy,Nikiforova:2018pdk,Barker:2020gcp}.

\item {\it Validity and appealing features.}
\indent 
Metric-affine gravity is often regarded as a UV modification of gravity, while the validity of metric-affine gravity as an IR modification of gravity has not been well investigated. It would be interesting to ask how the connection affects gravitational interactions and establish observational constraints on possible deviations from the Riemannian geometry. 

\item {\it Maturity and Calculability.}
\indent One can effectively analyse a class of metric-affine gravity by recasting the model into another class of modified gravity. As for a general metric-affine gravity, although cosmological solutions have been obtained at the background level in certain models, cosmological perturbations have not been well developed even at linear order due to technical difficulty in treating a large number of independent components of the connection.
\end{itemize}

\textbf{Cuscuton/Minimally-modified gravity}
\begin{itemize}
\item {\it Physical motivations.}
\indent Cuscuton and minimally modified gravity provide yet another way to modify gravity without introducing an additional degree of freedom. Lorentz-symmetry breaking is required according to the Lovelock theorem and a violation of a locality condition is also required to evade the uniqueness of the massless spin-2 field indicated by the scattering amplitude arguments~\cite{Benincasa:2007qj, Arkani-Hamed:2008bsc, Carballo-Rubio:2018bmu, Pajer:2020wnj}. They are achieved by a non-dynamical scalar field $\phi$ obeying an elliptic (instead of hyperbolic) equation. On the phenomenological side, various cosmological scenarios are possible without worrying about pathological instabilities and a strong coupling because of the absence of an additional dynamical degree of freedom.

\item {\it Validity and Appealing features.}
\indent These models can work as an alternative for dark energy. A certain model of cuscuton~\cite{Iyonaga:2021yfv} exhibits the same predictions as GR in weak gravitational fields and the propagation speed and is even identical to the $\Lambda$CDM model at the background level. Even so, the growth of perturbations on cosmological scales is modified by a single parameter $\beta$ (see Eqs.~\eqref{2DoF1}--\eqref{2DoF3}) and the CMB constraints read $\beta = -0.0388^{+0.011}_{-0.0083}$, indicating a $\sim 4\sigma$ deviation from GR $(\beta=0)$~\cite{Hiramatsu:2022ahs}. In addition, the authors in \cite{DeFelice:2020cpt} have claimed that the VCDM model, one of the minimally modified gravity theories, can resolve the $H_0$ tension.

\item {\it Maturity and Calculability.}
\indent Cosmological perturbations at linear level have been given in
\cite{Iyonaga:2018vnu,Aoki:2020oqc,DeFelice:2020cpt,DeFelice:2020prd,Iyonaga:2021yfv,Bartolo:2021wpt,Hiramatsu:2022ahs}, and the CMB constraints have been obtained by implementing a model in a Boltzmann solver~\cite{Aoki:2020oqc,DeFelice:2020cpt,Hiramatsu:2022ahs}. 
Non-linear calculations including the development of $N$-body code have not been done yet. 
\end{itemize}

Let us conclude this section. 
The Horndeski theory is one of the most well-developed theories of modified gravity.
Deviations from GR have been already tested at the level of linear perturbations.
Further studies about non-linear structure formation will enable more precise tests of the Horndeski theory with forthcoming observations. The DHOST theory is an interesting extension of the Horndeski theory as there are several features distinct from GR and the Horndeski theory. Although the linear perturbation theory has been investigated, studies about non-linear structure formation need to be developed. Theories of massive gravity are well-motivated candidates to explain the present cosmic acceleration as the small graviton mass is protected by quantum corrections.
Healthy cosmological models have been explored by extending the dRGT theory. 
The vector-tensor theories are analogous to the scalar-tensor theories, but they provide a different avenue for modifying gravity. The surviving parameter space would be wider than the scalar-tensor theories thanks to the presence of intrinsic vector modes that potentially prevents the enhancement of the gravitational coupling $G_{\rm matter}$. Theoretical and numerical tools for perturbative calculations are under development in the framework of EFT. Studies in this direction are expected to reveal distinct features of the vector-tensor theories. Metric-affine gravity modifies the geometry itself and there may be interesting features that are not seen in the other theories. Since metric-affine gravity is typically thought of as a UV modification of gravity, its consequence on an IR modification of gravity has been less explored.
The cuscuton/minimally modified gravity evades the uniqueness of the theory of the massless spin-2 field by introducing a non-dynamical field, yielding a rich phenomenology. It is possible to modify gravity only on cosmological scales and the modification is testable using cosmological observations.

All in all, cosmological probes of gravitational theories are essential to understanding the present cosmic acceleration as well as the nature of gravity. Although strong constraints on modifications of gravity have been obtained, there is a large number of viable theories and it would be remarkable that some theories are preferred over the $\Lambda$CDM cosmology at least for certain observations. The constraints will be improved by forthcoming observations and related theoretical developments are also indispensable to achieve it. In particular, aspects of non-linear structure formation have not been well studied even in the Horndeski theory, one of the most maturated theories of modified gravity. It would be interesting and important to study theories of modified gravity from both theoretical and observational viewpoints to connect the
recent theoretical developments to forthcoming observations.

\clearpage

\section{Summary}

In this paper, we 
have provided a comprehensive review of cosmological probes of gravity, to prioritise a subset of well-motivated gravity models along with rationales for developing a strategy for forthcoming cosmological observations.
We first discussed the theoretical aspects of modified gravity theories. 
We then have described observational aspects for testing gravity.
In particular, we presented cosmological observables of the cosmic microwave background and large-scale structure, and we have introduced phenomenological parametrisations commonly used for testing gravity at cosmological scales.
In the subsequent section, 
we have shown concrete analytical predictions of cosmological observables from well-motivated theories such as scalar-tensor theories, massive gravity theories, vector-tensor theories, 
as well as commenting on other theories. 
As numerical tools, we have introduced {\tt CMB2nd}, which is the only available CMB Boltzmann code for DHOST theory, 
and \texttt{DarkEmulator}, 
which is the emulator code providing theoretical templates for large-scale structure observations. 
Finally, we have provided an outlook of cosmological probes of gravity in terms of current studies of each gravity theory and identified a selection of well-motivated gravity models for developing an observational strategy.
We have concluded that the Horndeski theory is one of the most well-developed theories of modified gravity and the highest-priority model that we can focus on when discussing the observational strategy.
However, even in the Horndeski theories as the most maturated theories, there are remaining issues such as a deeper theoretical understanding, theoretical predictions of the non-linear structure formation, breaking the degeneracy between modifications of gravity and other physical effects, developments of efficient computational tools, etc.
Overall, other theories would require more studies even at linear orders in perturbations to be tested by precise cosmological observations.
To solve such issues, further collaborations among theoretical developments, forthcoming observations, and computational implementations as mentioned in this paper are needed.
We expect that testing gravity theory on cosmological scales makes progress further from the strong connection between theory and observations. 


\section*{Acknowledgment}

We would thank all the participants for monthly meetings to support the activity of Testing Gravity THxOBS working group.\footnote{\url{https://sites.google.com/view/testing-gravity-thxobs-japan/}}
This work was supported by JSPS KAKENHI Program, Japan Science and Technology Agency (JST) FOREST Program and AIP Acceleration Research Grant; JPMJFR2137 (S.A. and H.M.); JP20K14468 (K.Aoki); JP21K03585, JP19H00674, JP18K13558, JP18H05539 (Y.C.); JP22K03605 (R.Kimura); JP20K03936, JP21H05182, JP21H05189, JP21F21019 (T.Kobayashi); JP20H01932, JP21H05456, JP21H00070 (H.M.), JP20317829 (H.M. T.Nishimichi, and A.T.); JP19H01891, JP22K03627 (D.Y.); JP20H01932, JP20K03968 (S.Y.); JSPS Overseas Research Fellowships (K.Akitsu); JP21K03559 (T.H.); JP21H01080 (S.H.); JP20K14471(R.Kase);
the National Key R$\&$D Program of China (2021YFA0718500) (T.Katsuragawa); JP20H05859, JP22K03682 (T.Namikawa); JP19H00677, JP22K03634 (T.Nishimichi), JP20H05861 (T.Nishimichi and M.Shirasaki,A.T.), JP21H01081 (T.Nishimichi and A.T.); JP19K14718, JP20H05859 (M.Shiraishi); JP19K14767 (M.Shirasaki); JP20J01600, JP20H05855 (S.T.); JP21J00695 (K.T.); JP20J00912, JP21K13922, IBS under the project code, IBS-R018-D1(J.T.). Y.K. is supported in part by the David and Lucile Packard foundation. This work is also supported by the Ministry of Science and Technology of Taiwan under Grants Nos. MOST 110-2112-M-001-045- and 111-2112-M-001-061- and the Career Development Award, Academia Sinina (AS-CDA-108-M02) (T.O.).

\let\doi\relax
\bibliographystyle{ptephy}
\bibliography{bibthxobs} 

\end{document}